%% file: main.tex
\documentclass{acmtog}

\usepackage{graphicx}   
\usepackage{epsfig}

\usepackage{amsmath}
\usepackage{amssymb}
\usepackage{amsfonts}
\usepackage{fixmath}
\usepackage{graphicx}
\usepackage{algorithm}
\usepackage{algorithmic}
\usepackage{parskip}
\usepackage{color}
\usepackage{subfigure}

\def\al{et al. }
\def\ie{i.e. }

\def\mov{\frak{A}}
\def\fix{\frak{B}}
\def\q{\mathbf{q}}

\def\PD{\mathrm{PD}(\mov,\fix)}

\acmVolume{VV} \acmNumber{N} \acmYear{YYYY} \acmMonth{Month}
\acmArticleNum{XXX} \acmdoi{10.1145/XXXXXXX.YYYYYYY}

\begin{document}

\markboth{C. Je et al.}{PolyDepth: Real-Time
Penetration Depth Computation using Iterative Contact-Space
Projection}

\title{PolyDepth: Real-time Penetration Depth Computation using Iterative Contact-Space Projection}

\author{Changsoo Je
\affil{Ewha Womans University and Sogang University}
\and 
Min Tang, Youngeun Lee, Minkyoung Lee  and  Young J. Kim
\affil{Ewha Womans University}}







\maketitle

\begin{bottomstuff}
Email: vision@sogang.ac.kr (C. Je)\\
\{tangmin,kimy\}@ewha.ac.kr (M. Tang, Y. J. Kim)\\
\{youngeunlee,minkyounglee\}@ewhain.net (Y. Lee, M. Lee).

Authors' addresses: C. Je, Ewha Womans University, Seoul, Korea and
Sogang University, Seoul, Korea; M. Tang, Y. Lee, M. Lee, and Y. J. Kim
(corresponding author), Ewha Womans University, Seoul, Korea.
\end{bottomstuff}

\begin{abstract}
We present a real-time algorithm that finds the penetration depth
(PD) between general polygonal models
based on iterative and local optimization techniques. Given an
in-collision configuration of an object in configuration space, we
find an initial collision-free configuration using several methods
such as centroid difference, maximally clear configuration, motion
coherence, random configuration, and sampling-based search. We
project this configuration on to a local contact space using a
variant of continuous collision detection algorithm and construct
a linear convex cone around the projected configuration. We then
formulate a new projection of  the in-collision configuration on
to the convex cone  as a linear complementarity problem (LCP),
which we solve using a type of Gauss-Seidel iterative algorithm.
We repeat this procedure until a locally optimal PD is obtained.
Our algorithm can process complicated models consisting of tens of
thousands triangles at interactive rates.\end{abstract}




\input{intro}
\input{prev}
\input{overview}
\input{ccd}
\input{iteration}

\input{results}
\input{concl}

\begin{acks}
This research work was supported in part by the NRF grant funded
by the Korea government (MEST) (No. 2009-0086684) and IT R\&D
program of MKE/MCST/KOCCA (2008-F-033-02).
\end{acks}

\bibliographystyle{acmtog}
\bibliography{manocha,geom,robotics,mp,PD,add,Trans}
\end{document}

%% file: intro.tex
\section{Introduction}



Measuring the distance between geometric objects is an important
problem in computer graphics, virtual reality, geometric modeling,
computational geometry, CAD/CAM and robotics \cite{lm03p}. When
objects are disjoint, the Euclidean distance between their closest
points, also known as the separation distance, is an obvious
measure of distance. However, when objects overlap, the separation
distance is undefined and a different  measure is needed to
quantify the amount of interpenetration. Different penetration
measures have been introduced, including penetration depth
\cite{cc-dmtdb-86}, generalized penetration depth \cite{gpd},
pointwise penetration depth \cite{tang09sig}, growth distance
\cite{Ong93phd}, and penetration volume \cite{Weller-RSS-09}.


Penetration depth (PD) has been widely used by the computational
geometry community. It is the distance that corresponds to the
shortest translation required to separate two overlapping, rigid
objects \cite{dhks-cidp-93}.  Many applications can benefit from
PD computations. In physically-based animation, the PD can be used
to locate the point of application for impulses
\cite{Guendelmann03sig}, or used to find the time of contact in
time-stepping methods \cite{YoungKim02c}.
In constraint-based dynamics, when
penetration is unavoidable due to numerical errors, PD can be used
to roll back from an invalid state (penetration) to a valid one
(non-penetration) \cite{redon2004}. Moreover, PD can be used for
post-stabilization such as the Baumgarte stabilization in
rigid-body dynamics to enforce non-penetration constraints. In
virtual prototyping, the PD can be used to verify tolerances
\cite{youngkim03}. The length of a PD can be used to determine the
appropriate force feedback in six-degree-of-freedom haptic systems
\cite{KOLM03}. Retraction-based motion planning algorithms can use
PD to find a collision-free sample in the narrow passage amongst
obstacles \cite{Zhang08-ICRA}, and PDs can be used to determine
the existence of a path in planning scenarios \cite{Zhang08-IJRR}.
%

However, it is well-known that much more computation is generally
required to determine a PD than a separation distance. For general
polyhedral objects, with a total of $n$ faces, the computation
time is $O(n^6)$ \cite{Youngkim04}. This has motivated some
researchers to find a more tractable approximation of the PD
\cite{youngkim03,YoungKim02c,RedonLin06}. Unfortunately, these
methods are either too slow for interactive applications
\cite{youngkim03,YoungKim02c} or provide no guaranteed error bound
\cite{RedonLin06}.

{\bf Main Contributions:} We present a real-time algorithm to
approximate the PD between arbitrary, polygon-soup models.  Our
algorithm actually determines an upper bound on the PD and we show
empirically that this is a tight bound on the exact value obtained
from Minkowski sums. We also show how to obtain a set of
\emph{local} PD values from the PD solution. These local PDs
characterize the amount of local overlap in each of several
interpenetrating regions.

Our algorithm is based on iterative and local optimization
techniques. Given an in-collision configuration for an object, we
find an initial collision-free configuration in configuration
space. Then, we project this initial configuration on to a contact
space using a variant of a continuous collision detection
algorithm and construct a linear convex cone around the projected
configuration. We then formulate the projection of the
in-collision configuration onto this cone as a linear
complementarity problem (LCP), which we solve using a Gauss-Seidel
iteration. This runs until a locally optimal solution is obtained.

Because ours is an iterative algorithm, finding a good initial
configuration is crucial. We therefore propose several search
techniques to find the configuration from geometric properties
such as the distance between centroids, maximally clear
configuration, sampling-based search or random search, as well as
exploiting application-dependent information such as motion
coherence.

We have implemented our algorithm, \emph{PolyDepth}, and
benchmarked its performance in various complicated scenarios, such
as random and pre-calculated configurations, and configurations
governed by rigid-body dynamics. In all these scenarios, our
algorithm achieves a highly interactive performance while
providing a tight estimate of the PD value.

{\bf Organization:} The rest of the paper is organized as follows.
In Sec.2, we briefly survey topics relevant to PD computations. We
provide some preliminary information needed to understand our
algorithm in Sec. 3, and give an overview of the algorithm. In
Secs. 4 and 6 we explain the two central techniques of our
algorithm, out- and in-projection as well as local optimization
techniques. In Sec. 5 we explain how we find a good initial
collision-free configuration. We present our experimental results,
analyze the performance of PolyDepth in different benchmarks, and
discuss implementation issues in Sec. 7. We conclude the paper in
Sec. 8.

%% file: prev.tex
\section{Previous Work}

There are several different types of PD algorithms for different
model geometries and objectives.

{\bf Convex Polytopes:} a straightforward algorithm was presented
to compute the PD between two convex polytopes by computing their
Minkowski sums \cite{cc-dmtdb-86}. If the direction of motion is
known, a minimal translational motion in this direction can be
found using a multi-resolution mesh hierarchy \cite{dhks-cidp-93}.
A randomized algorithm was presented by
\cite{Agarwal00penetrationdepth}. These algorithms provide exact
solutions, but they are difficult to implement; in fact, no good
implementations are known. However, various approximate algorithms
have been developed based on upper and lower bounds on the PD
\cite{Cameron97}, expansion of polyhedral approximations
\cite{Gino01}, and dual-space expansion \cite{Youngkim04}. The
convex PD problem requires only $O(n^2)$ time in the worst case,
where $n$ is the total number of faces in the polytope.

{\bf Non-convex Polyhedra:} the computational complexity of PD for
non-convex polyhedra is $O(n^6)$, and thus no practical exact
algorithms exist. A hierarchical refinement technique combined
with GPU-assisted ray-shooting can provide an upper bound on the
PD \cite{YoungKim02c}. Redon and Lin \cite{RedonLin06} also
proposed a CPU/GPU hybrid method of computing a lower bound on the
PD, and this algorithm is applicable to polygon-soup models. Lien \cite{Lien08,lien09} presented a sampling-based
approach to PD computation based on approximate Minkowski sum.
More recently, Hachenberger \cite{Hachenberger09} presented a
method of obtaining an exact Minkowski sum based on convex
decomposition. However, these approaches are rather slow, and some
require an explicit boundary representation of Minkowski sums that
has to be re-evaluated whenever the orientation of an object is
changed.

A somewhat different route is to use distance fields that has the
advantage of being applicable to deformable models \cite{FL01b}.
Moreover, a GPU can be used to accelerate the computation of
distance fields \cite{Hoff02,sud2006fpc}. However, these
approaches only provide a lower bound on a PD. There is also a
variant of PD called a pointwise PD which is defined as the
distance between the points of deepest interpenetration between
two objects. Pointwise PDs can be found using Hausdorff distance
\cite{tang09sig}. However, a pointwise PD is merely a lower bound
on the PD. A lower bound on a PD cannot be used to achieve
separation between overlapping objects, but an upper bound can.
More recently, a penetration resolution technique based on volume
has been presented by \cite{Allard2010}. This method uses GPUs and
can handle deformable objects.

{\bf Generalized Penetration Depth:} the constraints of some
applications mean that a translation is not sufficient to separate
intersecting objects, and thus does not provide a useful measure
of inter-penetration. In these cases, more complicated motions
need to be considered. Zhang \al proposed a generalized
penetration depth, which is the minimal combination of
translational and rotational motion needed to separate two objects
\cite{gpd,fap}. More recently, kinematical geometry has been used
to compute generalized PDs \cite{Nawratil2009}. Non-Euclidean
distance, such as growth distance \cite{OG96,Ong93phd}, can also
be used as a measure of inter-penetration. Finally, some
researchers have investigated penetration volumes instead of PD
\cite{Weller-RSS-09}, but the relation between this and the PD is
questionable.

%% file: overview.tex
\section{Preliminaries}

We will now define the problem of penetration depth computation
between general polygonal models and give an overview of our
approach.

\begin{figure*}[tbp]
\centering \epsfig{file=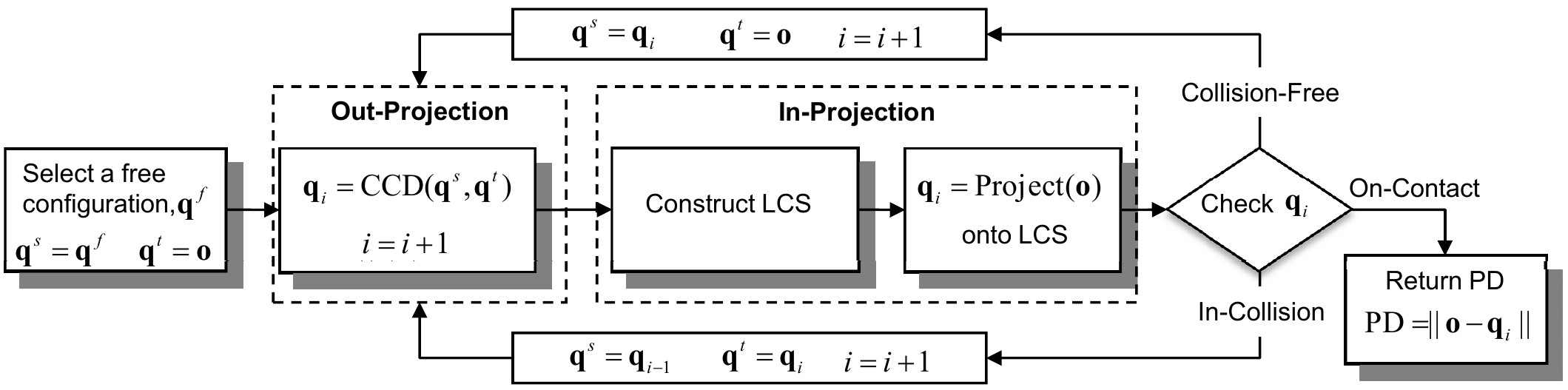,width=16cm} \caption
{{\bf PD computation pipeline.}}\label{fig:pipeline}
\end{figure*}

\subsection{Problem Formulation}

%

Suppose we have two objects $\mov$ and $\fix$ in $\mathbb{R}^3$.
Without loss of generality, we will assume that $\mov$ is movable
by translation and $\fix$ is stationary, and both objects have the
common global origin $\mathbf{o}$. If $\mov$ and $\fix$ are
polyhedral objects and are interpenetrated, their penetration
depth $\PD$ is formally defined as \cite{dhks-cidp-93}:
\begin{equation}\label{eq:pd1}
\PD=\min\{ \| \mathbf{d} \| \; | \;
\mathrm{interior}(\mov+\mathbf{d}){\cap}\fix=\emptyset, \forall
\mathbf{d} \in \mathbb{R}^3 \}.
\end{equation}


%

It is well known that the PD computation is closely related to the
Minkowski sum. Formally, the Minkowski sums $\mov \oplus \fix,
\mov \oplus -\fix$ between two compact sets, $\mov$ and $\fix$,
are defined \cite{b-egc-66,Cameron97} as follows:
\begin{eqnarray}
\mov \oplus
\fix=\{\mathbf{a}+\mathbf{b}|\mathbf{a}\in{\mov},\mathbf{b}\in{\fix}\}\\
\mov \oplus
-\fix=\{\mathbf{a}-\mathbf{b}|\mathbf{a}\in{\mov},\mathbf{b}\in{\fix}\}.
\end{eqnarray}
We can then reduce the problem of computing the PD between $\mov$
and $\fix$ expressed by Eq.~\ref{eq:pd1} to the search for the
minimum distance between $\mathbf{o}$ and the boundary surface of
their Minkowski sum, $\partial(\mov\oplus-\fix)$
\cite{dhks-cidp-93}. However, since the interior of the polygon-soup models that we wish to consider may not be closed and thus be undefined, we need to define our PD problem in another way.

Our definition of PD still follows the intuitive notion of using a
minimal translation to separate two overlapping models, even
though the inside and outside of polygon-soup models is not
properly defined. Thus, we define the PD to be the minimum
distance from $\mathbf{o}$ to the boundary of the Minkowski sum
$\mov\oplus-\fix$:
\begin{equation}\label{eq:pd2}
\PD=\min\{ \|\mathbf{d}\| \; | \; \mathrm{interior}(\mov \oplus
-\fix) \cap \{\mathbf{o}+\mathbf{d}\}
 = \emptyset, \forall
\mathbf{d} \in \mathbb{R}^3 \}.
\end{equation}
Essentially, the boundary of the Minkowski sum constitutes the
{\em translational} contact space
between $\mov$
and $\fix$, given that $\mov$ has a fixed orientation because it
can only undergo translational motion. If $\mov$ and $\fix$ are
polyhedra, then Eq.\ref{eq:pd1} is equivalent to Eq.\ref{eq:pd2}.

\begin{figure*}[htb] \centering
\subfigure[Local contact space and the
origin]{\epsfig{file=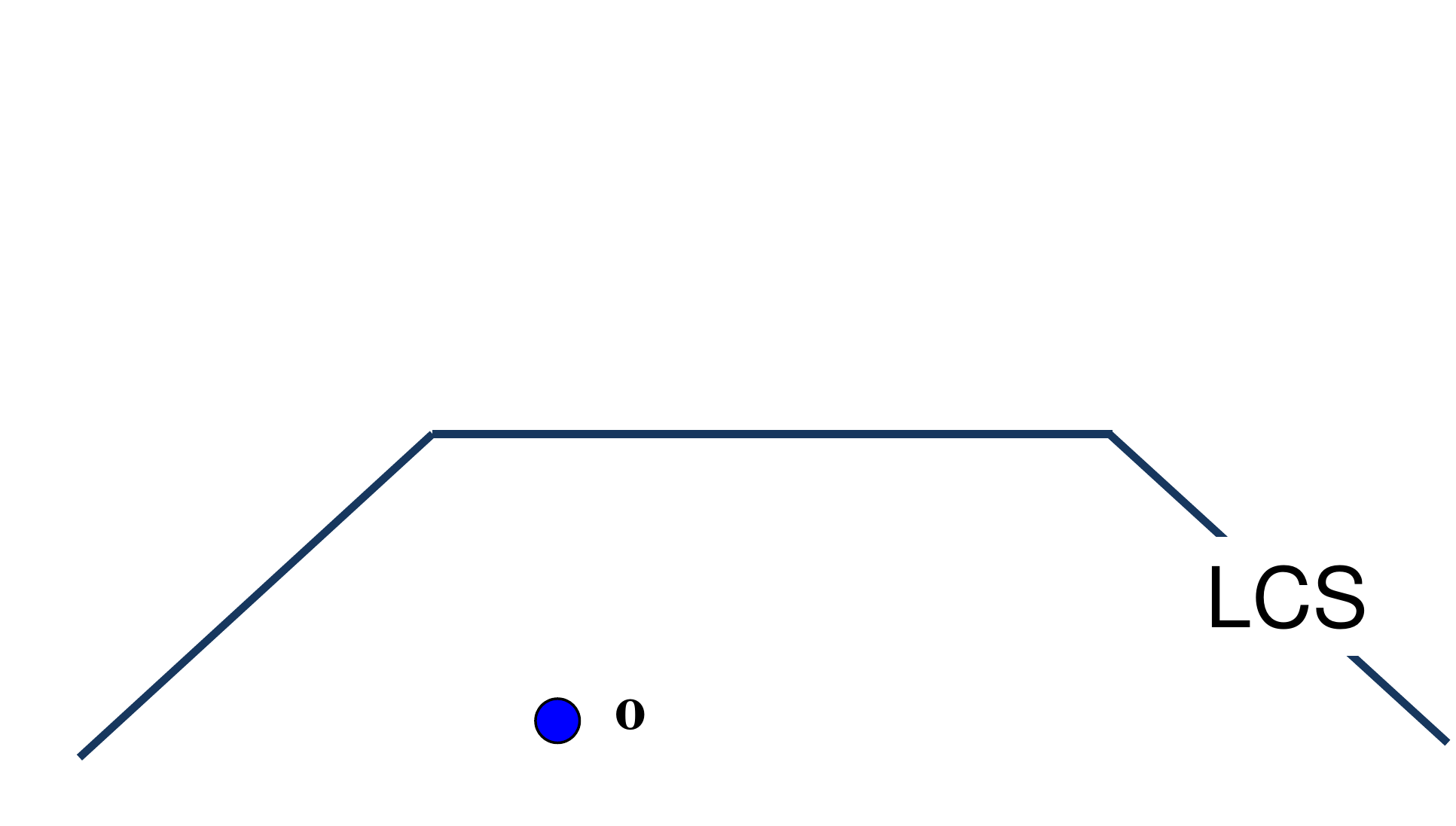,width=3.4cm}}
\subfigure[Collision-free configuration
$\mathbf{q}^f$]{\epsfig{file=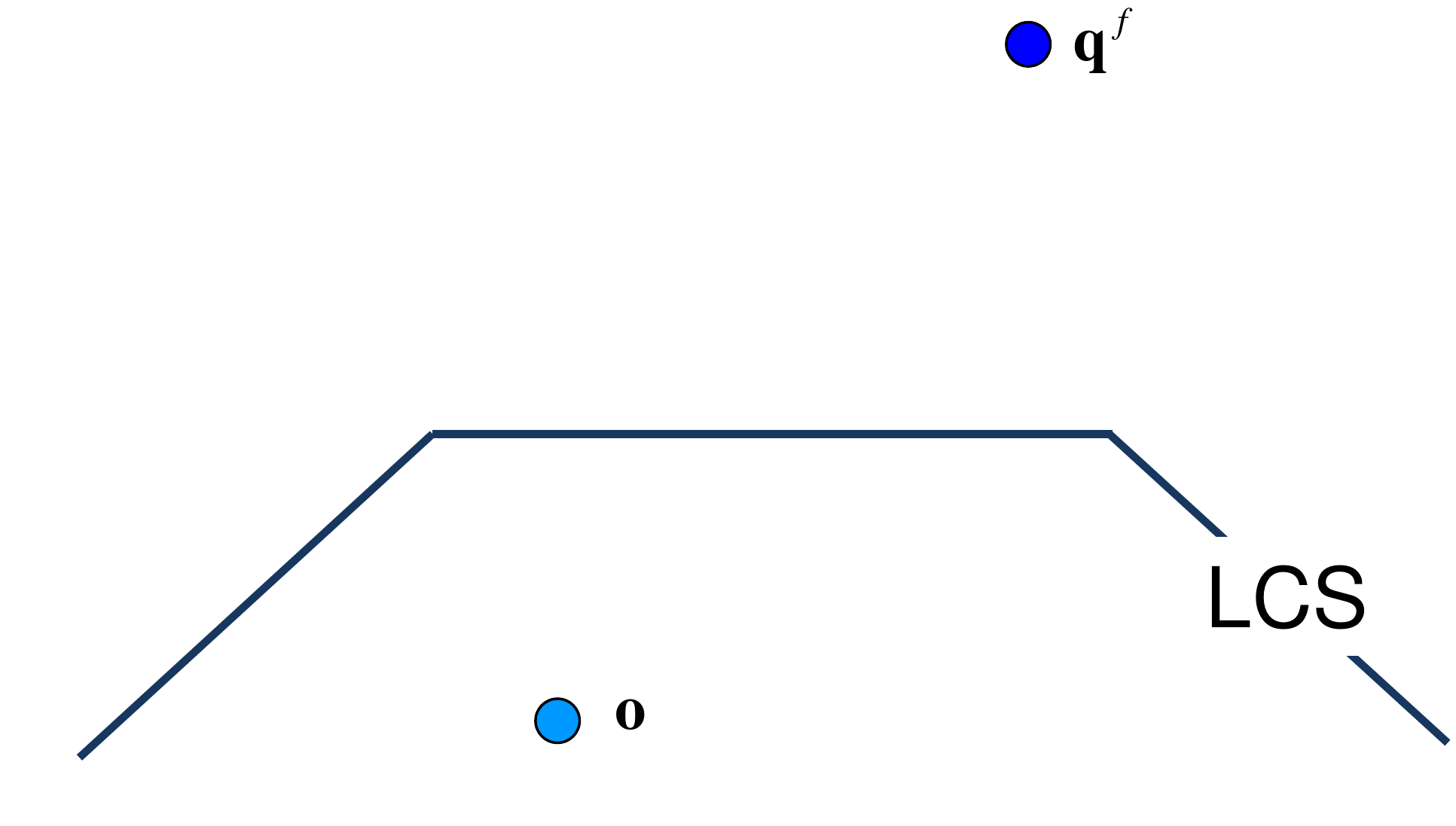,width=3.4cm}}
\subfigure[Out-projection from $\mathbf{q}^f$ to $\mathbf{o}$ to
obtain
$\mathbf{q}_0$]{\epsfig{file=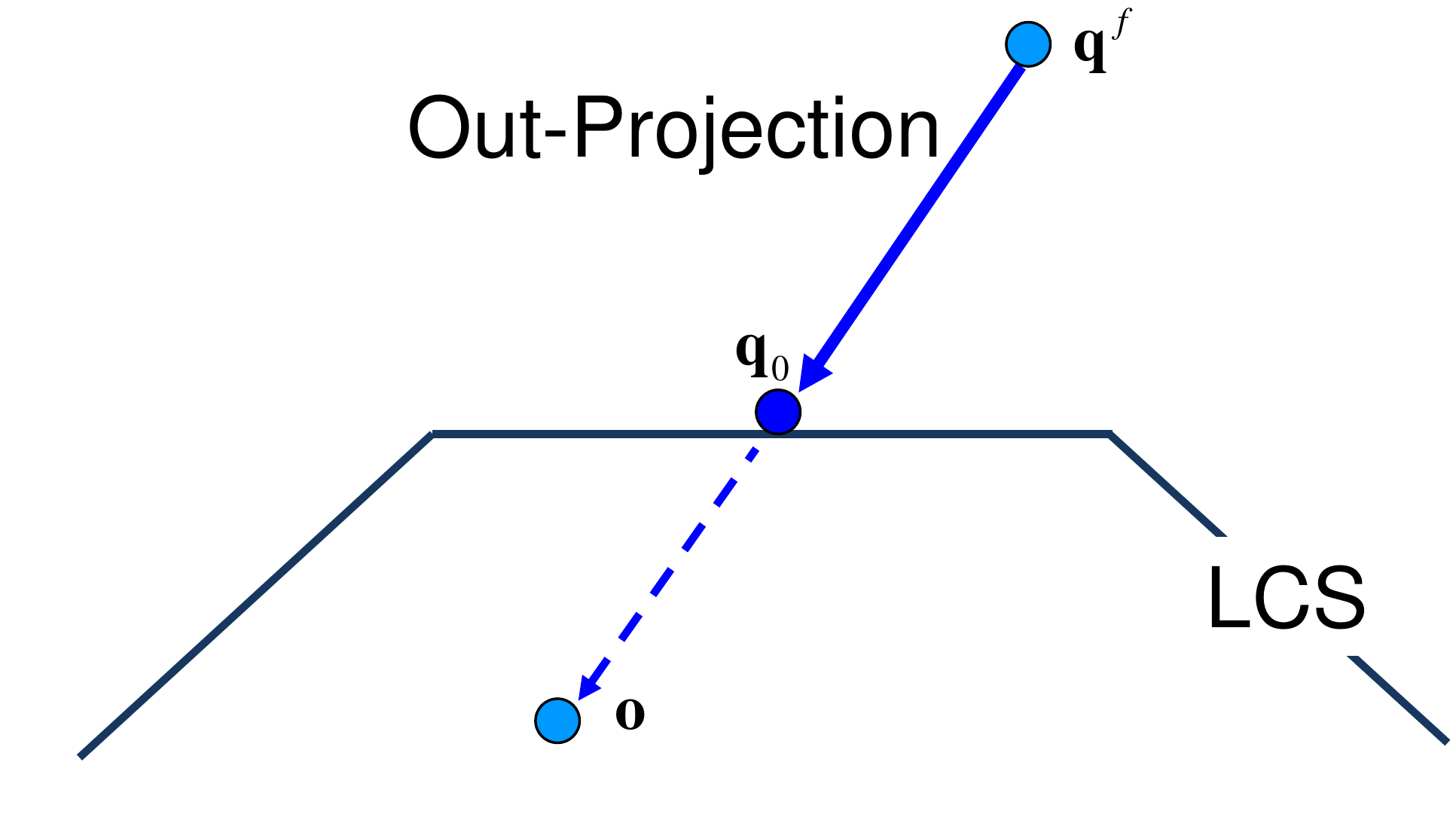,width=3.4cm}}
\subfigure[Construction of an LCS around
$\mathbf{q}_0$]{\epsfig{file=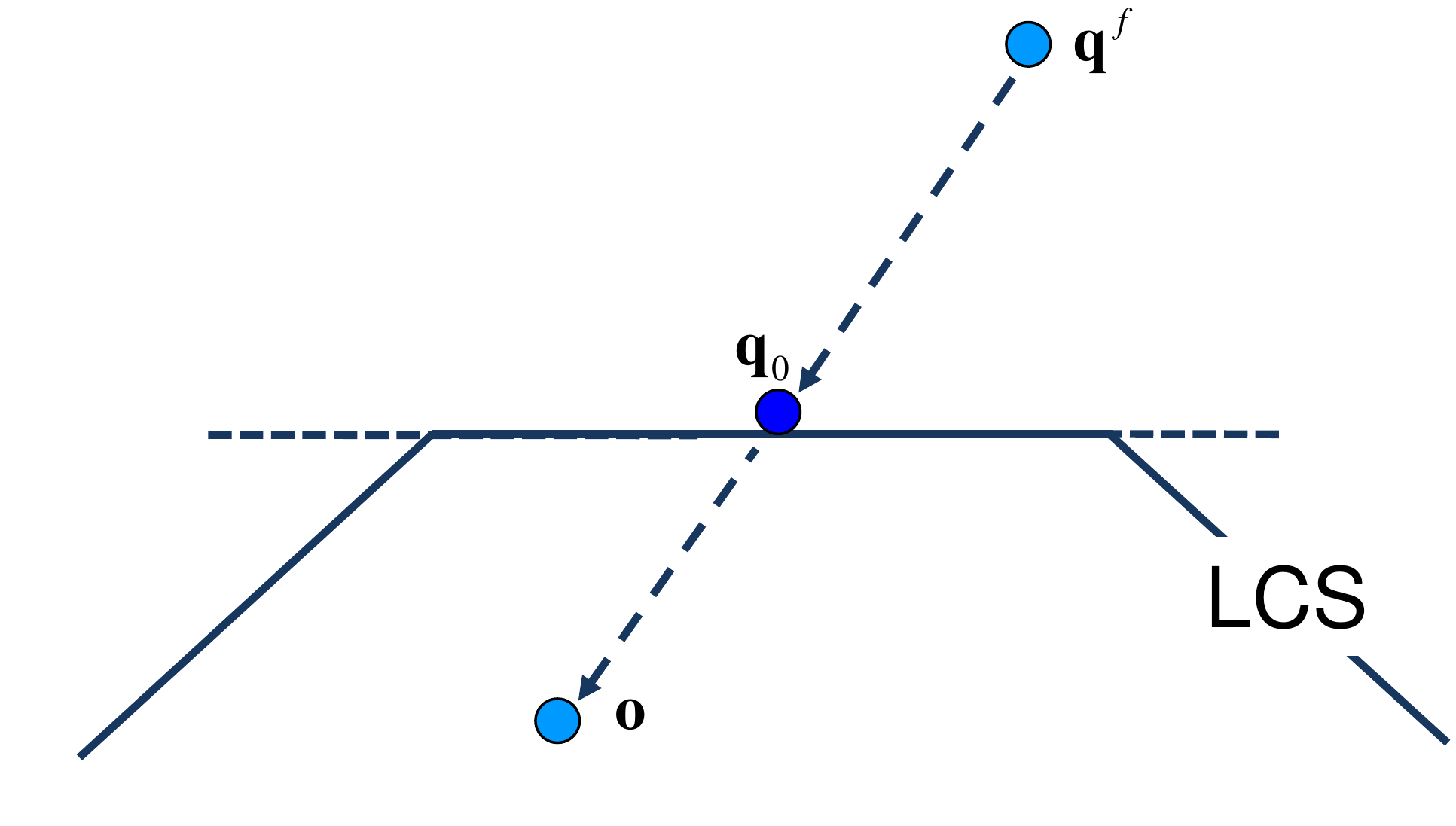,width=3.4cm}}
\subfigure[In-projection on to the LCS to obtain
$\mathbf{q}_1$]{\epsfig{file=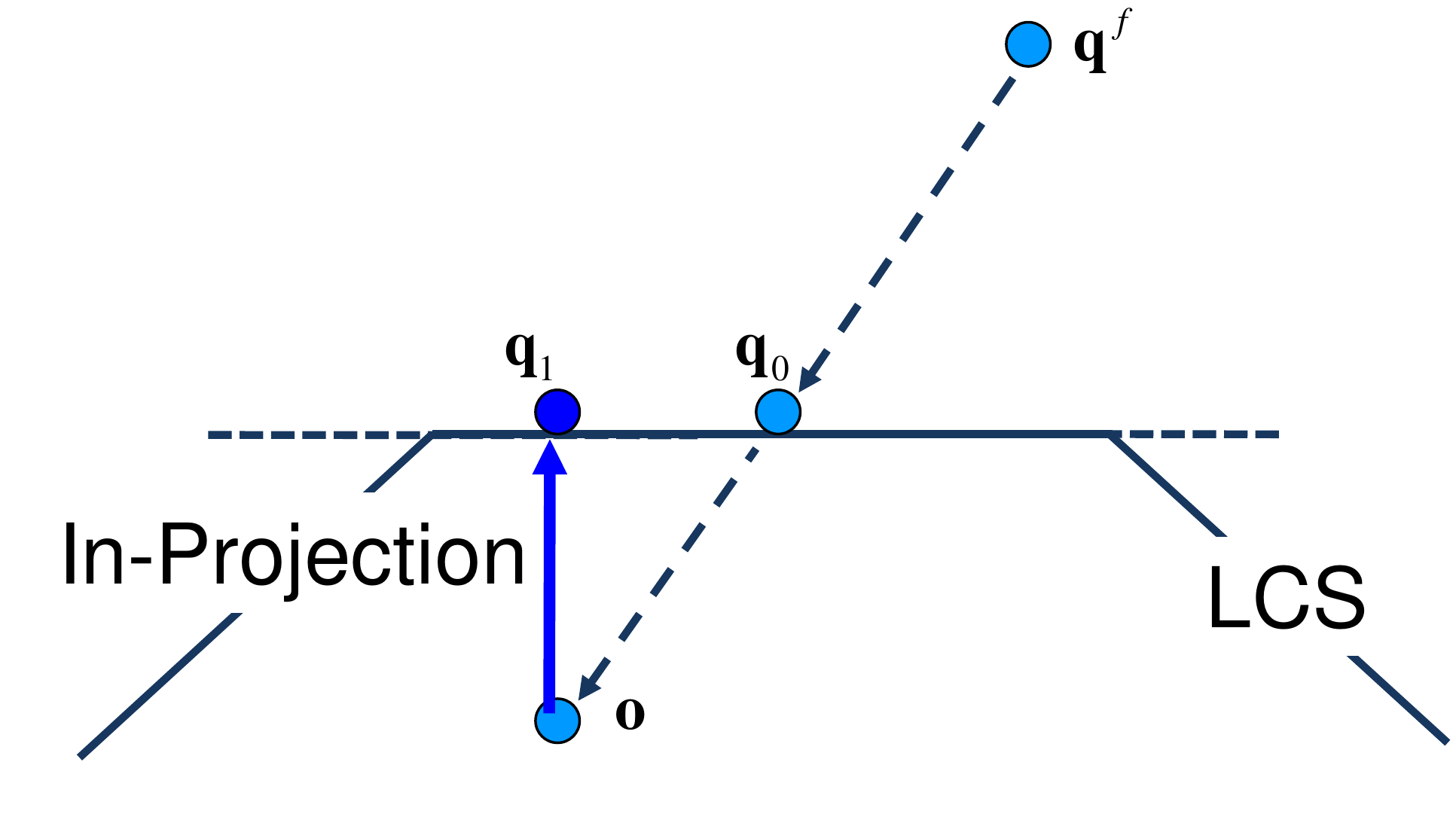,width=3.4cm}}
\caption{{\bf Iterative optimization of a PD for a simple case of
convex LCS.} In this simple case, the PD algorithm terminates
right after a single iteration of successive out- and
in-projections. Then, $PD=|| \mathbf{o} - \mathbf{q}_1
||$.}\label{fig:simple_lcs1}
\end{figure*}

\begin{figure*}[htb] \centering
\subfigure[Local contact space and the
origin]{\epsfig{file=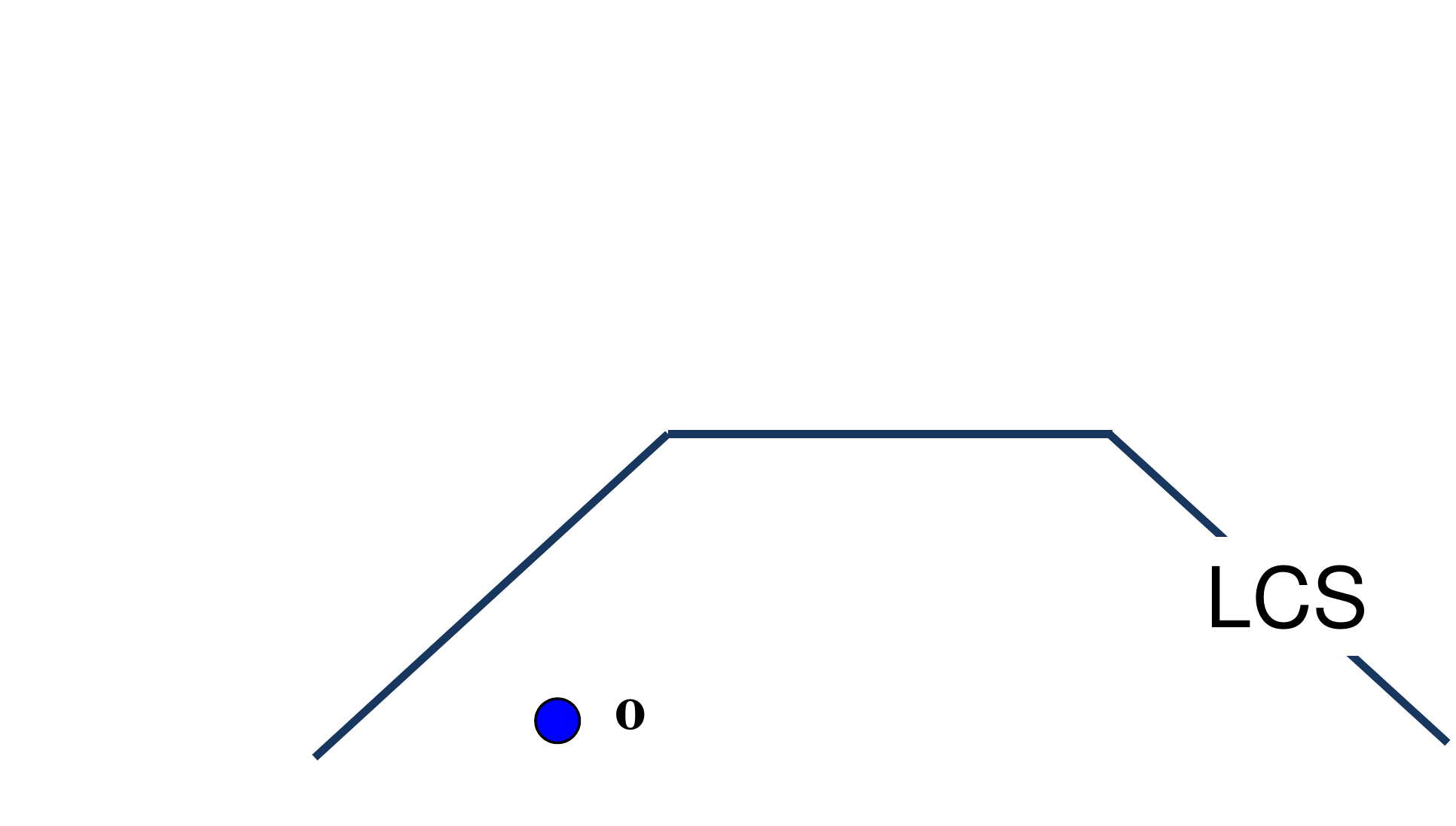,width=4.2cm}}
\subfigure[Collision-free configuration
$\mathbf{q}^f$]{\epsfig{file=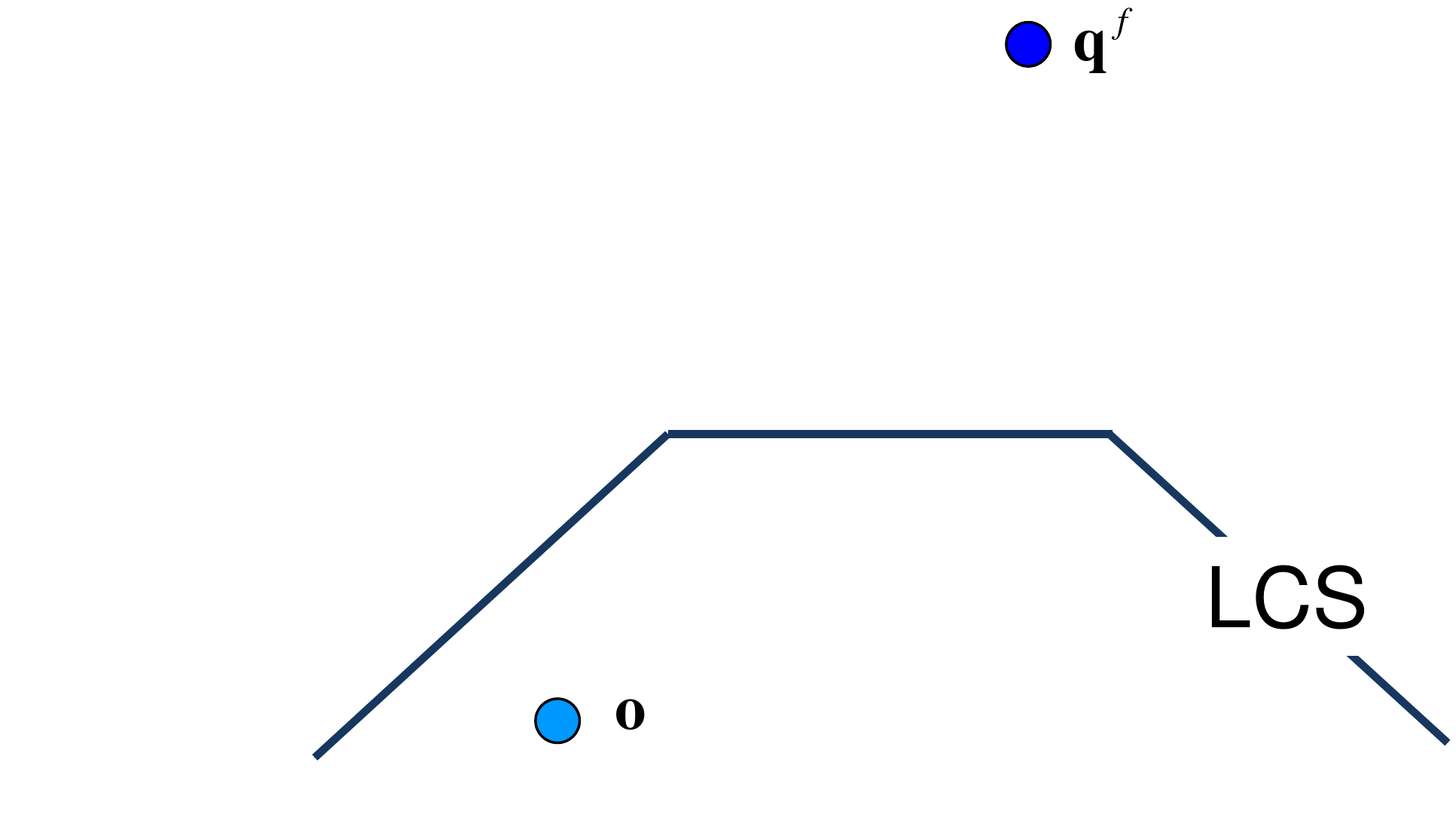,width=4.2cm}}
\subfigure[Out-projection from $\mathbf{q}^f$ to $\mathbf{o}$ to
obtain
$\mathbf{q}_0$]{\epsfig{file=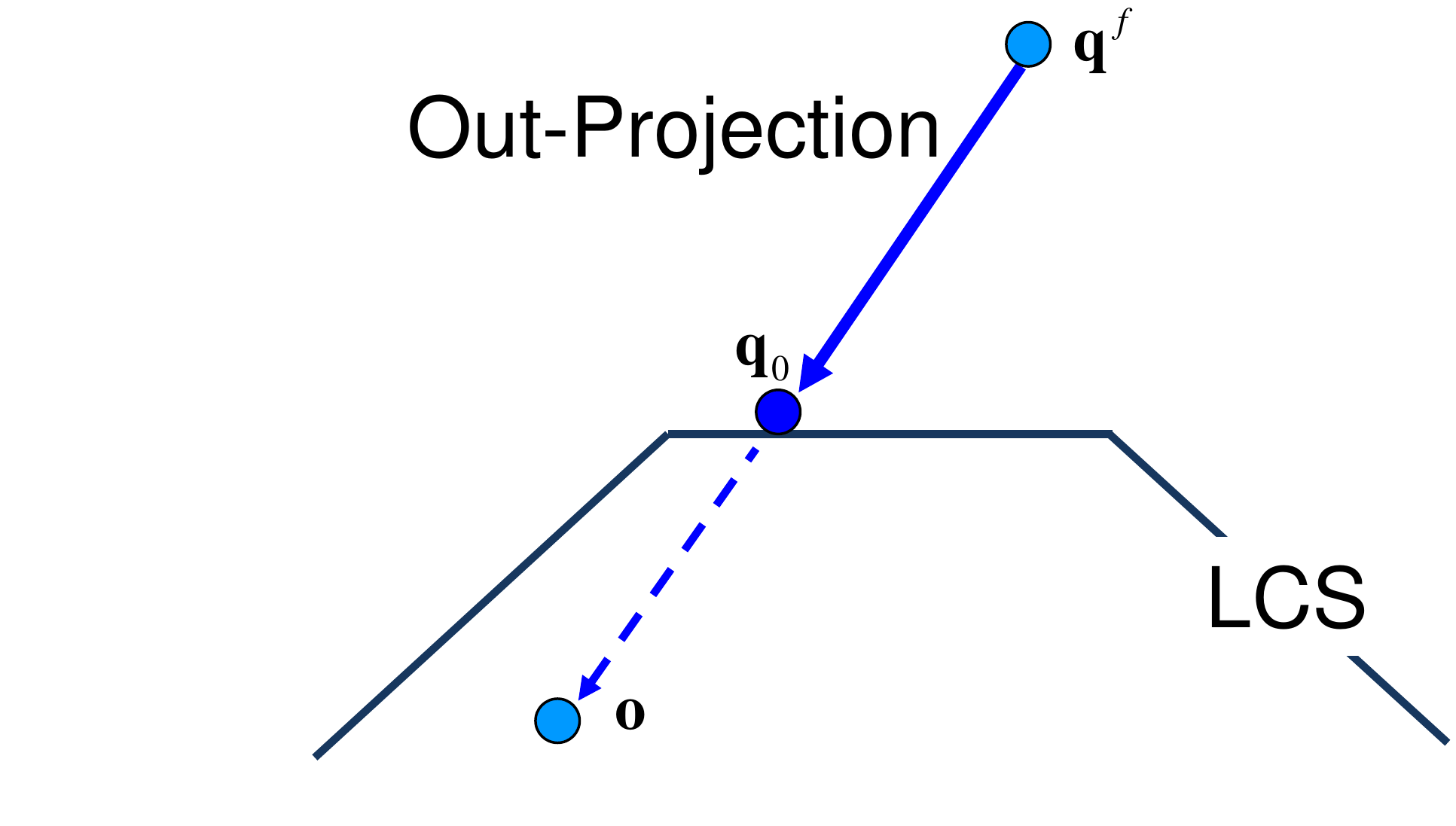,width=4.2cm}}
\subfigure[Construction of an LCS around
$\mathbf{q}_0$]{\epsfig{file=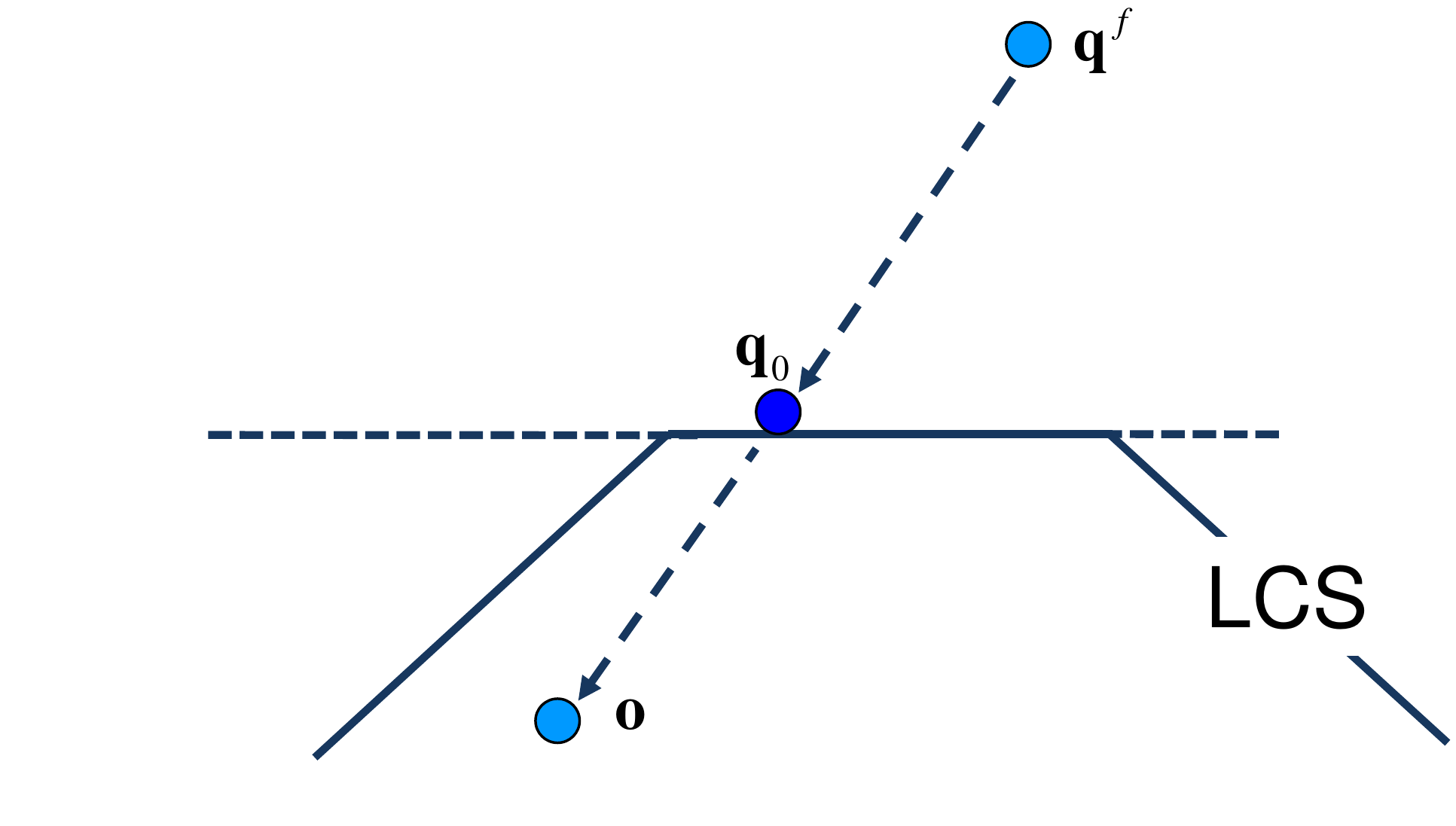,width=4.2cm}}
\subfigure[In-projection on to the LCS to obtain
$\mathbf{q}_1$]{\epsfig{file=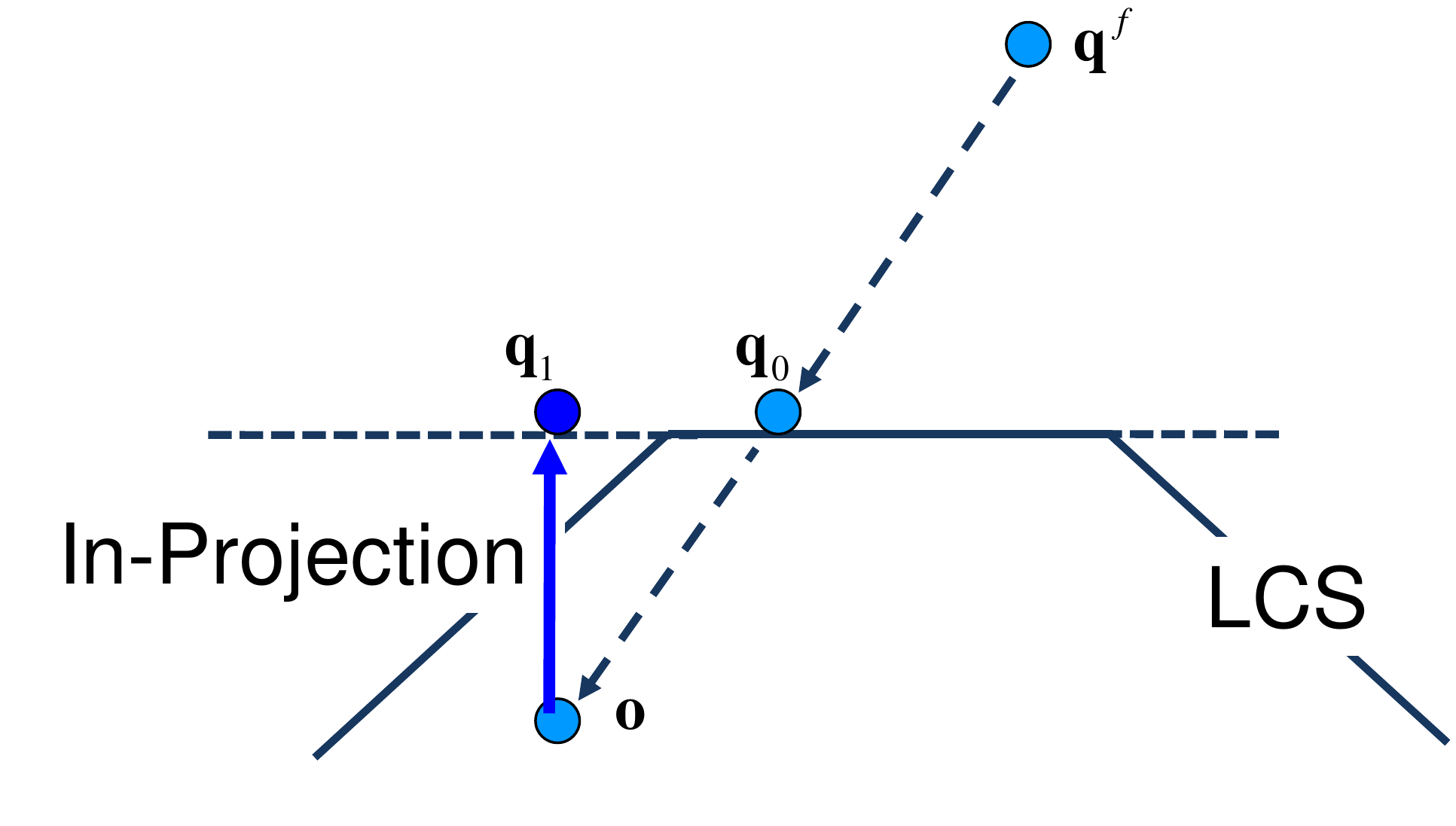,width=4.2cm}}
\subfigure[Out-projection from $\mathbf{q}_1$ to $\mathbf{o}$ to
obtain
$\mathbf{q}_2$]{\epsfig{file=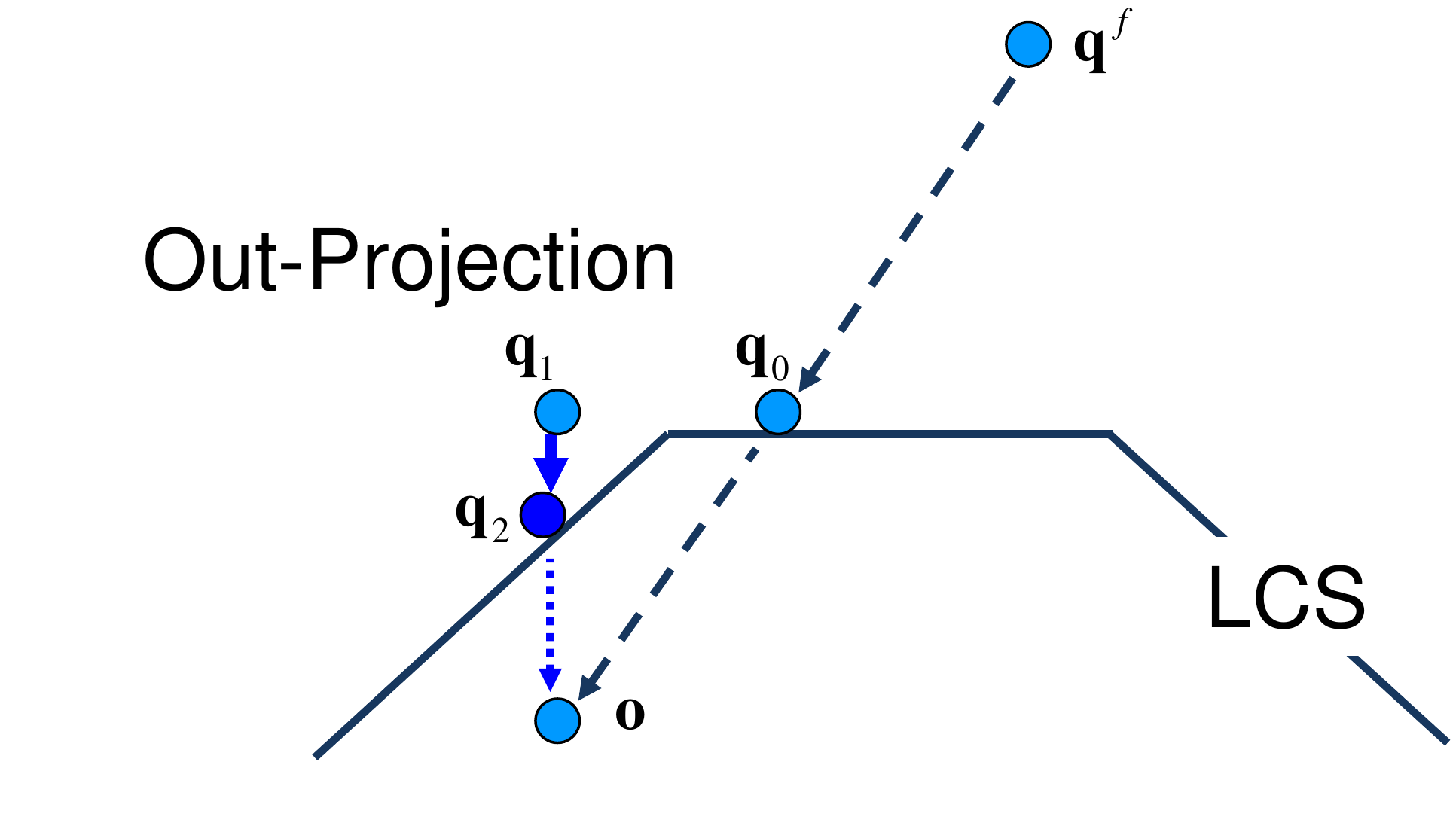,width=4.2cm}}
\subfigure[LCS around
$\mathbf{q}_2$]{\epsfig{file=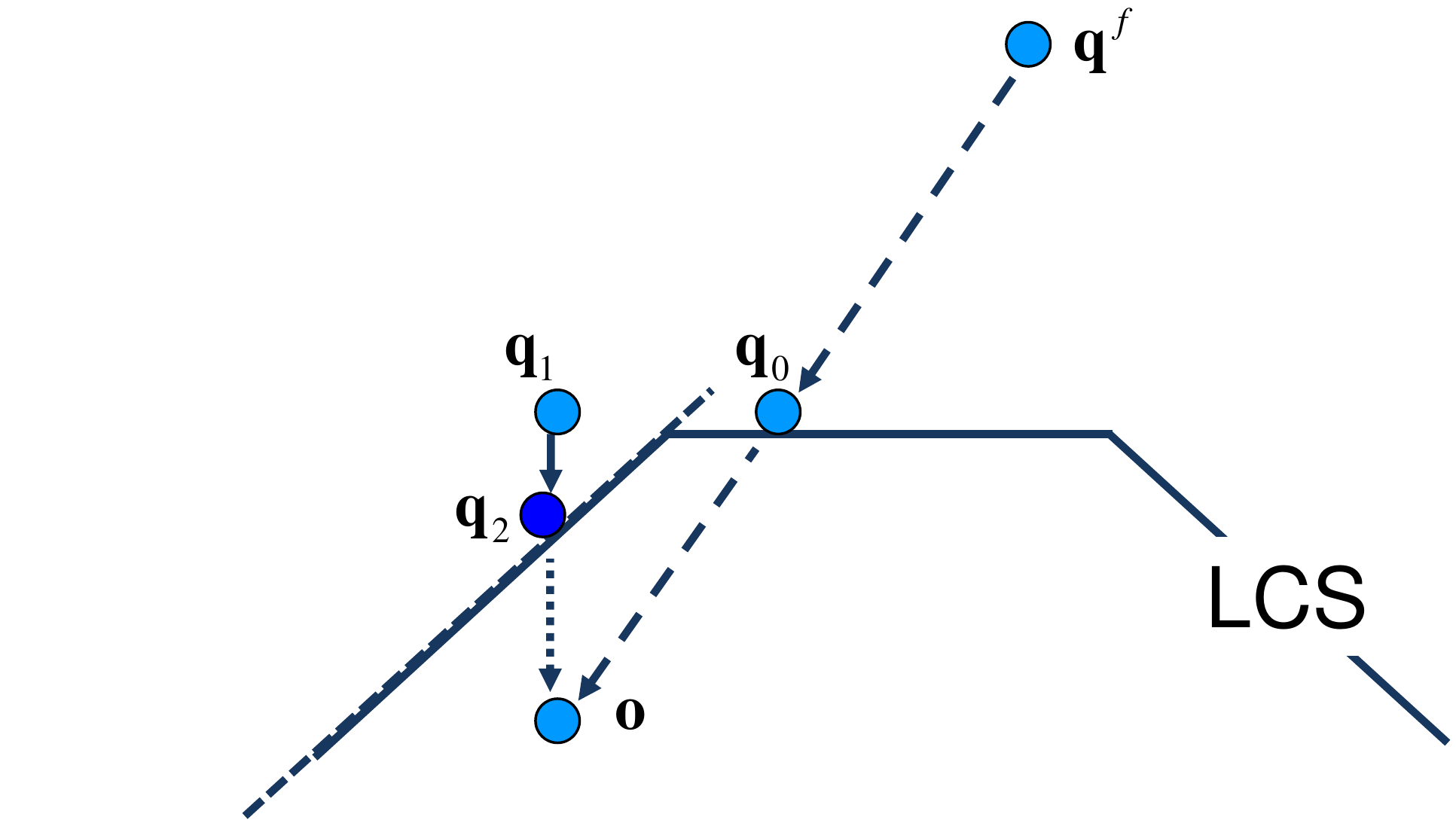,width=4.2cm}}
\subfigure[In-projection on to the LCS to obtain
$\mathbf{q}_3$]{\epsfig{file=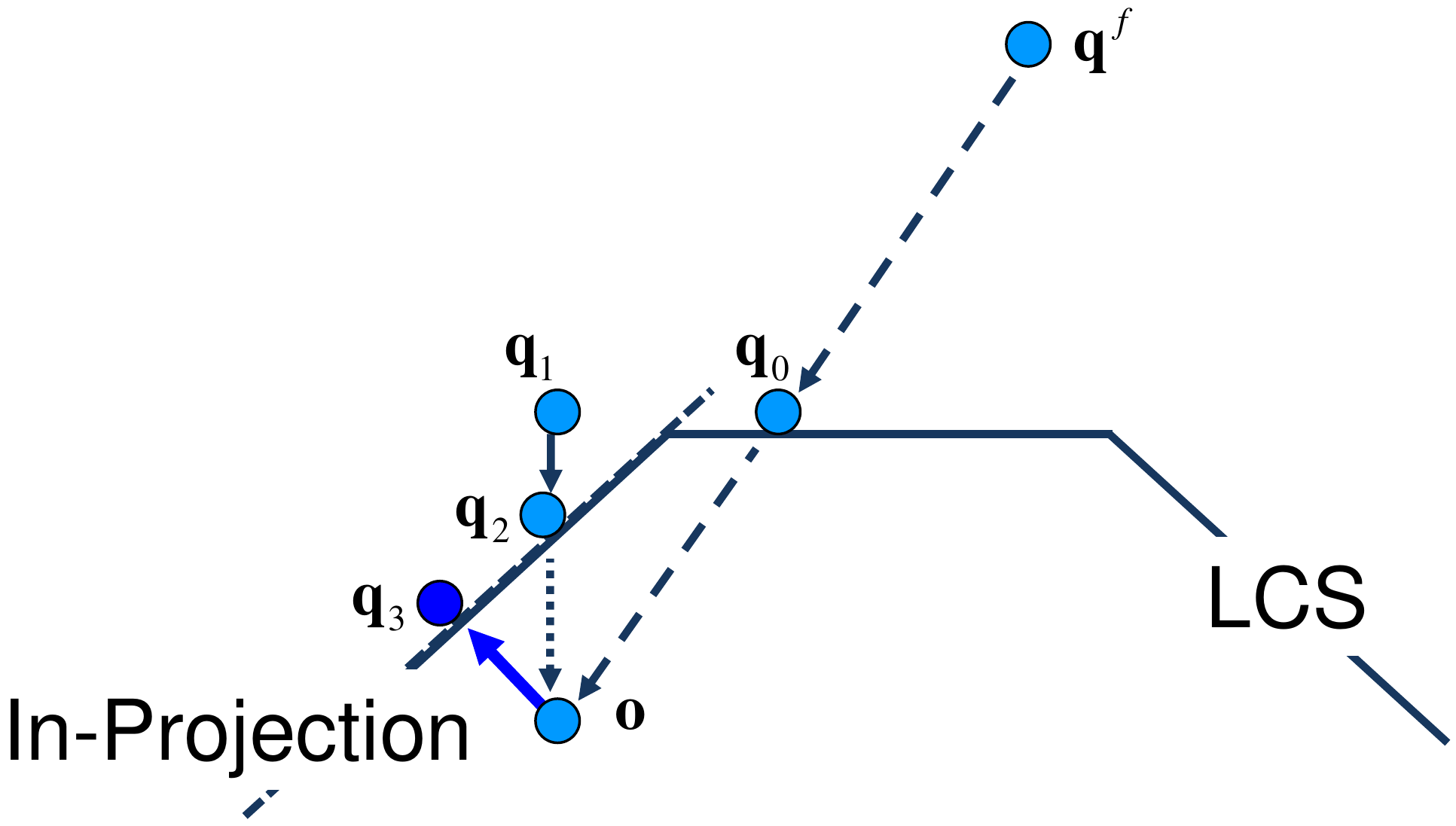,width=4.2cm}}
\caption{{\bf Iterative optimization of a PD for a slightly
complicated case of convex LCS.} In this case, PD requires two
iterations since $\mathbf{q}_1$ in (e) corresponds to separation.
Then, $PD=|| \mathbf{o} - \mathbf{q}_3
||$.}\label{fig:simple_lcs2}
\end{figure*}

\subsection{Overview}\label{sec:overview}


Although the exact PD can be computed from Minkowski sums, the
explicit computation of Minkowski sums between complicated
polygon-soup models is too slow for interactive applications.
Looking at Eq.~\ref{eq:pd2}, we see that finding the PD boils down
to the search for the closest point to the origin on the contact
space. Our algorithm approximates the contact space, which is the
Minkowski sum boundary, only as far as is necessary to locate the
closest point, refining its location until a locally optimal
solution is obtained.

The main computational components of our iterative algorithm are
two projections: (a) a projection from an in-collision
configuration on to the contact space (the \emph{in-projection});
and (b) a projection from a collision-free or contact
configuration on to the contact space (the \emph{out-projection}).
These two projections allow us to obtain a contact configuration
from an in-collision configuration or from a collision-free
configuration.

The in-projection itself consists of two steps: (1) the
construction of a local contact space (LCS) based on a linear
complementarity problem (LCP) formulation; and (2) projection on
to the LCS using a form of Gauss-Seidel iterative solver. The
out-projection step is implemented using translational continuous
collision detection (CCD). Both out- and in-projections are
explained in detail in Secs. \ref{sec:outprj} and \ref{sec:local}.
The overall flow of our iterative algorithm is as follows (see
Fig.~\ref{fig:pipeline}):

%
%
%
%

\begin{enumerate}

\item {\bf Free-configuration selection:} given two overlapping
polygon-soup models $\mov$ and $\fix$, their common origin
$\mathbf{o}$ should be inside the Minkowski sums $\mov \oplus
-\fix$. We select a collision-free configuration $\mathbf{q}^{f}$
in the configuration space (Sec. 5).

\item {\bf Out-projection:} then we perform out-projection from a
source configuration $\mathbf{q}^{s} \equiv \mathbf{q}^{f}$ to a
target configuration $\mathbf{q}^{t} \equiv \mathbf{o}$ using CCD
(Sec. 4). We call the configuration projected on to the contact
space the current configuration $\q_i$.

\item {\bf In-projection:}
\begin{enumerate}
\item We then find the contact features of $\q_i$, such as
vertex/face (VF) or edge/edge (EE) contacts. From these features,
we construct a local contact space (LCS) around $\q_i$, which is a
linear convex cone in configuration space. \item We perform
in-projection from $\mathbf{o}$ to the LCS by formulating this
problem as an LCP, which we solve using a form of Gauss-Seidel
algorithm (Sec. 6.1). Thus this in-projected configuration becomes
the new current projection $\q_i$, and $\q_{i-1}$ is set to the
configuration obtained from the previous out-projection.
\end{enumerate}

\item {\bf Sample classification:} since $\q_i$ was obtained from
the LCS, and not from the global contact space, $\q_i$ can be
in-collision, in-contact or collision-free. We further classify
the collision status of $\q_i$ by performing a static collision
query on $\q_i$ as follows:

\begin{enumerate}

\item If $\q_i$ is an in-contact configuration, we compute the
Euclidean distance from $\mathbf{o}$ to $\q_i$, and return it as
the PD; \ie $\PD = \| \mathbf{o}-\q_i \|$. The algorithm
terminates.

\item If $\q_i$ is a collision-free configuration, $\q_i$ becomes
the source configuration ($\q^s \equiv \q_i$), and the origin
becomes the target configuration ($\q^t \equiv \mathbf{o}$). Then
we go to step 2.

\item If $\q_i$ is an in-collision configuration, we obtain a
proper contact configuration by setting $\q_i$ to  a target
configuration ($\q^t \equiv \q_i$), and the previous contact
configuration $\q_{i-1}$ becomes the source configuration ($\q^s
\equiv \q_{i-1}$). Then we go to step 2.

\end{enumerate}


\item {\bf Iteration:} steps 2-4 are repeated until the algorithm
terminates.
\end{enumerate}

Note that this algorithm always terminates since it is a local
optimization on the LCS, and a locally optimal solution is
obtained whenever $\mathbf{q}_i$ is in-contact. More discussion on
termination conditions is provided in Sec. \ref{sec:terminate}. In
Figs. \ref{fig:simple_lcs1} and \ref{fig:simple_lcs2}, we
illustrate our iterative algorithm for convex LCSs, and in Figs.
\ref{fig:iteration} and \ref{fig:CompLCS} for more general cases.
These examples illustrate the cases of translational configuration
space for two dimensional polygonal objects.

%
%
%
%
%
%
%

%% file: ccd.tex
\section{Out-Projection Using Continuous Collision Detection}\label{sec:outprj}

Our algorithm iteratively updates a sample configuration on the
contact-space to find a locally optimal configuration. This update
process requires a method of projecting the current in-contact or
collision-free configuration on to another configuration on the
contact-space. This out-projection is achieved by a variant of
continuous collision detection (CCD).

\subsection{Continuous Collision Detection}

Let $\mov$ and $\fix$ be two polygon-soup models in 3D, where
$\mov$ is movable by a translation $\mathbf{M}(t)$ and $\fix$ is
fixed. The source and target configurations of $\mov$ are
$\mathbf{q}^{s}$ and $\mathbf{q}^{t}$ at times $t=0$ and $t=1$,
respectively. We also define $\mov(t) \equiv \mathbf{M}(t)\mov$,
and $\mathbf{q}(t)$ represents a configuration of $\mov(t)$. Then
the continuous collision detection (CCD) problem can be formulated
as a search for the first time of contact (ToC) $\tau$, between
$[0,1]$, if it exists:
\begin{equation}\label{eq:ccd}
\tau=\min\{t\in[0,1]\ |\ \mov(t)\cap\fix\neq\emptyset\}.
\end{equation}
There are many methods for CCD, but our choice is conservative
advancement (CA)
\cite{l-ecdar-93,Mirtich96phd,Zhang06,SIG07CATCH,mkm09} which is
known to be fastest in practice. Like other CCD algorithms, CA
takes the source $\q^s$ and target $\q^t$ configurations of an
object $\mov$, and computes the first time of contact when $\mov$
linearly interpolates from $\q^s$ to $\q^t$ in the configuration
space. For a convex polytope, CA computes a lower bound on $\tau$
by repeatedly advancing $\mov$ by $\Delta{t}$ toward $\fix$ until
collision occurs. The value of $\Delta{t}$ is chosen to correspond
to a lower bound on the closest distance $d(\mov(t),\fix)$ between
$\mov(t)$ and $\fix$, and an upper bound $\mu$ on the motion of
$\mov(t)$ projected on to the direction of $d(\mov(t),\fix)$ per
second:
\begin{equation}\label{eq:ca}
\Delta{t}\leq\frac{d(\mov(t),\fix)}{\mu}.
\end{equation}
The first time of contact $\tau$ is the sum of the time-steps
$\Delta{t}$ before collision; \ie $\tau = \sum \Delta{t}$. Whereas
the CA algorithm applies to convex polytopes, we need to use the
modified version of the C$^2$A algorithm proposed by Tang \al
\cite{mkm09} for polygon-soup models. The C$^2$A algorithm uses a
bounding volume hierarchy (BVH) based on swept sphere volumes
(SSVs) \cite{LGLM00} to control the depth of the recursive BVH
traversal during  iterations of the algorithm, which reduces the
computation time significantly. Since our PD problem is restricted
to translational motion, we can simplify C$^2$A and make it
faster. We present more details of this technique in Sec.
\ref{sec:transccd}.

When we have found $\tau$, we can perform a static proximity query
\cite{LGLM00} to find all the contact features between
$\mov(\tau)$ and $\fix$, such as VF and EE contacts. These will be
used to construct the boundary of the local contact space in Sec.
\ref{sec:local}.


\subsection{Translational Continuous Collision Detection}\label{sec:transccd}


\begin{figure}[htb]
\centering\vspace{0em}\hspace{0em}
\epsfig{file=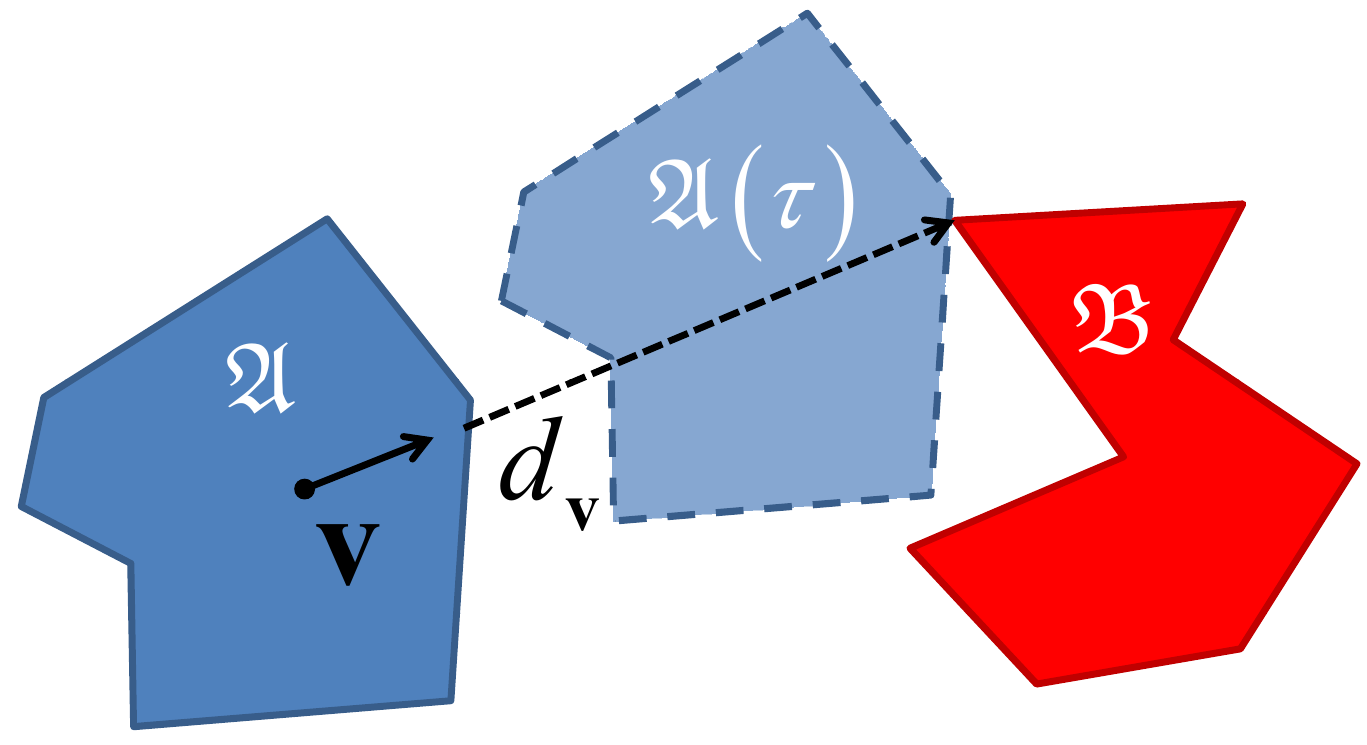,width=5cm}
 \caption{\bf{Minimal directional distance along $\mathbf{v}$ between two objects.}} \label{fig:MDD}
\end{figure}

If $\mov$ is moving with a constant translational velocity
$\mathbf{v}$, then we can call the shortest distance between
$\mov$ and $\fix$ in the direction of $\mathbf{v}$ the minimal
directional distance (MDD) $d_{\mathbf {v}}(\mov,\fix)$, as
illustrated in Fig.~\ref{fig:MDD}. The time at which $\mov$
contacts $\fix$ can be calculated as follows:
\begin{eqnarray}\label{eq:trandist}
\tau=\frac{d_{\mathbf {v}}(\mov,\fix)}{\|\mathbf{v}\|}.
\end{eqnarray}
If $\tau<1$, then $\mov$ and $\fix$ will collide at $\tau$;
otherwise, they are collision-free during the entire time-step.

\cite{Choi06} presented a method of computing the MDD between
convex polytopes based on Minkowski sums and ray-shooting.
However, no algorithm for computing the MDD between polygon-soup
models has been reported. We propose a simple method in which we
construct the BVHs of polygon-soup models and recursively compute
the MDD between pairs of nodes pairs in the BVHs; this is similar
to the computation of Euclidean distance based on BVHs. Thus
computing $d_{\mathbf {v}}(\mov,\fix)$ boils down to computing the
MDD between pairs of nodes in the BVHs (\ie these nodes are
bounding volumes (BVs) or triangles). We will explain how to
compute the MDD between two triangles $\triangle_\mov$ and
$\triangle_\fix$; and it should be apparent that a similar method
can be used between BVs.


\begin{figure}[htb]
\centering\vspace{0em}\hspace{0em}
\epsfig{file=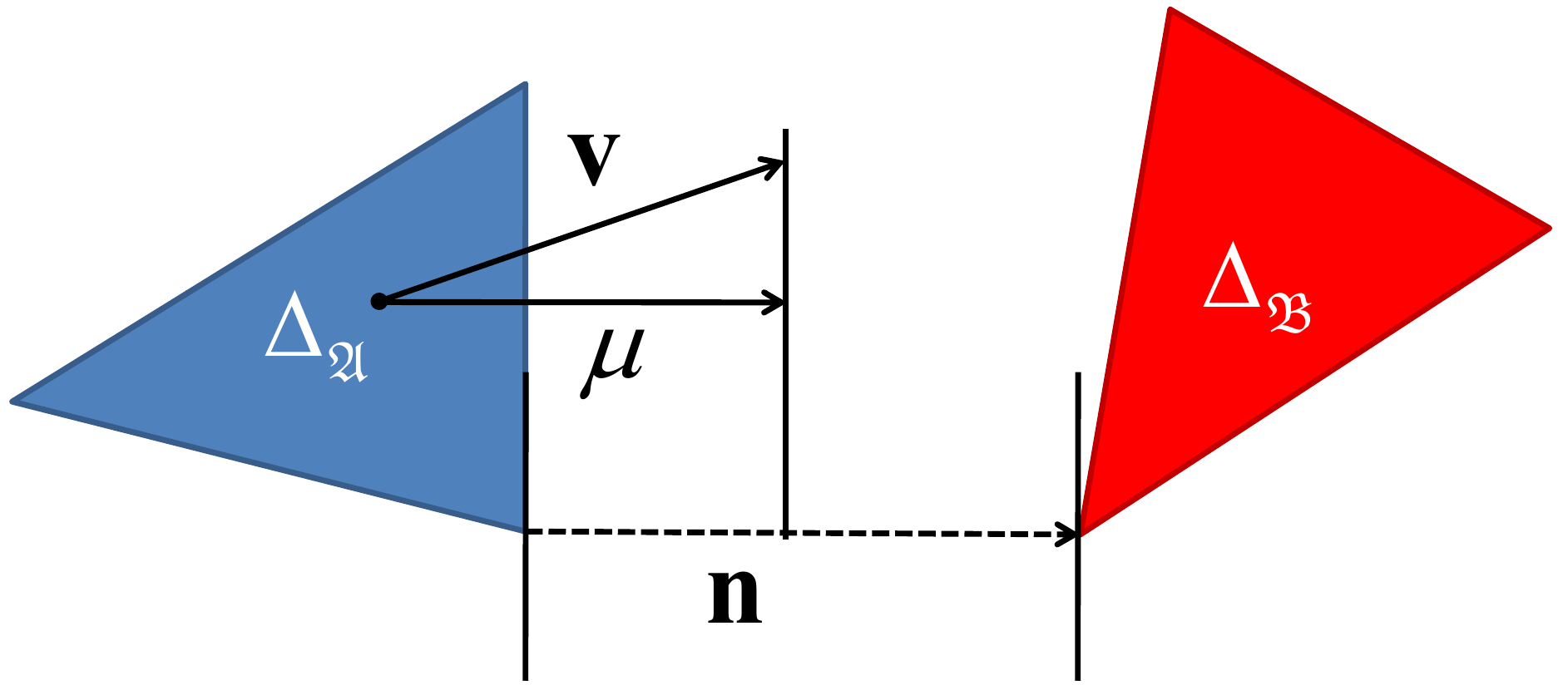,width=5cm}
 \caption{\bf{Translational CA for triangles.}} \label{fig:CAtri}
\end{figure}

The ToC $\tau'$ of two triangles under translational motion
$\mathbf{v}$ can be found by repeated application of Eq.
\ref{eq:ca}. The motion bound ${\mu}$ is simply
$\mathbf{v}\cdot{\mathbf{n}}$, where $\mathbf{n}$ connects two
closest points on the triangles, as illustrated in
Fig.~\ref{fig:CAtri}. Then the MDD between $\triangle_{\mov}$ and
$\triangle_{\fix}$ is written
$d_{\mathbf{v}}(\triangle_{\mov},\triangle_{\fix})=\tau'\|\mathbf{v}\|$.
We recursively compute the MDD between BV pairs or triangles pairs
in the BVHs, and use Eq.~\ref{eq:trandist} to obtain the ToC
between the polygon-soup models $\mov$ and $\fix$.

\section{Finding a Collision-Free Configuration}\label{sec:selecseed}

To obtain a non-trivial solution to Eq.~\ref{eq:ccd}, the source
configuration $\q^s$ needs to be collision-free; otherwise, $\tau$
is trivially zero. In the context of our PD computation, $\q^s$ is
unknown, whereas the target configuration $\q^t$, which is an
in-collision configuration, is an input to the PD problem. We will
introduce several methods of finding a collision-free source
configuration $\q^s$. This is crucial to our PD computation, since
our algorithm is a local optimization and starts from the contact
configuration computed by CCD.

\subsection{Centroid Difference}\label{sec:dir}

\begin{figure}[htb]
\centering \epsfig{file=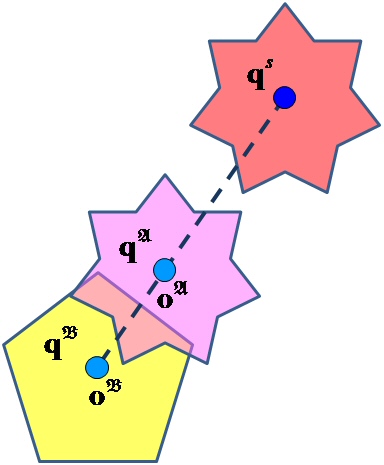,height=5cm}
\caption {{\bf A collision-free configuration from the centroid
difference.} Obstacle $\fix$ (yellow), initial in-collision
configuration of $\mov$ (magenta), and the collision-free
configuration of $\mov$ (red).}\label{fig:dragon_free}
\end{figure}

When no prior information is available about the interpenetration
of the objects, the penetration direction can be estimated from
the centroids of the objects. This direction can then be used to
place the moving object in an interpenetration-free configuration,
which can be expressed as follows:

\[\mathbf{q}^{s}=\mathbf{q}^{\mov}+(r_{\mov}+r_{\fix})\frac{\mathbf{o}^{\mov}-\mathbf{o}^{\fix}}{||\mathbf{o}^{\mov}-\mathbf{o}^{\fix}||},\] where
$\mathbf{q}^{\mov}$ is the initial configuration of $\mov$,
$\mathbf{o}^{\mov}$ and $\mathbf{o}^{\fix}$ are the centroids of
$\mov$ and $\fix$ respectively, and $r_{\mov}$ and $r_{\fix}$ are
the diameters of spheres enclosing each object.
Fig.~\ref{fig:dragon_free} shows a collision-free configuration
determined from the centroid difference.

\begin{figure*}[htb] \centering
\subfigure[]{\epsfig{file=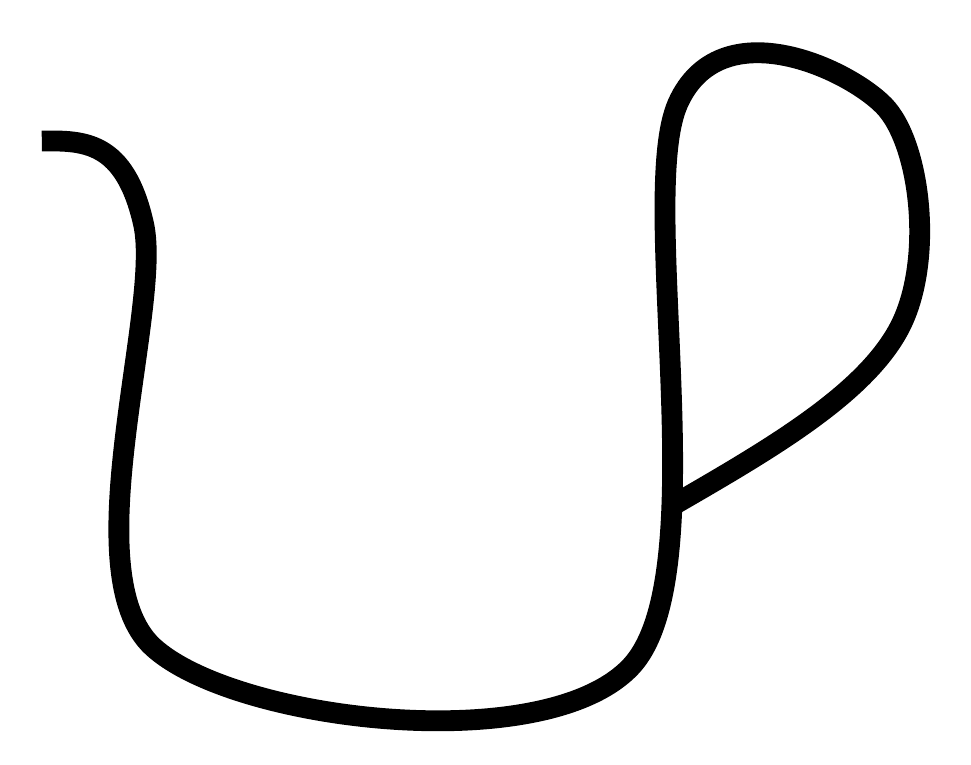,width=2.25cm}}
\subfigure[]{\epsfig{file=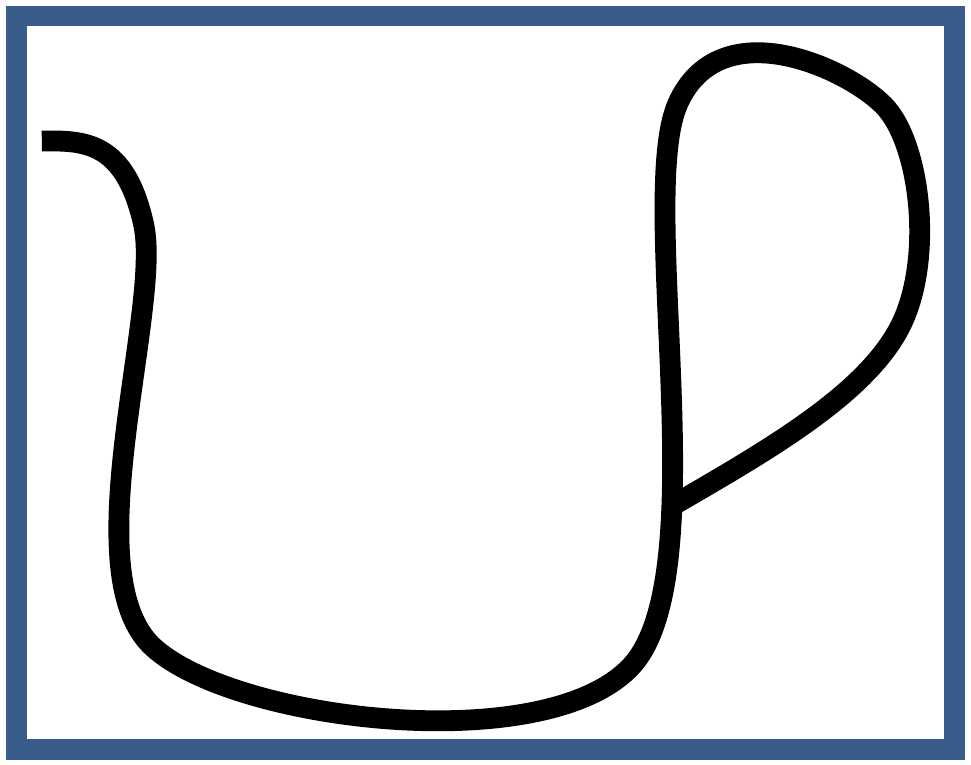,width=2.25cm}}
\subfigure[]{\epsfig{file=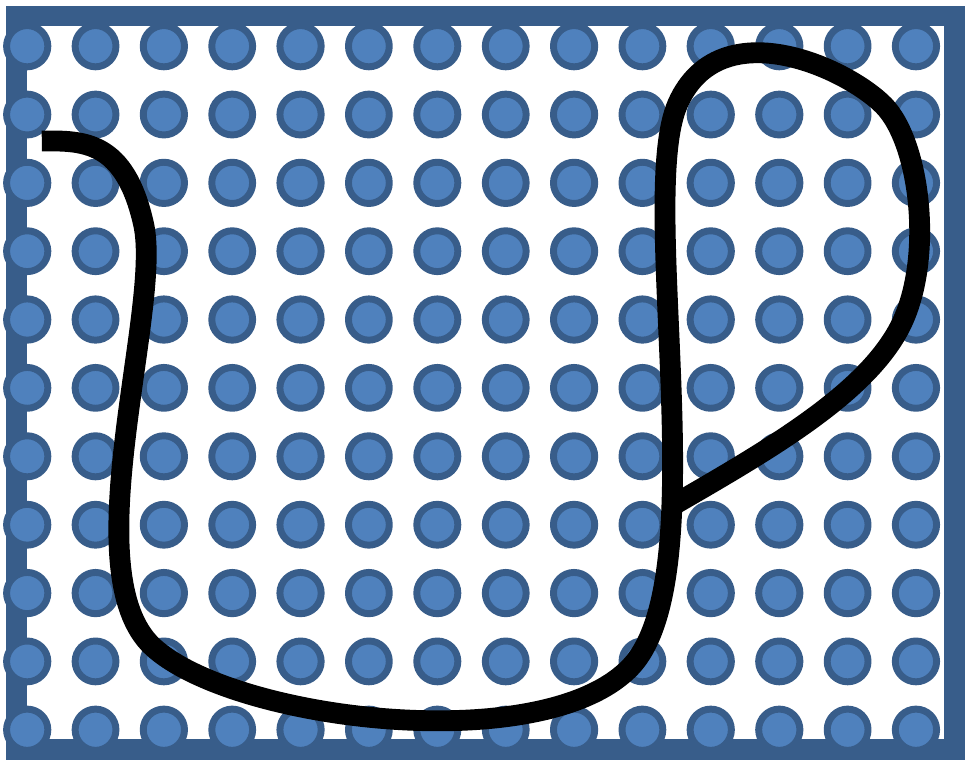,width=2.25cm}}
\subfigure[]{\epsfig{file=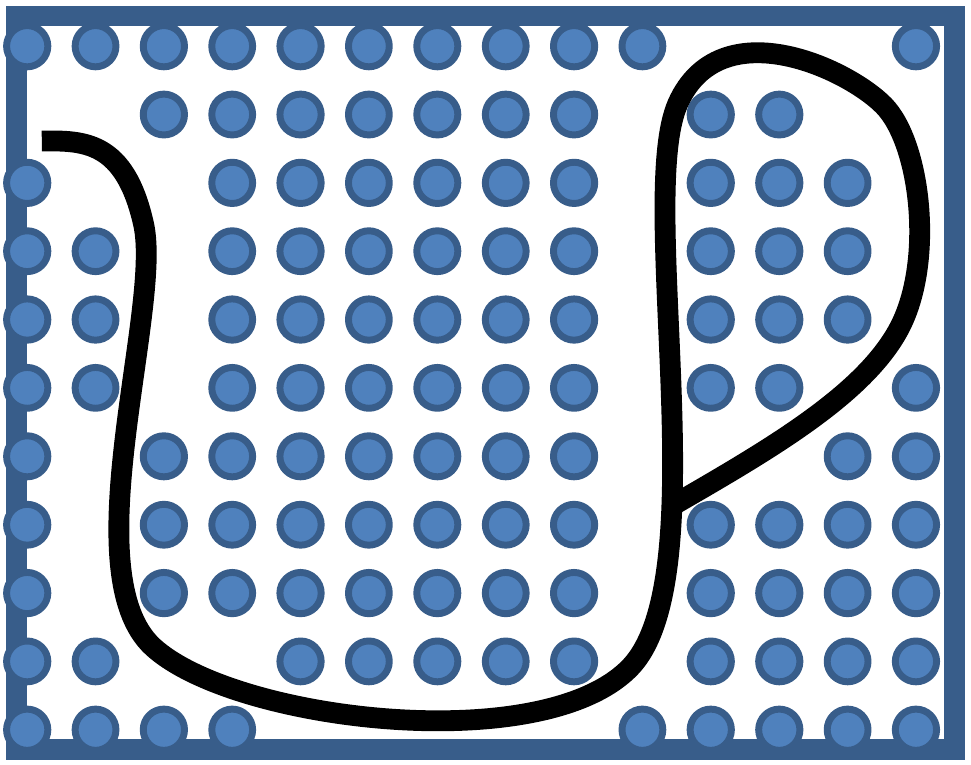,width=2.25cm}}
\subfigure[]{\epsfig{file=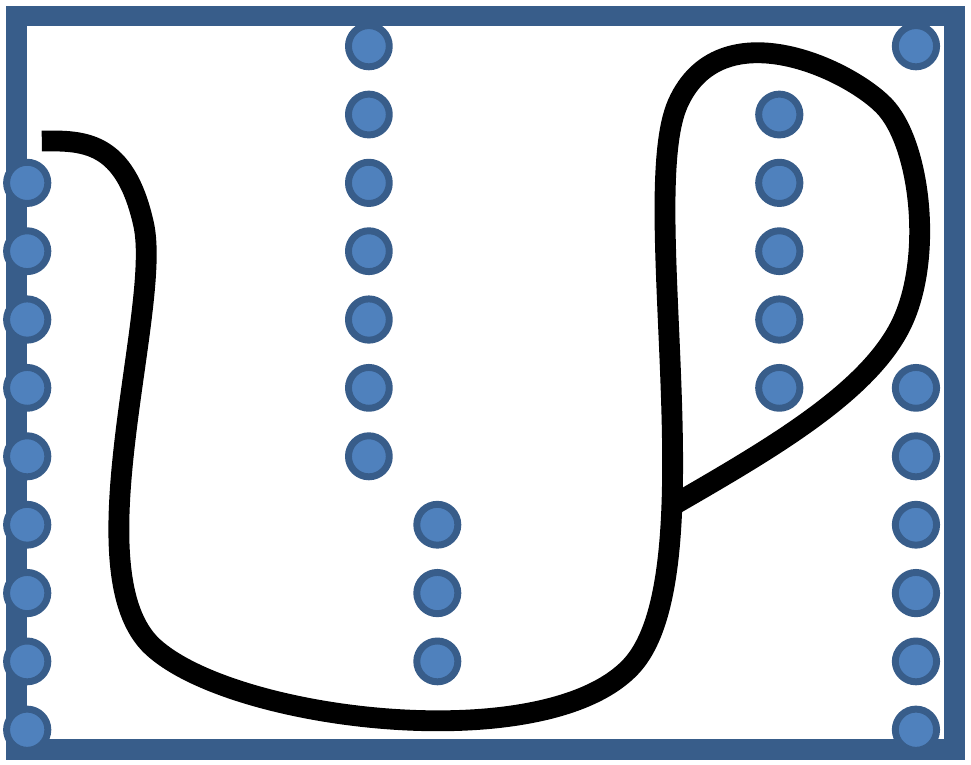,width=2.25cm}}
\subfigure[]{\epsfig{file=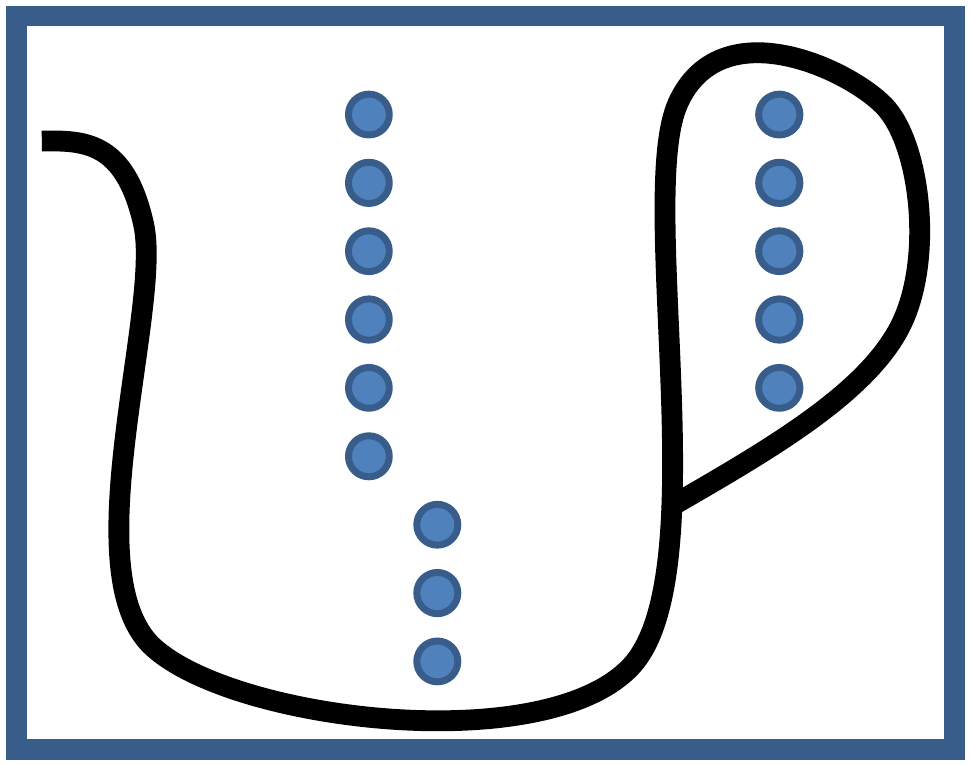,width=2.25cm}}
\subfigure[]{\epsfig{file=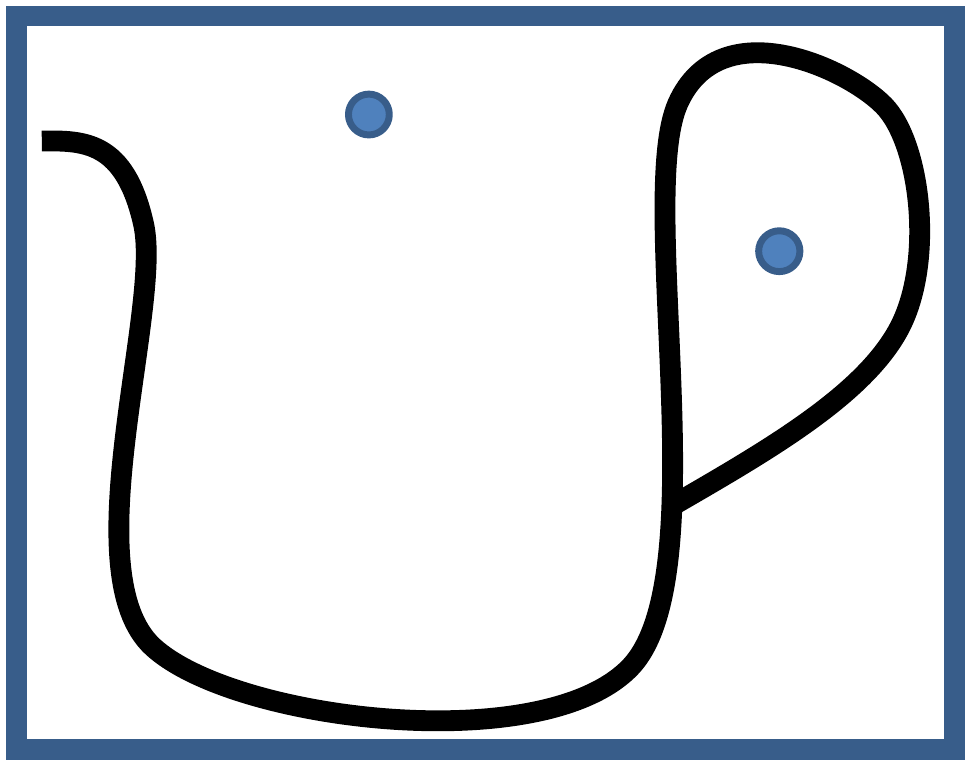,width=2.25cm}}
\caption{{\bf Finding maximally clear configurations.}  (a) The
given model. (b)  AABB of the model. (c) Voxelize the AABB and
compute the distance to the model from each grid position. (d)
Remove the grid points corresponding to small distances. (e)
Compare neighboring grid points and remove those with smaller
distances by scanning along the $x$- and $y$-directions. (f)
Remove the grids on the boundary of AABB. (g) Compare neighboring
grid points to remove those with smaller distances, and find
maximally distant configurations in the $x$-, $y$- and
$z$-directions.}\label{fig:preprocess}
\end{figure*}

\subsection{Maximally Clear Configuration}

\begin{figure}[htb] \centering
\subfigure[]{\epsfig{file=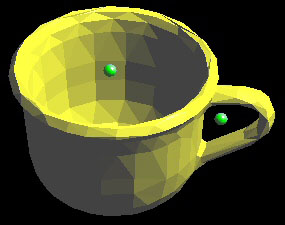,height=2.5cm}}
\subfigure[]{\epsfig{file=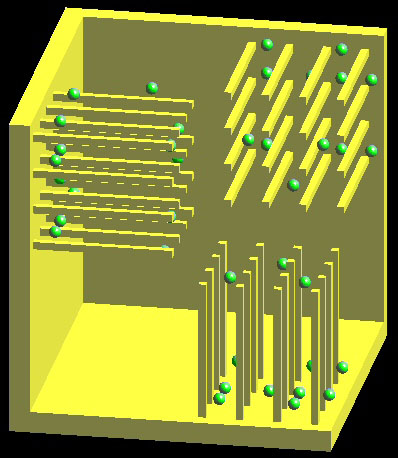,height=2.5cm}}
\subfigure[]{\epsfig{file=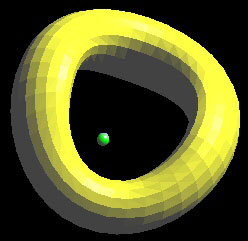,height=2.5cm}}
\caption{{\bf Maximally clear configurations for the Cup, Grate,
and Distorted-torus models.} The small green spheres show the
positions of the maximally clear
configurations.}\label{fig:pre_free}
\end{figure}

Although the centroid difference allows us to obtain a
collision-free configuration, that configuration may be quite
different from the optimal PD configuration for a complicated
object with a lot of concavities or many holes. To obtain a
collision-free configuration for objects of this sort, we find
points in space that have maximal clearance. This preprocess
allows objects to be positioned without interpenetration. These
points correspond to the points on the boundary of external
Voronoi regions of the object, which are equidistant from at least
two points on the surface of the model. These maximally clear
configurations are appropriate candidates for collision-free
configurations since they correspond to locally maximum
likelihoods of a collision-free state (see
Fig.~\ref{fig:pre_free}). A similar strategy has been used in
workspace sampling in motion planners \cite{l-rmp-91}. Since the
construction of a generalized Voronoi diagram for a polygonal
model is quite hard \cite{HCKLM99} we propose a simple algorithm
based on a voxelization of space (also see Fig.
\ref{fig:preprocess}):

\begin{enumerate}

\item Compute the axis-aligned bounding box (AABB) of the static
object, and voxelize the AABB with a grid.

\item Calculate distance to the object from each point in the
grid.

\item Eliminate configurations with small distance values, since
these nearly correspond to collisions.

\item Compare neighboring configurations and eliminate those with
shorter distances; and find maximally distant configurations in
one and two dimensions.

\item Remove any configurations remaining on the boundary of the
AABB.

\item Again compare neighboring configurations and eliminate those
with shorter distances, thus identifying the maximally distant
configurations in full dimensions.
\end{enumerate}

Note that the above procedure only computes a subset of the points
on the Voronoi boundary, which are those likely to have maximal
clearance.

\subsection{Motion Coherence}

Applications of our PD algorithm are likely to exhibit motion
coherence: for instance, in rigid or articulated body dynamics, a
series of collision-free or contact configurations are calculated
as a function of time; we can exploit the underlying motion
coherence to find a good initial source configuration. A simple
way of doing this is to cache a sequence of collision-free
configurations and choose the one closest to a given in-collision
configuration.
  More precisely, in our implementation, we store the last three in-contact configurations
  obtained from PD computations, and choose the one $\q_{c}$ that
  has the minimum translational difference from the input in-collision
   configuration $\mathbf{o}$.
   Then, we change the orientation of $\q_{c}$ to be the same as that
   of $\mathbf{o}$, and if this does not create any collision,
   we use $\q_{c}$ as a starting configuration for the iteration;
   otherwise, we slightly move $\q_{c}$ to the opposite direction
   toward $\mathbf{o}$, and see if it causes collision again.
   If not, we use $\q_{c}$ as an initial configuration.
   Otherwise, we abandon the motion coherence strategy and switch
   to other methods (e.g. centroid difference).

\subsection{Random Configuration}

In a more general setting, in which we do not have any knowledge
of $\q^s$, we can randomly sample $\q^s$ in the free configuration
space \cite{fap} or simply locate $\q^s$ at infinity.

\subsection{Sampling-based Search}

\begin{figure}[htb]
\centering \epsfig{file=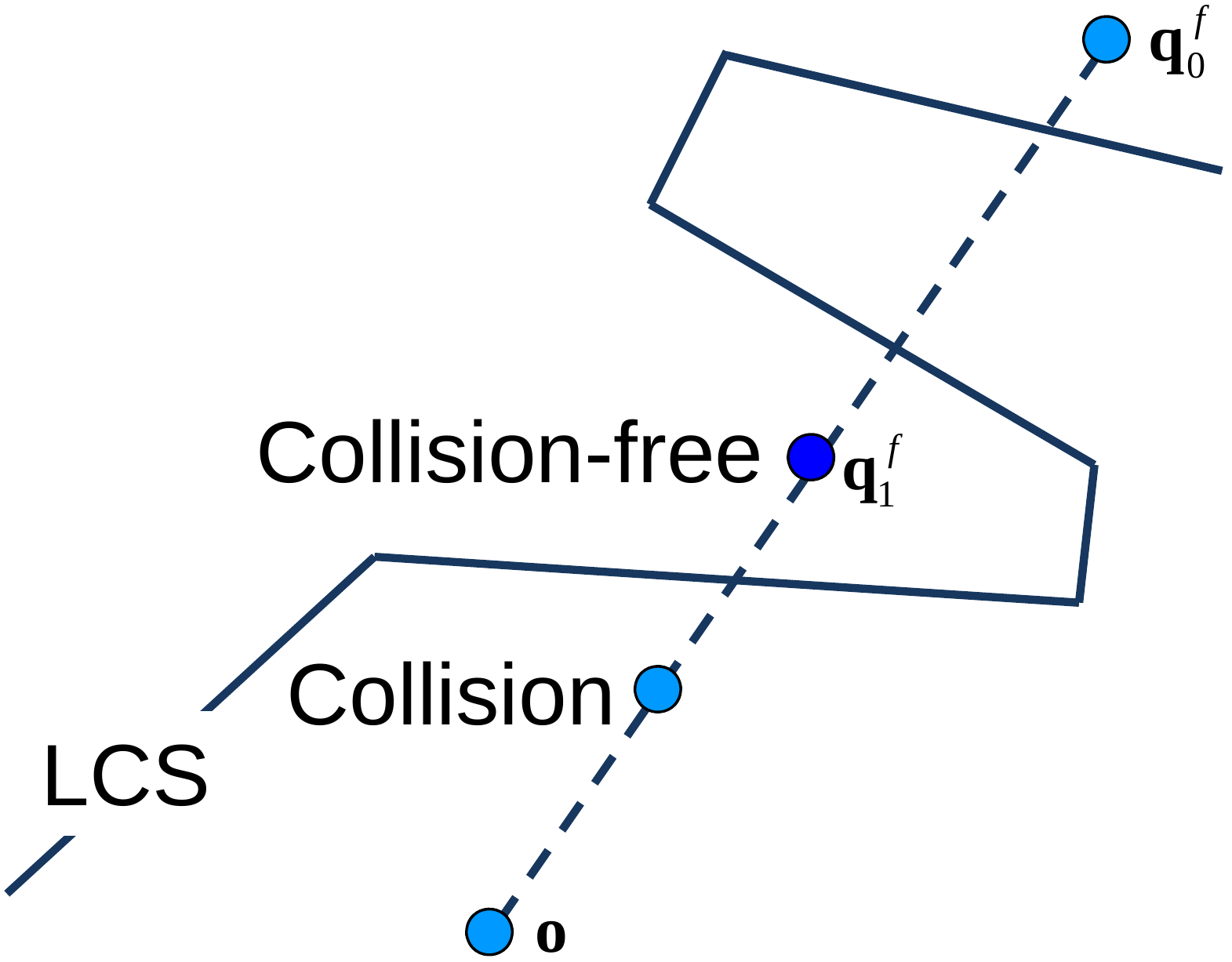,width=5cm} \caption
{{\bf Sampling-based search for a collision-free configuration.}
The input collision configuration is $\mathbf{o}$ and
$\mathbf{q}^{f}_{0}$ is an initial collision-free configuration.
The first sample on the line from $\mathbf{o}$ to
$\mathbf{q}^{f}_{0}$ is determined to be in-collision, but the
second sample $\mathbf{q}^{f}_{1}$ is collision-free.
}\label{fig:search}
\end{figure}

The methods of finding a suitable collision-free configuration
from which to begin out-projection that we have just described may
still generate a configuration far from the optimal.
A sampling-based search algorithm can be employed to refine an
initial collision-free configuration.
%
By performing collision detection on a series of configurations
sampled on a line from the given collision configuration
$\mathbf{o}$ to $\mathbf{q}^{f}_0$, as illustrated in Fig.
\ref{fig:search}, this sampling process 
terminates whenever a collision-free configuration is found. For
collision detection, we use \cite{LGLM00} in our implementation.

%

%% file: iteration.tex
\section{Iterative Optimization}\label{sec:local}

Any sampled configuration $\q$ in the contact space can be used as
an estimate of the PD by computing its distance from the origin;
\ie $\PD=\|\mathbf{o}-\q\|$. When we have to deal with a highly
non-convex object, finding a good candidate for $\q$ taxes the
strategies suggested in Sec. \ref{sec:selecseed}. Thus we need to
refine the sample to get a PD closer to the optimum, which we
achieve by a \emph{walk} in contact space. We use the result found
by one of the techniques described in Sec. \ref{sec:selecseed} as
the start of this walk, and then build a local approximation to
the contact space, and refine the configuration by in-projection.
We repeat this process until a locally optimal solution is
obtained.

%
%

\begin{figure*}[htb] \centering
\subfigure[Local contact space and the
origin]{\epsfig{file=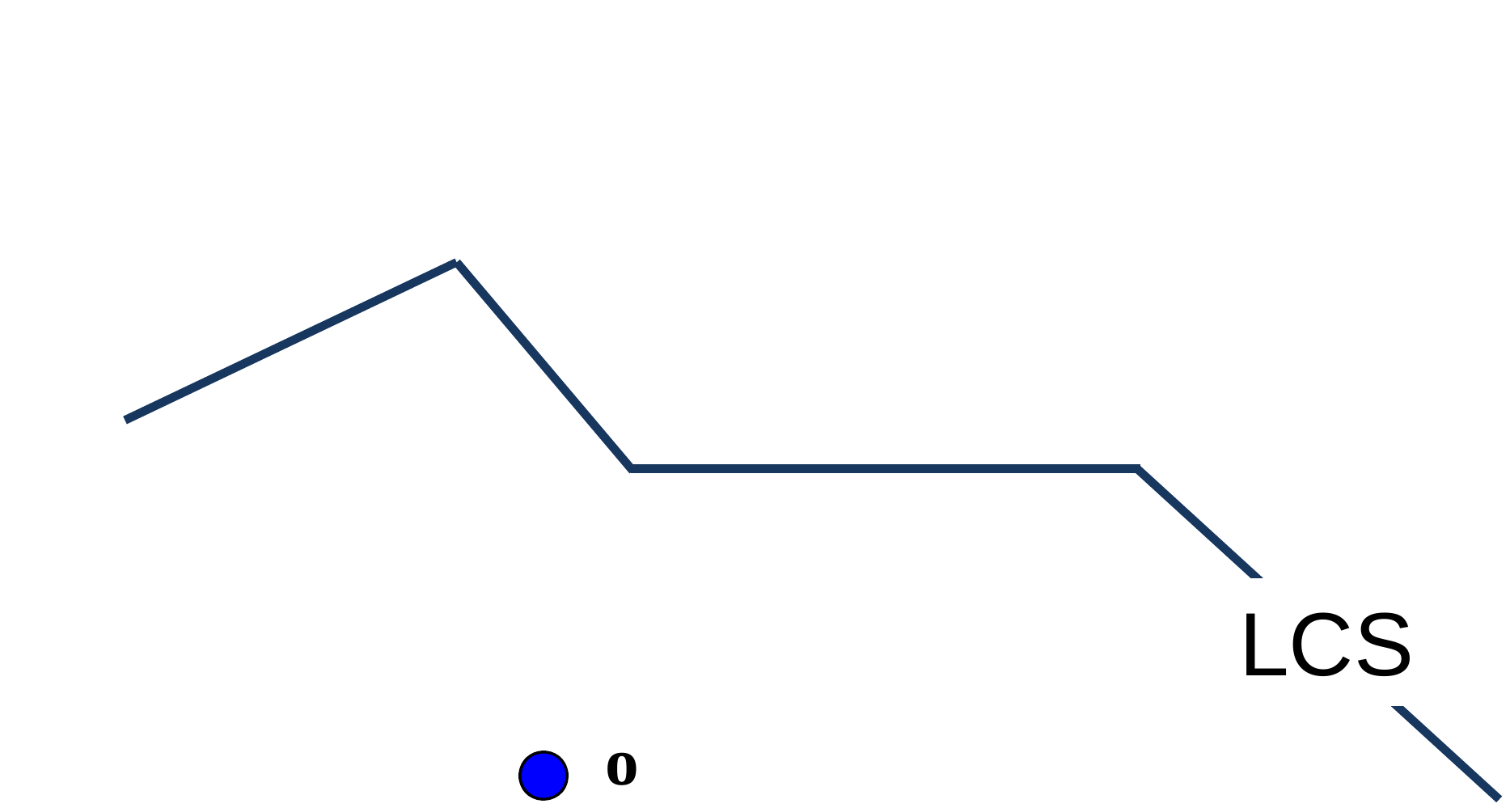,width=4.2cm}}
\subfigure[Collision-free configuration
$\mathbf{q}^f$]{\epsfig{file=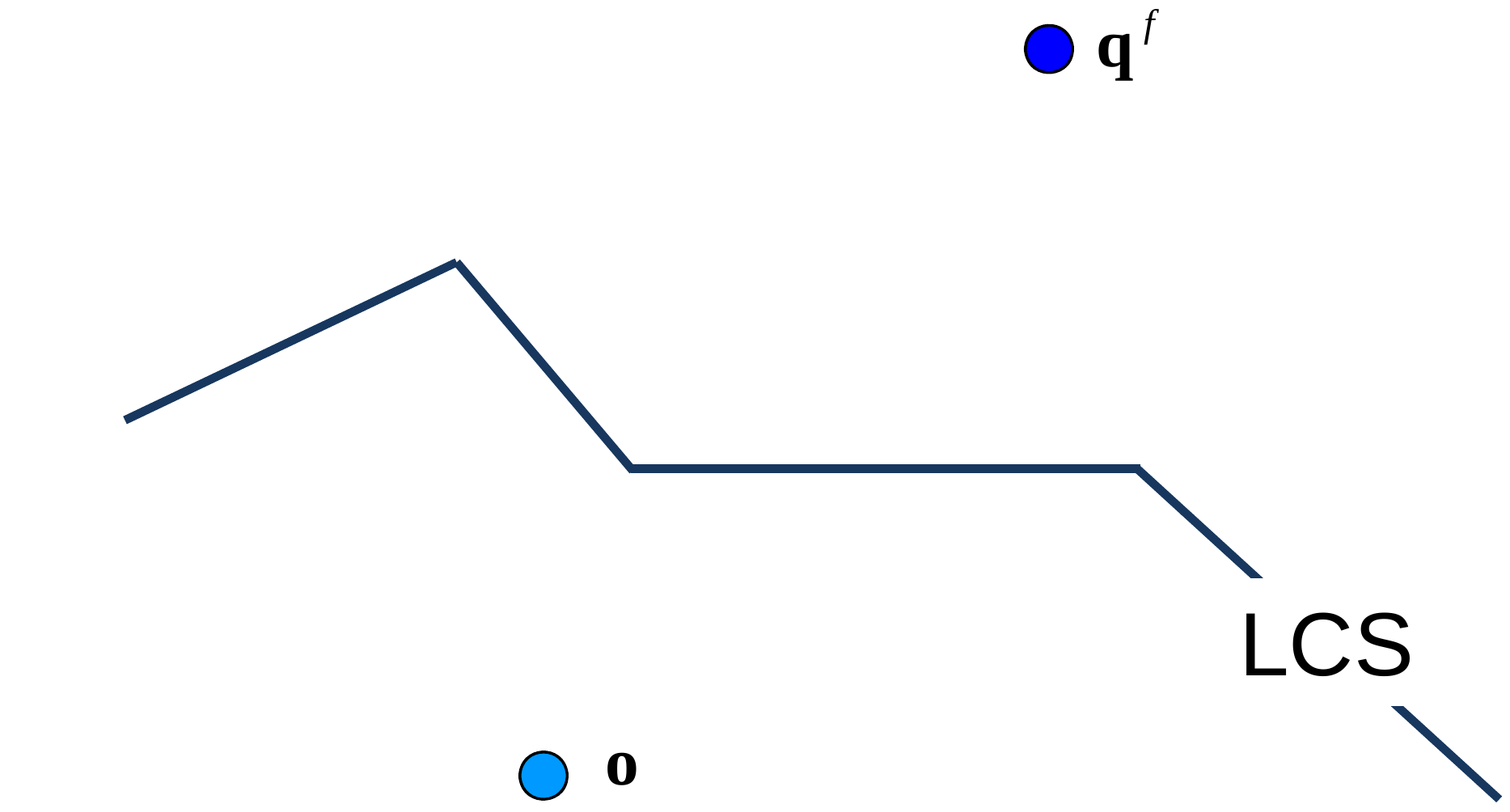,width=4.2cm}}
\subfigure[Out-projection from $\mathbf{q}^f$ to $\mathbf{o}$ to
obtain
$\mathbf{q}_0$]{\epsfig{file=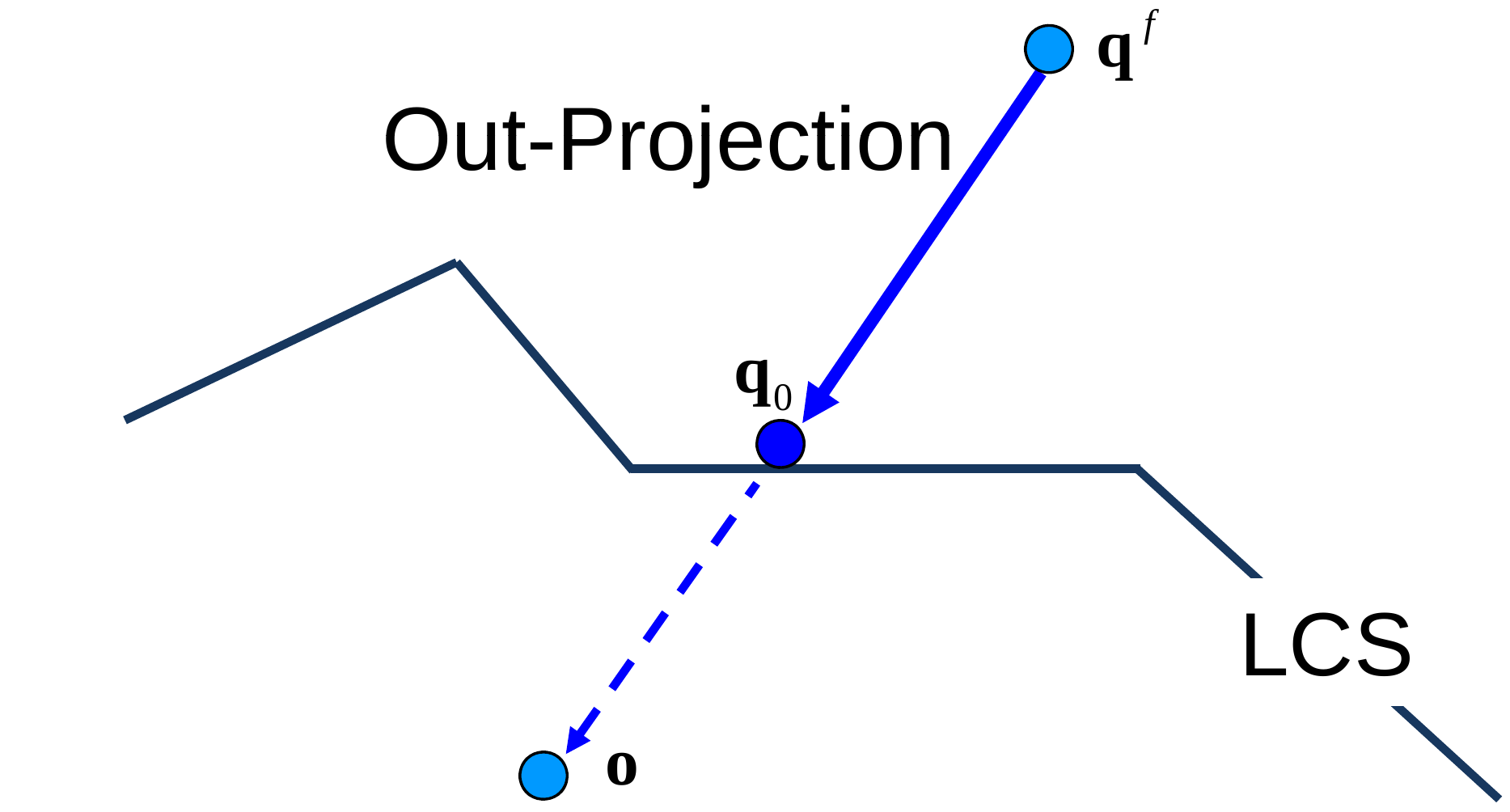,width=4.2cm}}
\subfigure[Construction of an LCS around
$\mathbf{q}_0$]{\epsfig{file=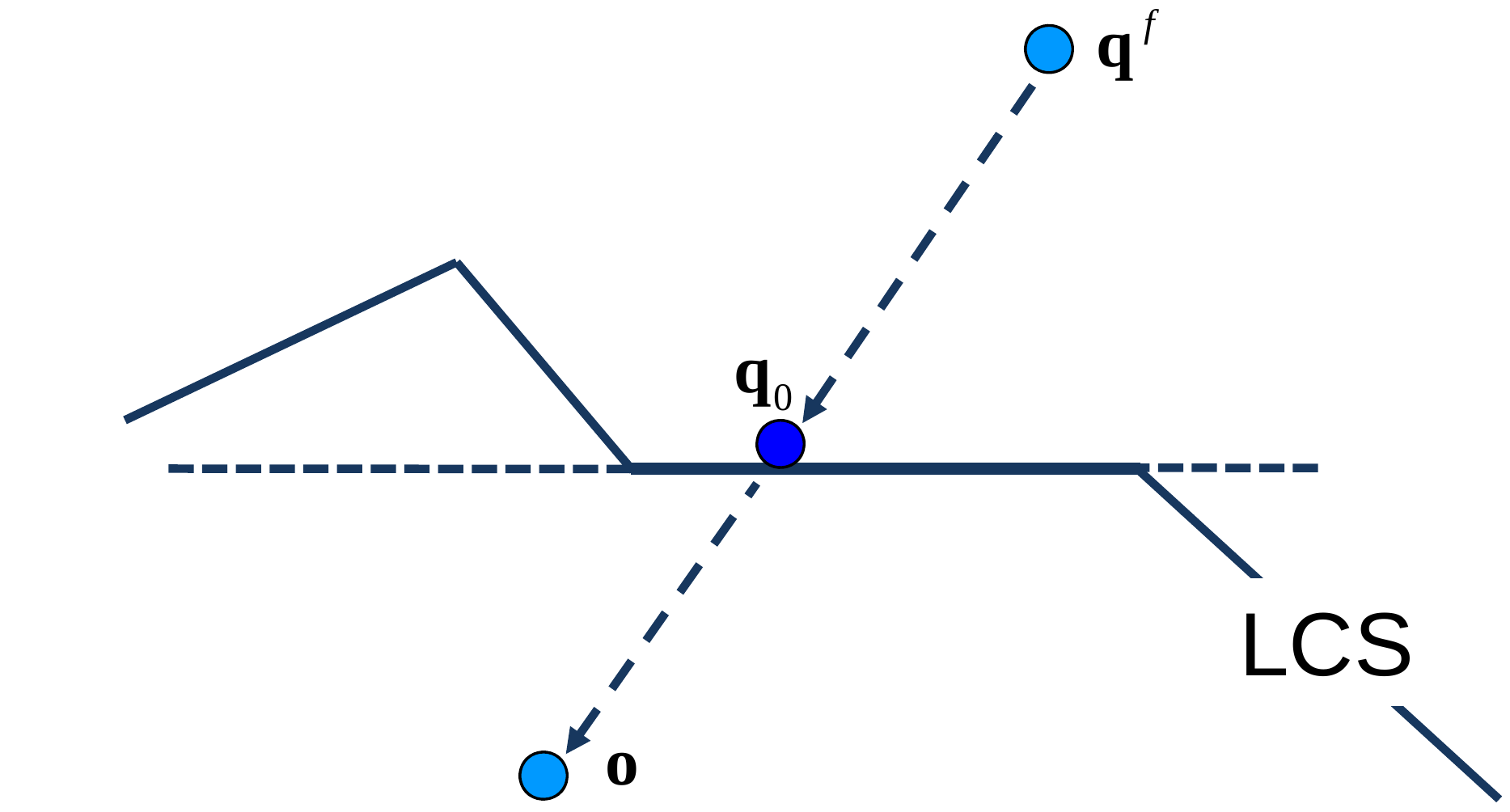,width=4.2cm}}
\subfigure[In-projection on to the LCS to obtain
$\mathbf{q}_1$]{\epsfig{file=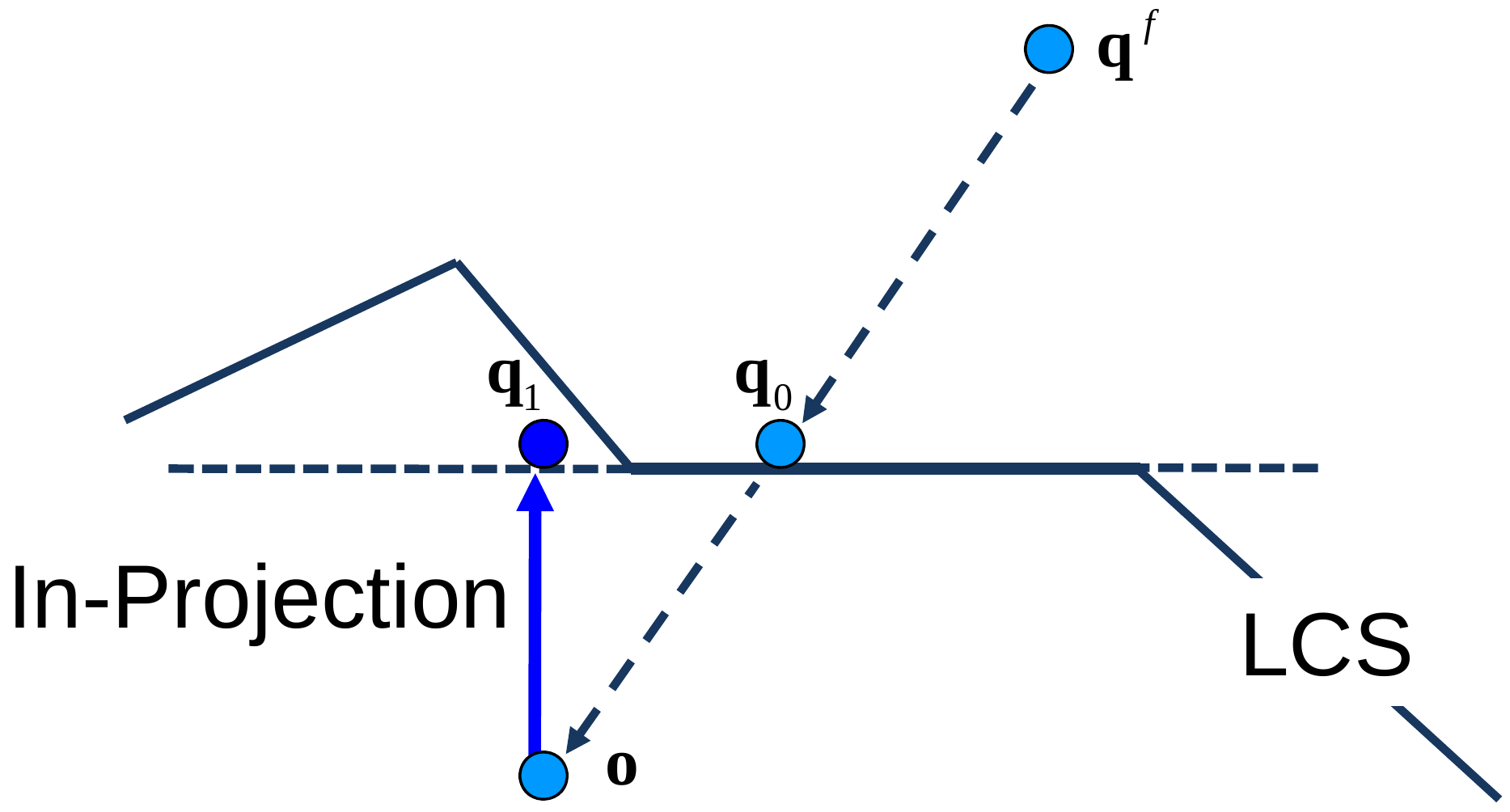,width=4.2cm}}
\subfigure[Out-projection from $\mathbf{q}_0$ to $\mathbf{q}_1$ to
obtain
$\mathbf{q}_2$]{\epsfig{file=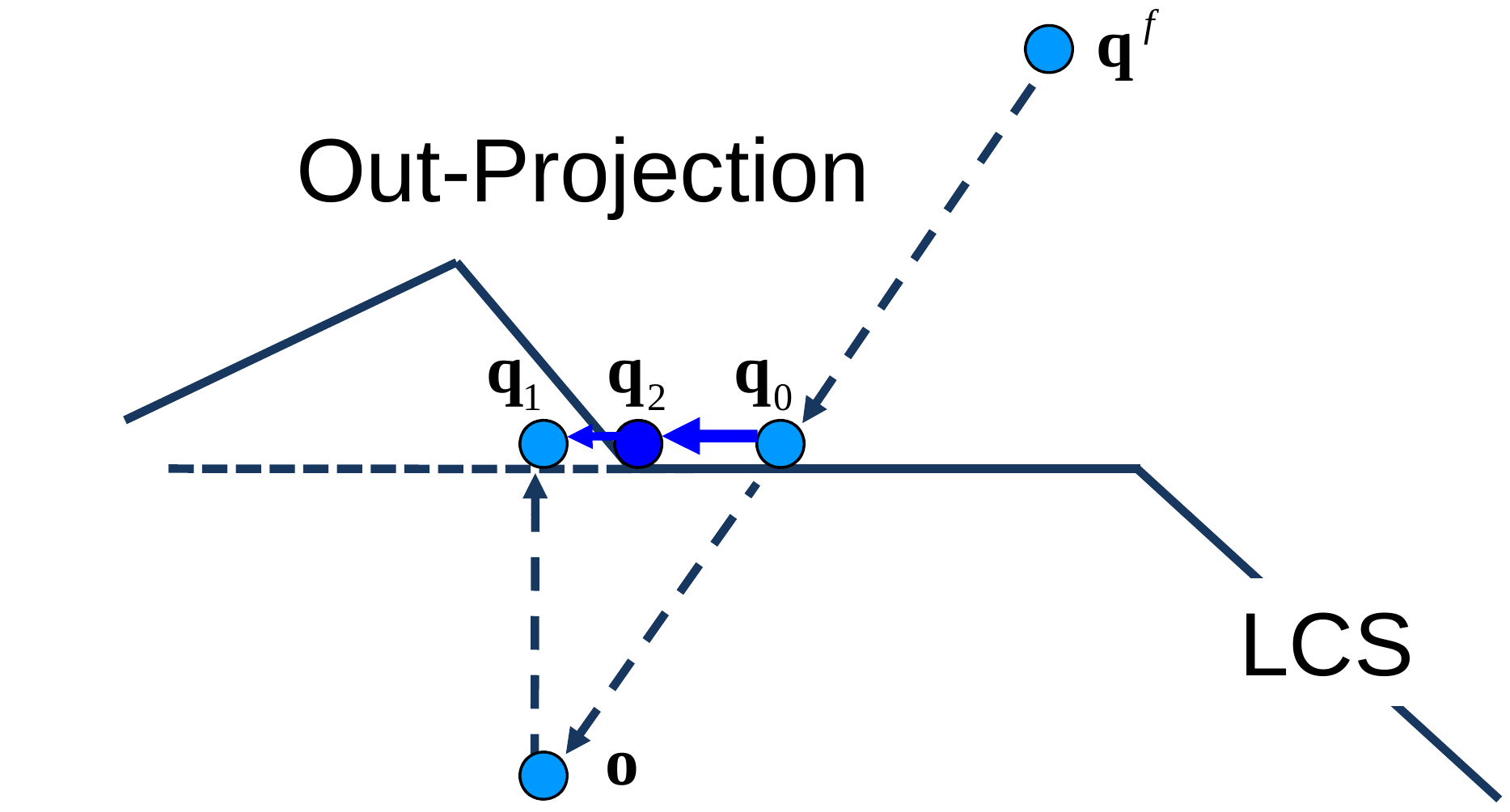,width=4.2cm}}
\subfigure[LCS around
$\mathbf{q}_2$]{\epsfig{file=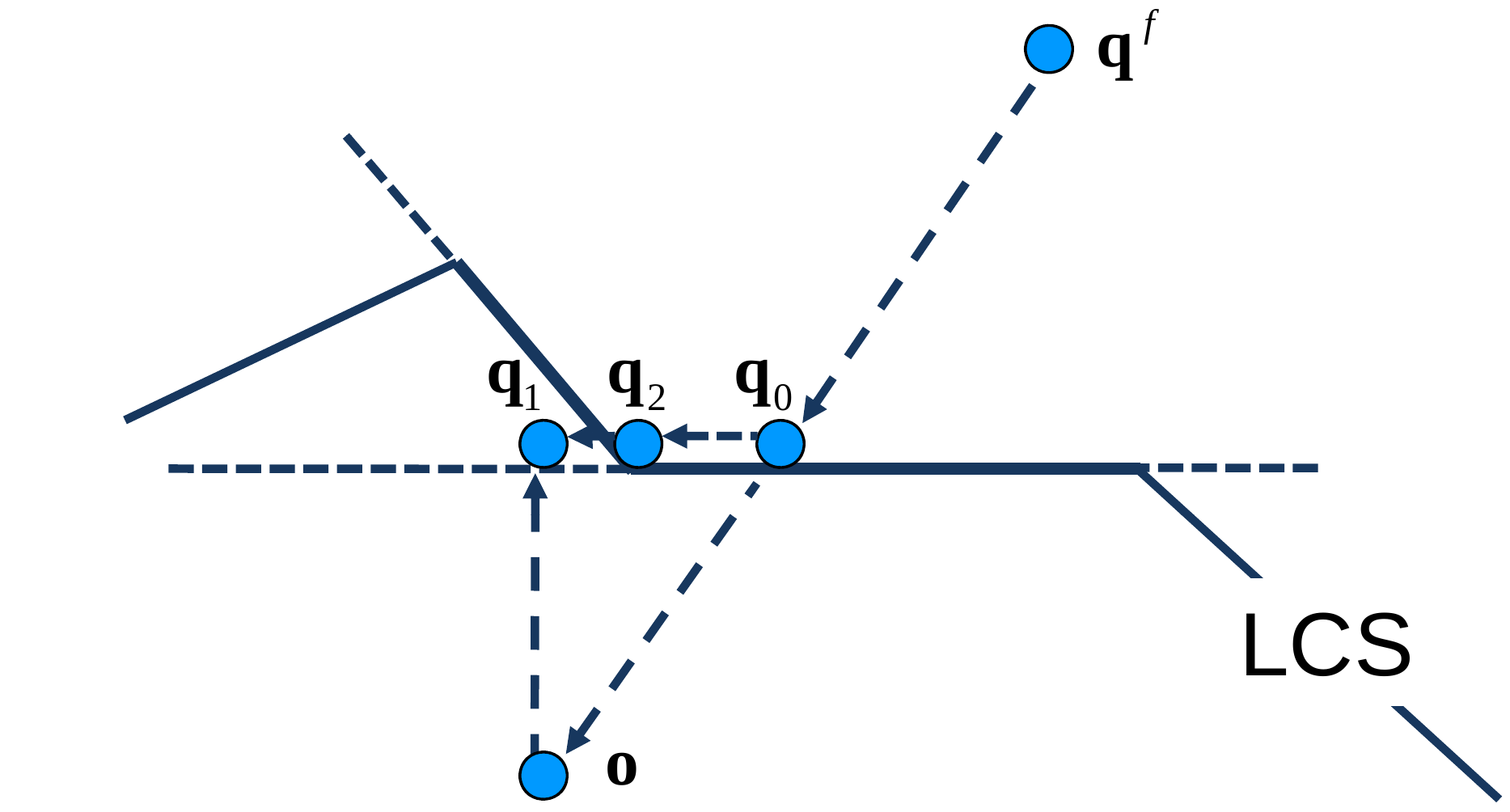,width=4.2cm}}
\subfigure[In-projection on to the LCS to obtain
$\mathbf{q}_3$]{\epsfig{file=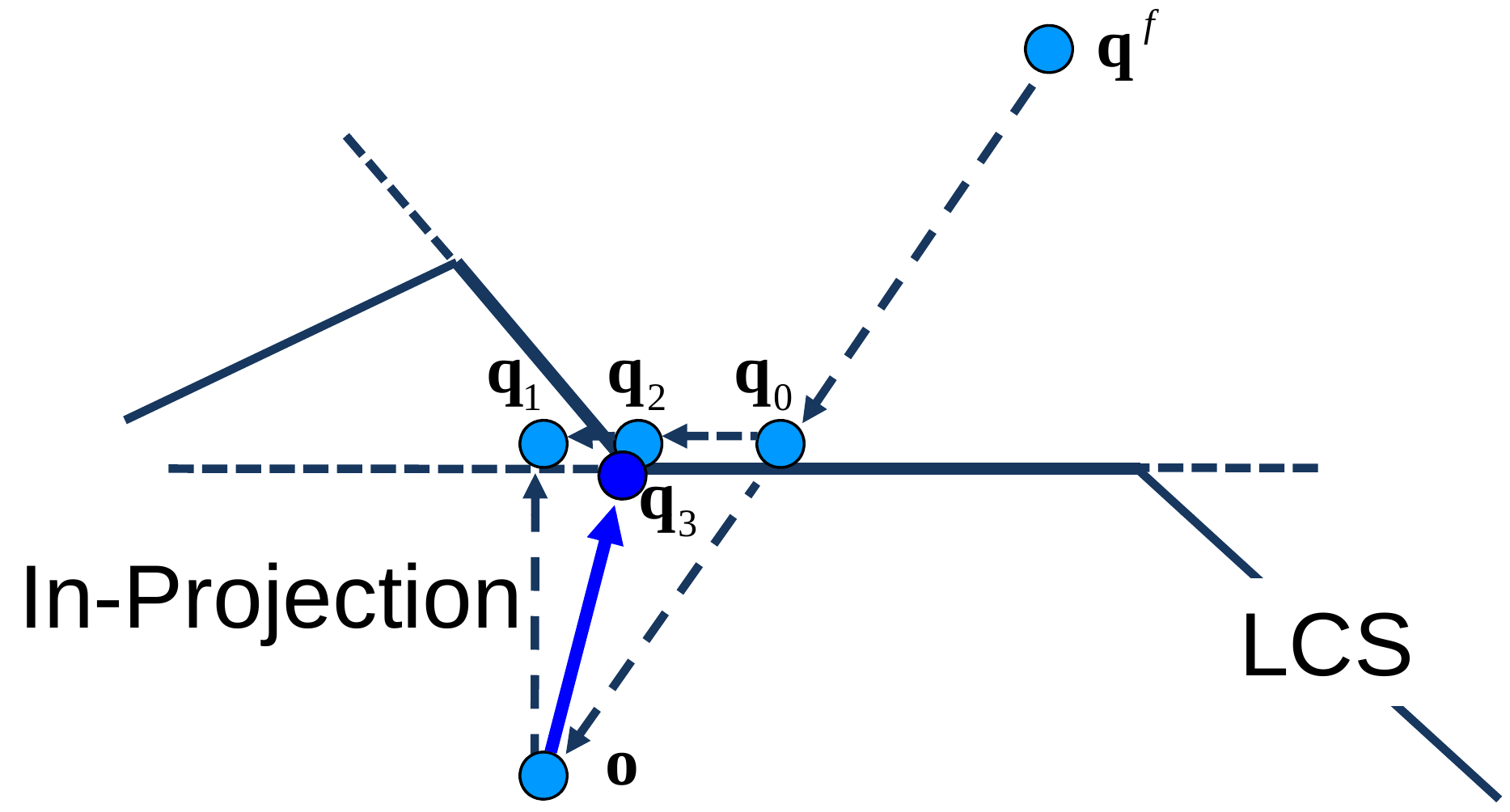,width=4.2cm}}
\caption{{\bf Iterative optimization of a PD for a non-convex
LCS.} The PD is computed after two iterations since $\mathbf{q}_1$
in (e) is in-collision. $PD=|| \mathbf{o} - \mathbf{q}_3
||$.}\label{fig:iteration}
\end{figure*}

\begin{figure*}[htb] \centering
\subfigure[Collision at
$\q_3$]{\epsfig{file=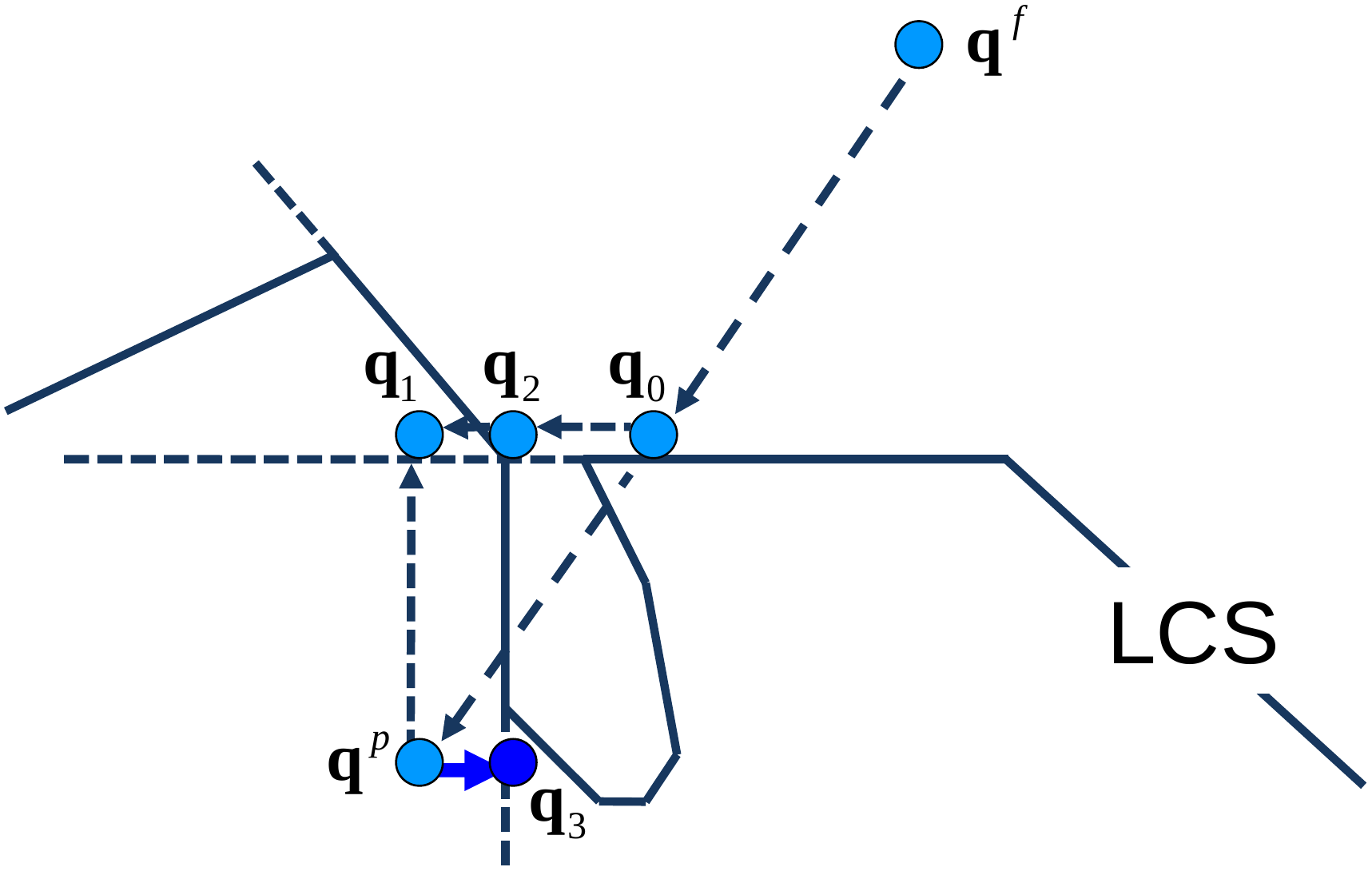,width=5cm}}
\subfigure[Out-projection to obtain
$\q_4$]{\epsfig{file=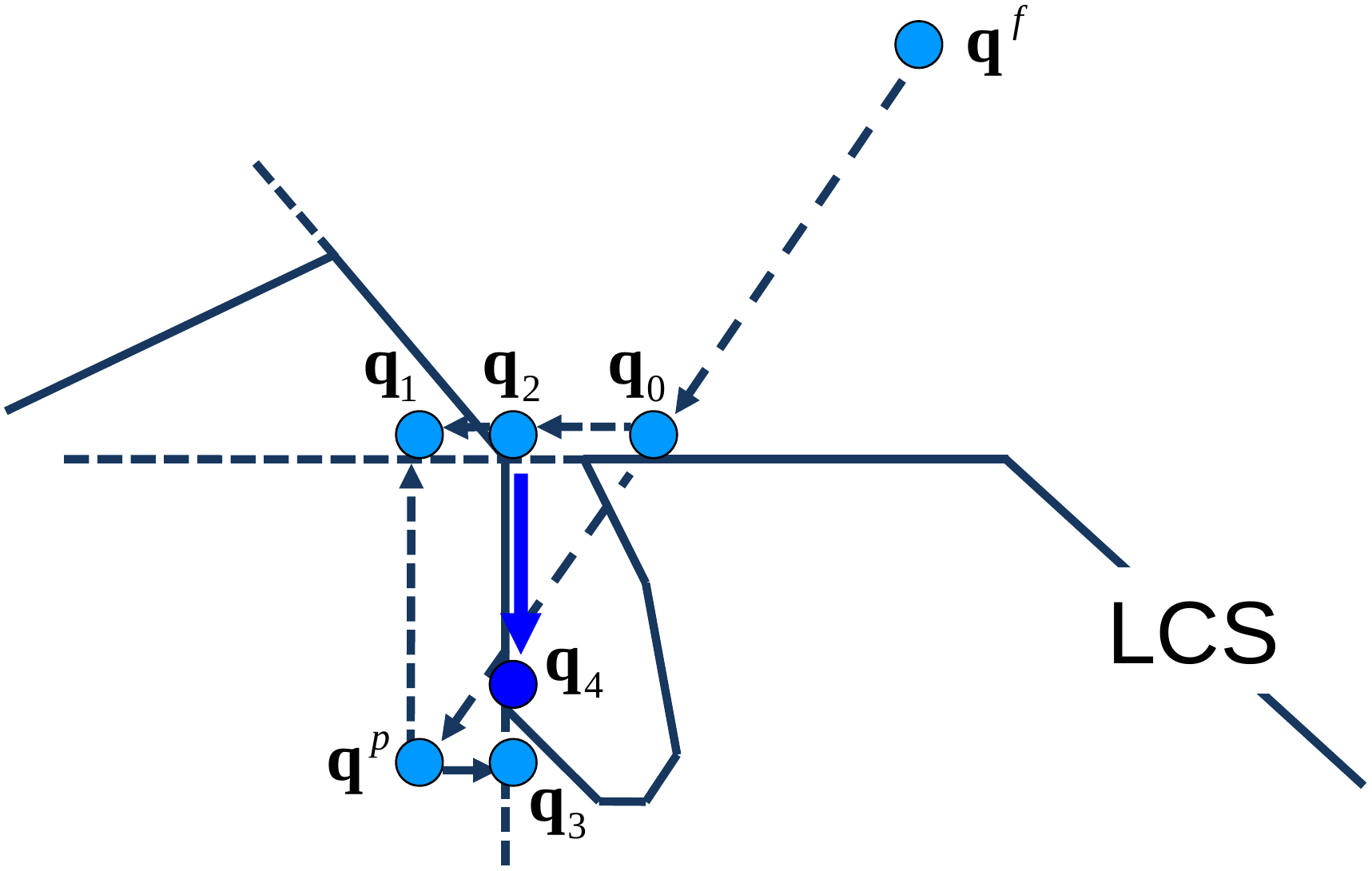,width=5cm}}
\subfigure[In-projection to obtain
$\q_5$]{\epsfig{file=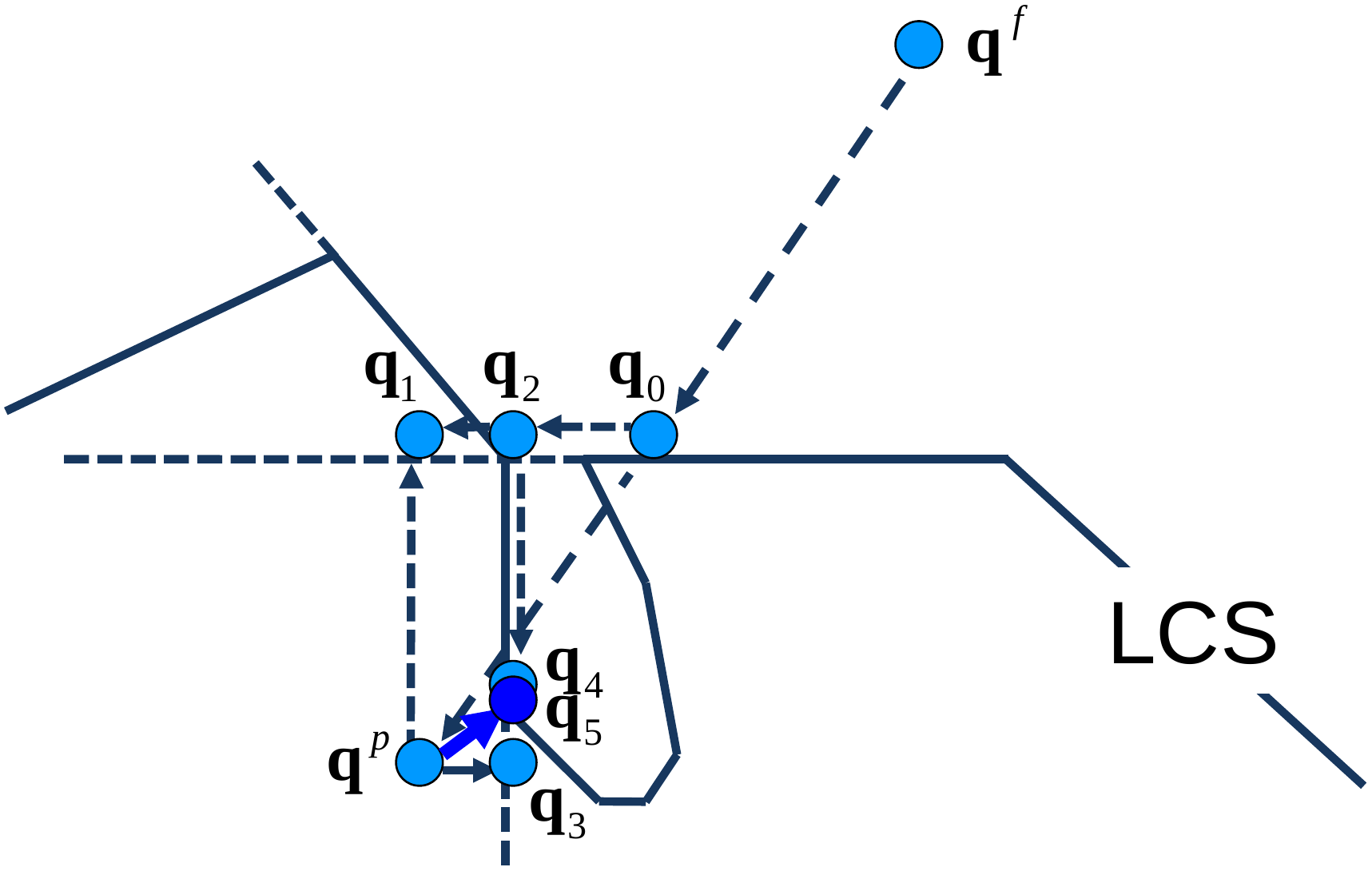,width=5cm}}
\caption{{\bf Iterative optimization of a PD for a more
complicated LCS.} Three iterations are required to obtain the PD
since $\mathbf{q}_3$ in (a) is in-collision. $PD=|| \mathbf{o} -
\mathbf{q}_5 ||$.}\label{fig:CompLCS}
\end{figure*}

\subsection{Contact-Space Localization and In-Projection}

The out-projection process explained in Sec. \ref{sec:outprj}
generates an in-contact configuration, which corresponds to a pair
of contact features, one in each of $\mov$ and $\fix$. In general,
this feature pair forms a vertex/face (VF), a face/vertex (FV) or
an edge/edge (EE) primitive contact, from which we can construct a
local contact space (LCS). Then, for the $i$th primitive contact,
we determine the equation of a contact plane,
$\mathbf{j}_i\mathbf{q}=c_i$ where $\mathbf{j}_i$ and $c_i$
respectively are the row vector of coefficients and the scalar
bias of the $i$th plane equation.

In some cases, this contact plane may degenerate to
a lower-dimensional subset of configuration space, such as a line
or a point. For example, contact between two collinear edges or a
vertex/edge (VE) contact produces a contact line equation, and
vertex/vertex (VV) contact produces a point. Our algorithm simply
ignores these VV or collinear contacts, since projection on to a
space of reduced dimensionality is unlikely to improve the PD.
In the case of a VE contact, we generate several VF primitive
contacts for all the faces that share the edge in the VE, and
intersect them. The result is likely to be over-constrained since
these contacts should be combined using the union operator
\cite{fap}. However, VE occurs rarely and the use of intersection
simplifies implementation and improves performance.
All other non-primitive contacts are also converted to several
primitive contacts.

Then, by stacking all the contact equations into a matrix, we can
construct a system of linear equations:
\begin{equation}\label{eq:ContactPlane}
\mathbf{J}\mathbf{q}=\mathbf{c}.
\end{equation}

We can then formulate in-projection as a minimization of the
squared Euclidean distance between the origin $\mathbf{o}$ and a
configuration $\mathbf{q}$ on the LCS, under the contact
constraints:
\begin{equation}\label{eq:Minimization}
\mathrm{Minimize} \; \|{\mathbf{q}}\|^{2}\ \; \mathrm{subject} \;
\mathrm{to}\ \; \mathbf{J}\mathbf{q}{\geq}\mathbf{c},
\end{equation}
where $\mathbf{J}\mathbf{q}{\geq}\mathbf{c}$ constrains
$\mathbf{q}$ to lie either on the boundary of the LCS or outside
it since  the system of equations
$\mathbf{J}\mathbf{q}=\mathbf{c}$ represents the LCS.

It is known \cite{Redon02gauss'least} that
Eq. \ref{eq:Minimization} is equivalent to a linear complementarity
problem (LCP), formulated as a search for a value of $\lambda$
which satisfies the following complementarity condition:
\begin{equation}\label{eq:LCP}
\left\{\begin{array}{c}
-\frac{1}{4}\mathbf{J}\mathbf{J}^{T}\lambda+\mathbf{c}\geq\mathbf{0}\\
\lambda\geq\mathbf{0}\\
(-\frac{1}{4}\mathbf{J}\mathbf{J}^{T}\lambda+\mathbf{c})^{T}\lambda=0
\end{array}\right.
\end{equation}
Then, $\mathbf{q}=\frac{1}{4}\mathbf{J}^{T}\lambda$.

There are several methods of solving an LCP such as Lemke's and
Dantzig's algorithm \cite{cottle2009linear}; however we choose a
projected Gauss-Seidel method for simplicity and stability
\cite{jourdan1998gsl}.
Then the LCP reduces to the search for a solution of the following
linear system:
\begin{equation}\label{eq:LinSys}
\mathbf{J}\mathbf{J}^{T}\lambda=4\mathbf{c}.
\end{equation}
And the Gauss-Seidel iteration is:
\begin{equation}\label{eq:GSIter}
\lambda_{k+1}=(\mathbf{D}+\mathbf{L})^{-1}(4\mathbf{c}-\mathbf{U}\lambda_{k}),
\end{equation}
where $\mathbf{D}+\mathbf{L}+\mathbf{U}=\mathbf{J}\mathbf{J}^{T}$,
and the matrices $\mathbf{D}$, $\mathbf{L}$ and $\mathbf{U}$
respectively represent the diagonal, strictly lower triangular,
and strictly upper triangular parts of the coefficient matrix
$\mathbf{J}\mathbf{J}^{T}$, and $k$ denotes the iteration number.
When a component of $\lambda$ becomes negative during the
iteration, we set it to zero: this is the {\em projection} step in
the projected Gauss-Seidel (PGS) algorithm.

A Gauss-Seidel iteration with an arbitrary symmetric positive
definite matrix \cite{matrix-computations1996} is known to
converge. $\mathbf{J}\mathbf{J}^{T}$ is positive semidefinite
since it is symmetric and
$\mathbf{x}^{T}\mathbf{J}\mathbf{J}^{T}\mathbf{x}=(\mathbf{J}^{T}\mathbf{x})^{T}\mathbf{J}^{T}\mathbf{x}=||\mathbf{J}^{T}\mathbf{x}||^{2}\geq0$
for $\mathbf{x}\in\mathbb{R}^{n}$. We remove the redundant rows of
$\mathbf{J}$ to make $\mathbf{J}$ full rank, and thereby promote
$\mathbf{J}\mathbf{J}^{T}$ from semidefinite to positive definite
since now
$\mathbf{x}^{T}\mathbf{J}\mathbf{J}^{T}\mathbf{x}=(\mathbf{J}^{T}\mathbf{x})^{T}\mathbf{J}^{T}\mathbf{x}=||\mathbf{J}^{T}\mathbf{x}||^{2}>0$
for all non-zero vectors $\mathbf{x}\in\mathbb{R}^{n}$. Therefore,
the Gauss-Seidel component of our algorithm always converges.

We store the contact features as a list, sorted by distance from
$\mathbf{o}$ to the corresponding contact plane. If a model is
complicated, the number of contact features can become excessively
high, and we select between only 10 and 30 contact feature pairs
to reduce the complexity of solving Eq.~\ref{eq:LCP}. This
produces satisfactory results in practice.

\subsection{Iteration}\label{sec:iteration}

As described in the overview presented in Sec. \ref{sec:overview},
we iteratively optimize a sample to refine the penetration depth.
Now we describe more details of this process, which relates to the
five steps in Sec. \ref{sec:overview} and Fig.\ref{fig:iteration}:

\begin{enumerate}

\item[(a)] An in-collision configuration (or the origin
$\mathbf{o}$) is given as input (Fig.~\ref{fig:iteration}(a)).

\item[(b)] Select a collision-free configuration $\q^f$ using the
technique described in Sec. \ref{sec:selecseed} (step 1 in Sec.
\ref{sec:overview} and Fig.~\ref{fig:iteration}(b)).

\item[(c)] Using the configurations $\mathbf{q}^f$ as source and
$\mathbf{o}$ as target, perform out-projection to obtain a contact
configuration $\q_0$ (step 2 in Sec. \ref{sec:overview} and
Fig.~\ref{fig:iteration}(c)).

\item[(d)] Construct an LCS around the contact configuration
$\mathbf{q}_0$ (step 3(a) in Sec. \ref{sec:overview} and
Fig.~\ref{fig:iteration}(d)). In this example, the LCS consists of
only a single contact plane.

\item[(e)] Perform in-projection on to the LCS to obtain $\q_1$
(step 3(b) in Sec. \ref{sec:overview} and
Fig.~\ref{fig:iteration}(e)). A proximity query is used to
classify the current configuration $\mathbf{q}_1$ as contact,
separation, or penetration. In this example, $\mathbf{q}_1$
corresponds to penetration; so the next step is to find a valid
in-contact configuration (step 4 in Sec. \ref{sec:overview}).

\item[(f)] Again perform out-projection to obtain $\q_2$ (step 2
in Sec. \ref{sec:overview} and Fig.~\ref{fig:iteration}(f)). This
requires  a non-colliding source configuration as well as an
in-collision target configuration.   The source configuration (\ie
$\q_0$) was obtained from the previous out-projection, and the
current configuration $\q_1$ can be used as the target
configuration.

\item[(g)] Construct the LCS around $\q_2$ (step 3(a) in Sec.
\ref{sec:overview} and Fig.~\ref{fig:iteration}(g)). This is a
convex cone (the intersection of two contact planes), since $\q_2$
consists of two contact feature pairs.

\item[(h)] In-projection is performed again (step 3(b) in Sec.
\ref{sec:overview} and Fig.~\ref{fig:iteration}(h)). The new
configuration $\q_3$ is in-contact and the PD that will be
returned is $||\q_3-\mathbf{o}||$ (step 4 in Sec.
\ref{sec:overview}).

\end{enumerate}

Fig.~\ref{fig:CompLCS} shows a more complicated scenario. The
first two steps obtain $\q_1$ and $\q_2$, as before. But this time
the proximity query at $\q_3$ detects penetration
(Fig.~\ref{fig:CompLCS}(a)). Thus we perform out-projection to
obtain $\q_4$ (Fig.~\ref{fig:CompLCS}(b)), construct the LCS, and
run in-projection again to get the final configuration $\q_5$
(Fig.~\ref{fig:CompLCS}(c)).

\subsection{Algorithm Termination}\label{sec:terminate}

Our algorithm is based on an iterative optimization technique
using in- and out-projections. The out-projection ensures that our
algorithm maintains an in-contact sample such that a proper upper
bound of PD is always obtained at any time of iteration. The
in-projection ensures that our algorithm always reduces the PD
value during iteration. These projections are iterated until a
result from in-projection corresponds to an in-contact
configuration so that no further reduction is possible. This way,
our algorithm finds a locally-optimal solution for PD.

Since the complexity of the local contact space (LCS) is bounded
above by $O(n^{6})$ where $n$ is the number of polygons in
polygon-soup models, the number of iterative projections in our
algorithm is finite. Moreover, since our algorithm reduces a PD
value at every iteration, no projection can be repeated for the
same LCS. Therefore, our algorithm always terminates after a
finite number of steps, which is typically 2 \~{} 3 in practice.

\subsection{Estimation of Local Penetration Depth}\label{sec:localpd}

When two objects intersect in multiple disjoint regions,
applications may require separate information about all the
penetrations, \ie local PDs instead of a single global PD. For
instance, in rigid-body dynamics, a local PD, defined as the
locally deepest penetrating points of two colliding objects
\cite{Guendelmann03sig,tang09sig}, is required to compute torques
or to stabilize the simulation. \cite{Guendelmann03sig} compute
local PDs from distance fields, whereas \cite{tang09sig} use
two-sided Hausdorff distance.

We propose an alternative method to derive local PDs from the
global PD information that our algorithm computes. Our method of
local PD estimation is based on the hypothesis that a pair of
locally deepest penetrating points, one from each object, coincide
when they are translated by the amount of the global PD.
In other words, the locally deepest penetrating
points are the ones that would be resolved later than any other
points; thus after the resolution by global PD, these points will
just touch each other. Since PD allows only rigid translational
motion, the amount of motion to resolve deepest penetrating points
is obtained by projecting the PD translation onto the penetrating
direction of the penetrating points, which is identical to the
normal of LCS corresponding to these points.

\begin{figure}[t] \centering
\subfigure[Global PD, $\mathbf{d}$, and two contact normals
$\mathbf{n}_1,\mathbf{n}_2$]{\epsfig{file=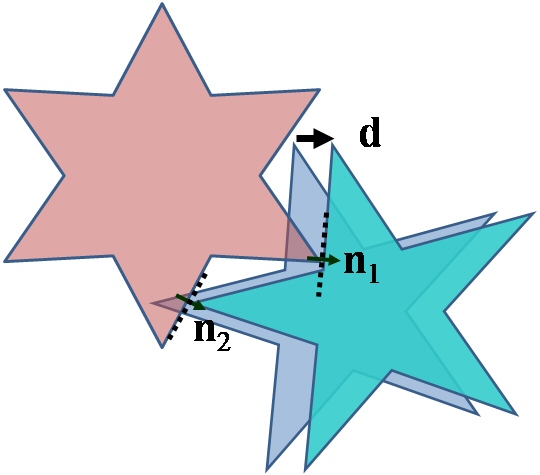,width=4cm}}
\subfigure[Two local PDs, $\mathbf{d}_1$ and
$\mathbf{d}_2$]{\epsfig{file=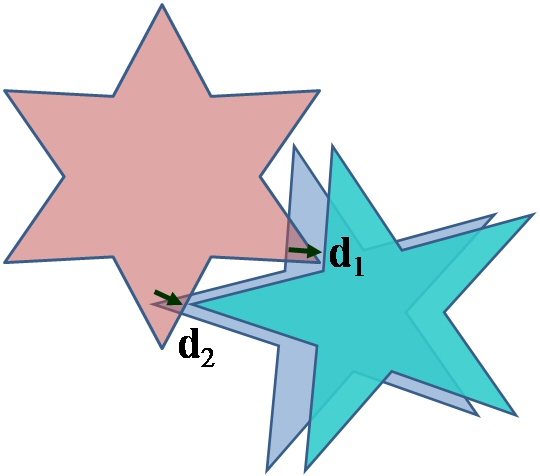,width=4cm}}
\caption{{\bf Estimation of local PDs from the global PD.} The red
($\fix$) and blue ($\mov$) objects are placed in a penetrating
configuration. The blue object is translated to the cyan one by
the global PD $\mathbf{d}$, and the two contact normals
$\mathbf{n}_1$ and $\mathbf{n}_2$ are obtained at the contact.
Projecting $\mathbf{d}$ on to $\mathbf{n}_1$ and $\mathbf{n}_2$
gives the two local PDs $\mathbf{d}_1$ and $\mathbf{d}_2$.
}\label{fig:lpd}
\end{figure}

More specifically, we compute the local PD $\mathbf{d}_i$ of the
$i$th interpenetrating region as follows (also see
Fig.~\ref{fig:lpd}):
\begin{enumerate}
\item Given two overlapping models $\mov$ and $\fix$, compute
their global PD $d\equiv\PD$ using the PD algorithm already
presented.

\item Translate $\mov$ by $\mathbf{d}$ to become
$\mov(\mathbf{d})$, and compute the contact normal
$\mathbf{n}_{i}$ for each contact $i$ between $\mov(\mathbf{d})$
and $\fix$.

\item Project $\mathbf{d}$ on to each $\mathbf{n}_{i}$ to get the
local PD $\mathbf{d}_{i}$; \ie
$\mathbf{d}_{i}=(\mathbf{d}\cdot\mathbf{n}_{i})\mathbf{n}_{i}$.
\end{enumerate}

Note that our definition of local PD is not equivalent to previous
definition of \cite{Guendelmann03sig,tang09sig}, but it is
computationally efficient. Our local PD algorithm is a simple
by-product of our global PD algorithm, and does not require any
costly and memory-intensive precomputation of distance fields
unlike \cite{Guendelmann03sig}, and no expensive Boolean
intersection is required unlike \cite{tang09sig}.
Also note that, according to our definition of local
PD, even if $\mathbf{d}$ is not zero but if $\mathbf{d}$ is normal
to $\mathbf{n}$, the local PD becomes zero. This is the case where
the corresponding contact features are sliding.

\subsection{Computational Complexity of Algorithm}

Our algorithms consist of two main procedures: out- and
in-projections. In our algorithm, the out-projection is
implemented using translational CCD. The computational complexity
of translational CCD is similar to that of distance computation
using bounding volume hierarchies (BVHs), since our translational
CCD is equivalent to computing the minimal directional distance
(MDD) between  BVHs.  According to \cite{LGLM00}, the cost
function $T$ for distance calculation can be analyzed as $T=N_{bv}
\times C_{bv}+N_{p} \times C_{p}$, where $N_{bv}$ is the number of
bounding volume pair operations, $C_{bv}$  is the cost for
pairwise distance computation for BVs, $N_{p}$ is the number of
primitive pair operations, and $C_{p}$ is the cost for pairwise
distance computation for primitives. Translational CCD has a
similar cost function, but $C_{bv}$ and $C_{p}$ now correspond to
the costs to compute the MDD for BVs and primitives, respectively.
Moreover, we use an iterative method to calculate MDD, and each
iteration requires Euclidean minimum distance calculation between
a BV or triangle pair, which is a constant. In practice, the
average number of iterations for MDD calculation is a small
constant; for example, for two bunny models with 40K triangles
each, as shown in Fig. \ref{fig:models}, it is 1.7. This means
that the computational cost for translation CCD is only 1.7 times
the cost for a minimum distance query (the function $T$ above).

The main computational cost for in-projection lies in solving
Eq.~\ref{eq:LinSys}. Since $\mathbf{J}\mathbf{J}^{T}$ is an $n
\times n$ matrix where $n$ is the number of contact features
returned by out-projection, solving the linear system takes
$O(n^{3})$ time, but the Gauss-Seidel iterative solver would take
$O(N_{G}n^{2})$ time where $N_{G}$ is the number of Gauss-Seidel
iterations. Thus, the total computational complexity of our
algorithm is $(T+O(N_{G}n^{2}))N$ where $N$ is the total number of
iterations for in- and out-projections.

%% file: results.tex
\begin{figure}[htb]
\centering \epsfig{file=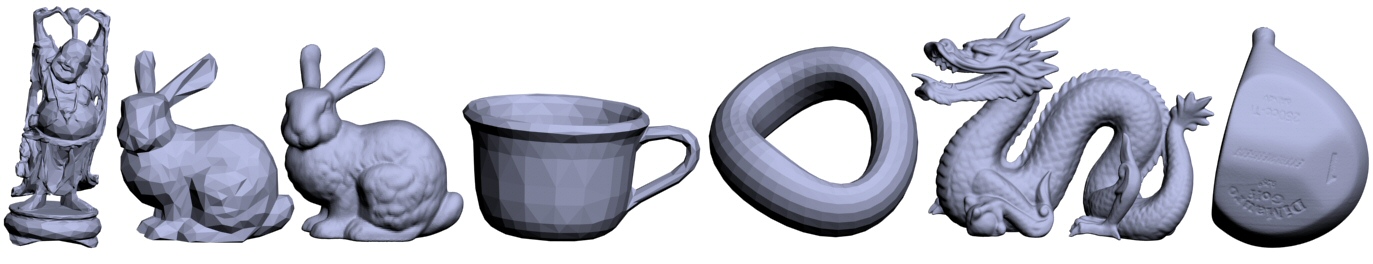,width=8.5cm}
\epsfig{file=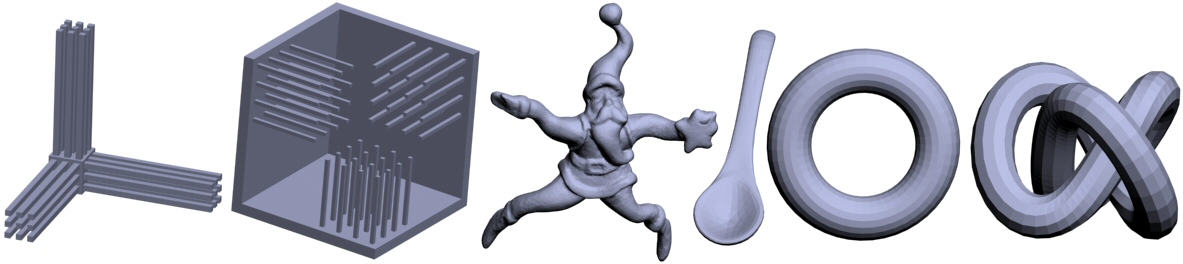,width=8.5cm} \caption{{\bf
Benchmarking models with triangle counts.} {\bf Top row:} Buddha
(10K), Bunny1 (1K), Bunny2 (40K), Cup (1K), Distorted-torus
(1.33K), Dragon (174K), Golf-club (105K) {\bf Bottom row:} Grate1
(0.54K), Grate2 (0.94K), Santa (152K), Spoon (1.34K), Torus (2K),
and Torusknot (3K).}\label{fig:models}
\end{figure}

\section{Results and Discussion}


We now present the results obtained by our PD computation
algorithm in various scenarios, and discuss some of the
implementation issues.

\subsection{Implementation and Benchmarks}

%

We implemented our PD computation algorithm\footnote{The source
codes are available for download at
\url{http://graphics.ewha.ac.kr/PolyDepth}} using Microsoft Visual
C++ 8.0 under the Windows XP operating system. We tested the
algorithm on a PC equipped with an AMD 2.22GHz CPU and 2Gb of RAM.
The benchmarking models that we used in these experiments are
shown in Fig.~\ref{fig:models}. The complexities of these models
range from 0.54K to 174K triangles, and their topologies contain
many holes and self-intersections (\ie polygon-soups). We
constructed a number of benchmarking scenarios for these models,
including random configurations, predefined sequences, and
collisions within a rigid-body dynamics simulation.

\subsubsection{Random Configuration Scenario}\label{sec:rand}

\begin{figure}[htb] \centering
{\epsfig{file=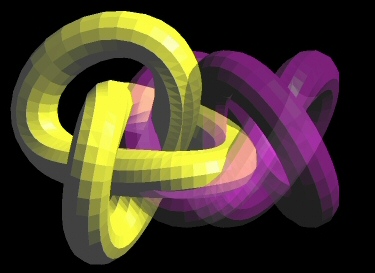,height=2.5cm}}
{\epsfig{file=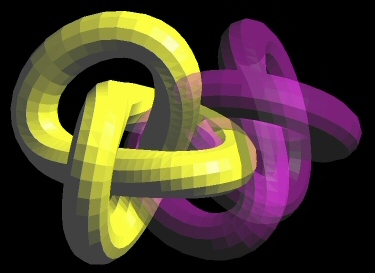,height=2.5cm}}
\caption{\bf Two Torusknot models in a random
configuration.}\label{fig:torusknot_rot}
\end{figure}

We created 100 random configurations for several of the models,
including the Torusknot, Buddha, Bunny2 and Dragon. We place a
model at a fixed configuration, and translated a copy of the same
model through a distance equal to half the size of its bounding
volume, while changing its orientation, all at random creating
many deeply penetrating configurations.
Fig.~\ref{fig:torusknot_rot} shows an example in which two copies
of the Torusknot model intersect.

\begin{figure}[htb]
\centering \subfigure[Number of contact
features]{\epsfig{file=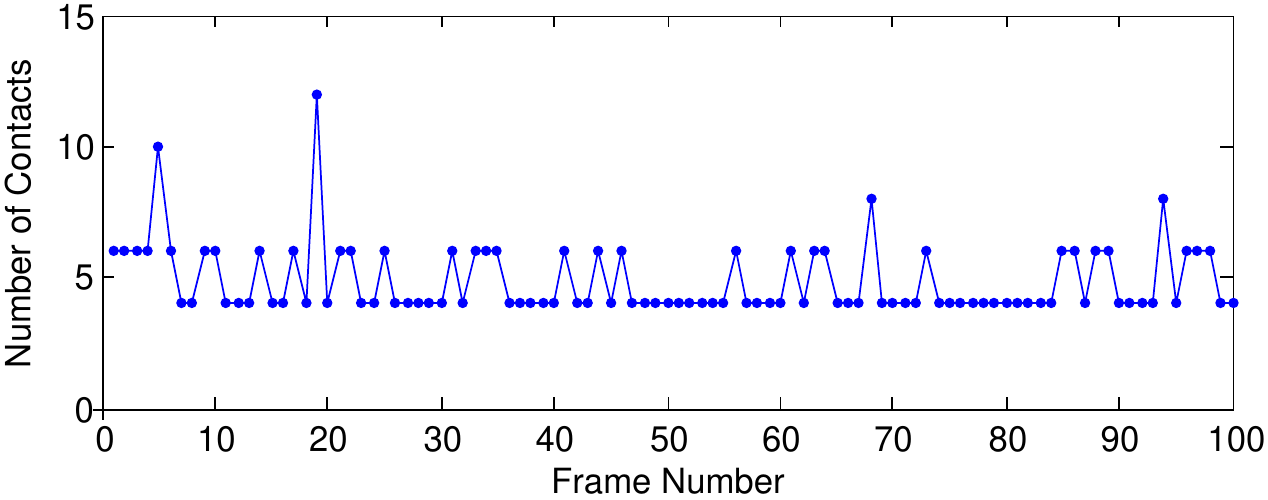,width=7cm}}
\subfigure[Number of
iterations]{\epsfig{file=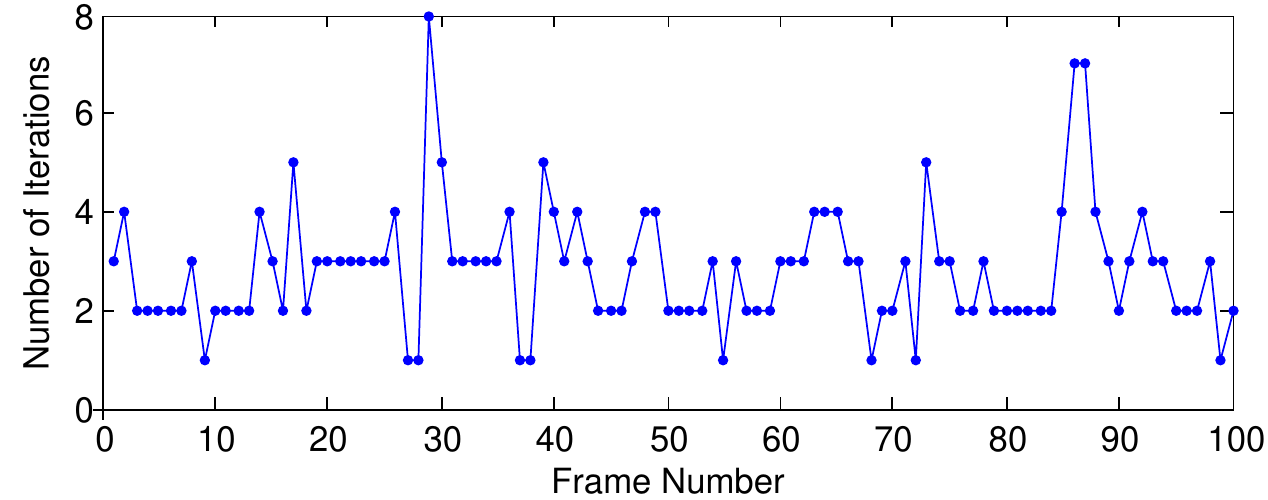,width=7cm}}
\subfigure[Computation time
(msec)]{\epsfig{file=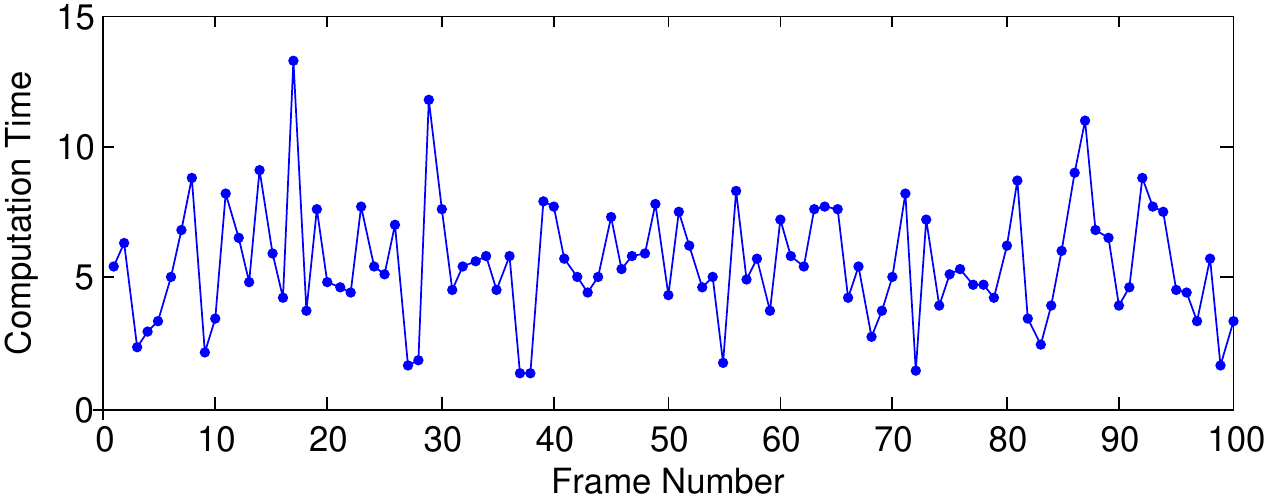,width=7cm}}
\caption{{\bf Graphs of PD computation for two Torusknot models in 100
random configurations.}}\label{fig:graph_torusknot_rotation}
\end{figure}

We use a centroid difference to find the initial collision-free
configuration. The performance of our algorithm is characterized
in Table \ref{tab:random}. These values are averaged for all the
collision frames; the number of contacts means the number of
contact feature pairs used to construct the LCS and the number of
iterations is the total number of in- and out- projections for
computing the PD. In Fig. \ref{fig:graph_torusknot_rotation}, we
also show the number of contact feature pairs, number of
iterations, and computation time for two intersecting Torusknots
in random configurations.
\begin{table}[htb]
\tbl{{\bf Performance of our algorithm with random
configurations.}}{
 \centering 
\begin{tabular}{c||c|c|c}
  \hline
  Model & Time (msec) & No. of Contacts & No. of Iterations \\
  \hline
  \hline
  Torusknot & 5.53 & 4.84 & 2.81 \\
  \hline
  Buddha & 6.23 & 5.04 & 1.92 \\
  \hline
  Bunny2 & 7.55 & 8.25 & 2.16 \\
  \hline
  Dragon & 12.76 & 11.92 & 2.29 \\
  \hline
\end{tabular}}
\label{tab:random}
\end{table}

\begin{figure*}[htb] \centering
{\epsfig{file=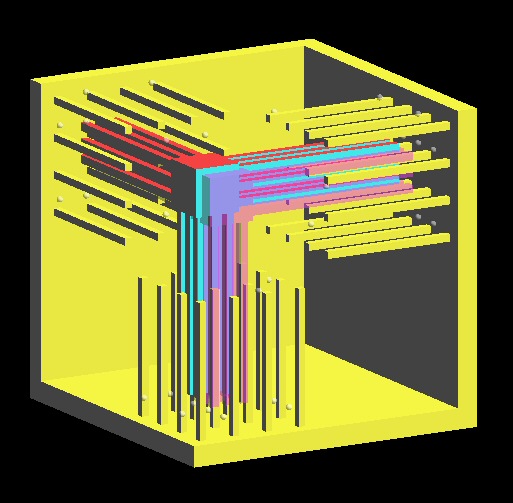,height=2.3cm}}
{\epsfig{file=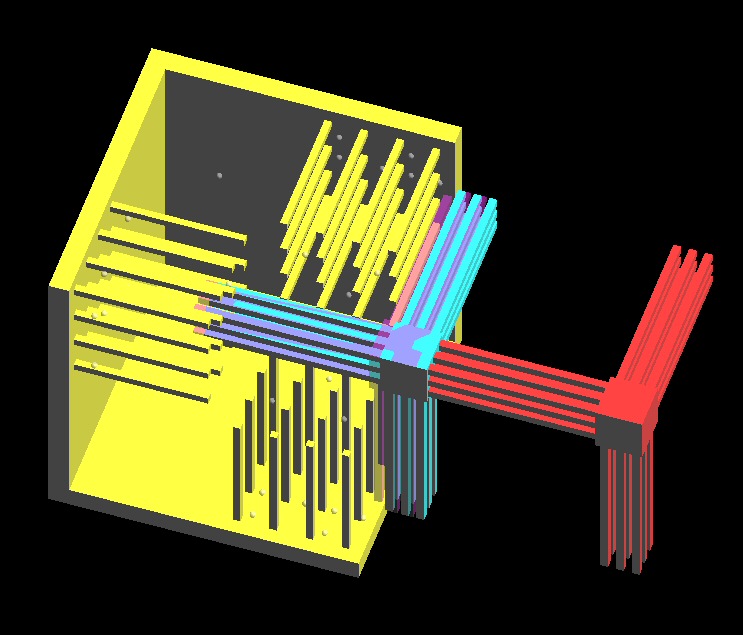,height=2.3cm}}
{\epsfig{file=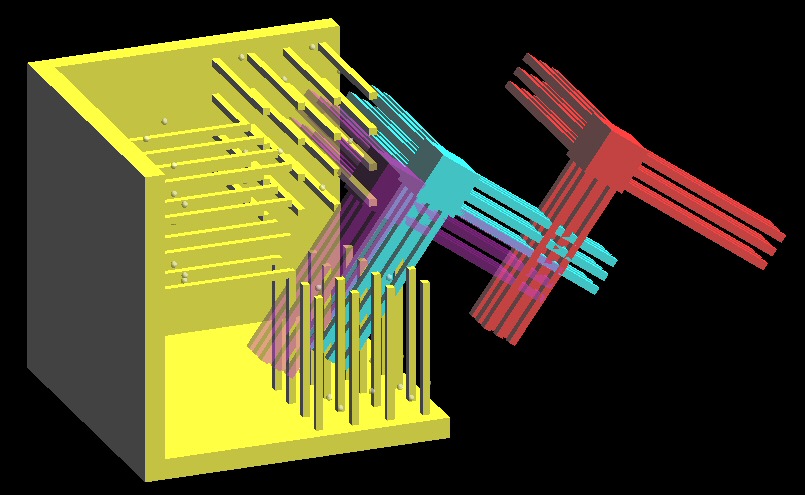,height=2.3cm}}
{\epsfig{file=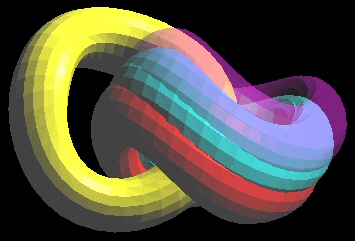,height=2.3cm}}
{\epsfig{file=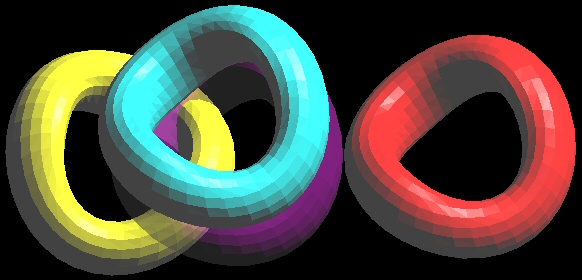,height=2.3cm}}
\caption{{\bf Using maximally clear configurations and
sampling-based search to determine collision-free configurations
for the Grate and Distorted-torus models.} Obstacle $\fix$ is
yellow, and the input in-collision configuration of $\mov$ is
semitransparent magenta. The initial collision-free configuration
of $\mov$ is red, and the configuration of $\mov$ translated by PD
is shown in cyan.}\label{fig:bisec_grate}
\end{figure*}

For the more topologically challenging models such as Grate1,
Grate2 and Distorted-torus, we use maximally clear configurations
and sampling-based search scheme to find the initial
collision-free configurations, because of their very concave
geometries. Results are shown in Fig. \ref{fig:bisec_grate}.

\begin{figure}[htb]
\centering \subfigure[PD Magnitude
(Bunny1)]{\epsfig{file=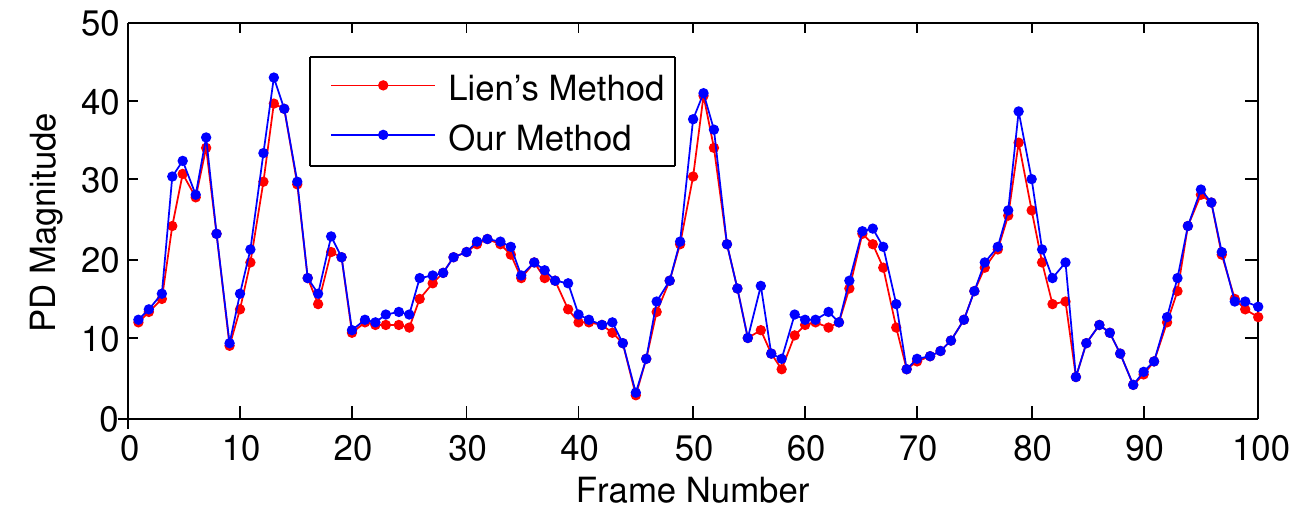,width=8cm}}
\subfigure[Computational Time
(Bunny1)]{\epsfig{file=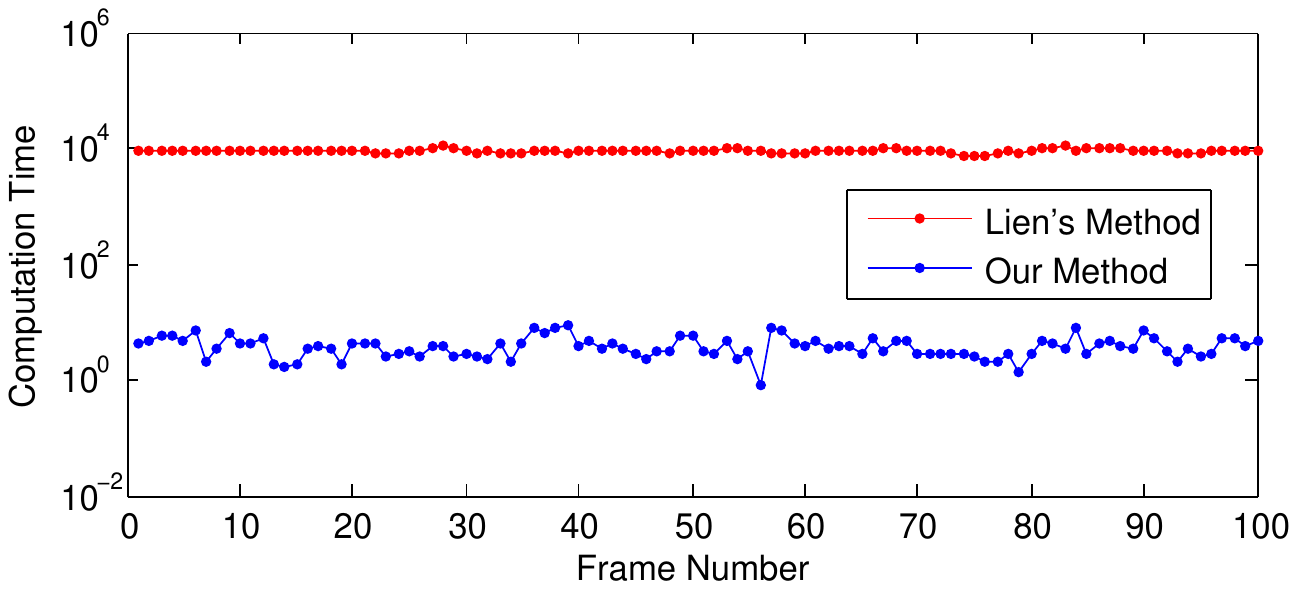,width=8cm}}
\subfigure[PD Magnitude (Distorted
Torus)]{\epsfig{file=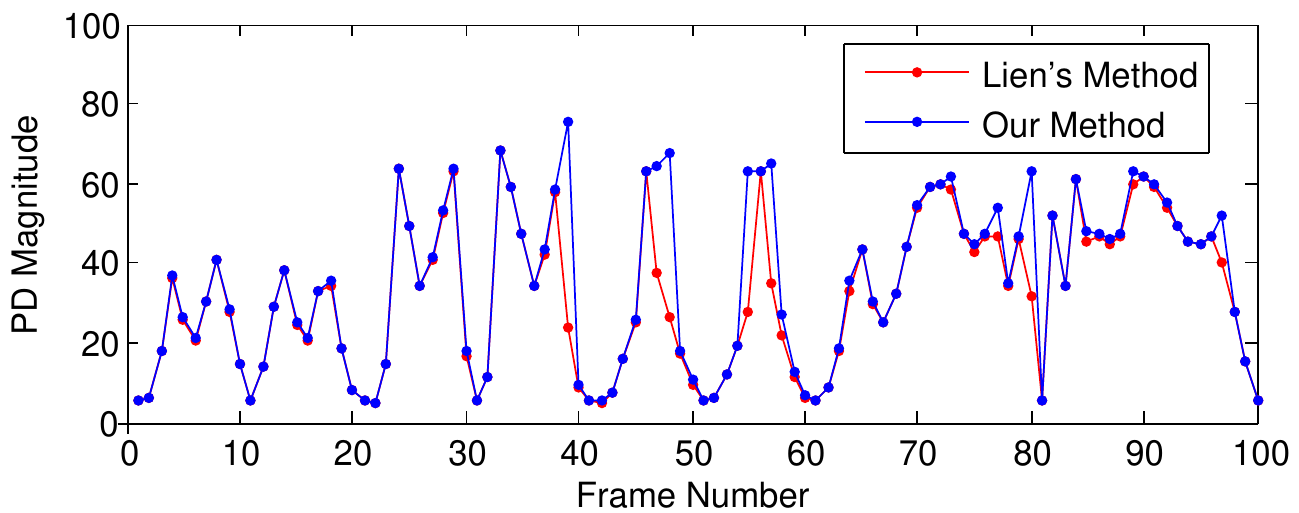,width=8cm}}
\caption{{\bf Graphs of PD comparisons using our method and Lien's
method \protect\cite{lien09} at random
configurations.}}\label{fig:comparison_bunny}
\end{figure}

In order to assess the accuracy of our PD algorithm, we used
Lien's open source library \cite{lien09} to compute
near-exact Minkowski sums and PDs for the same models. Lien's
method is near-exact in the sense that it ignores low-dimensional
boundaries of the Minkowski sums which have a volume close to
zero. Fig. \ref{fig:comparison_bunny} shows the PDs and the
computational times for two Bunny1 models and two Distorted-torus
models using our method and Lien's.


The results are very close, even though our method is about 2000
times faster than Lien's. A quantitative comparison can be
obtained by defining a metric for the relative error between
approximate and exact PDs as follows:
\[\mathbf{e}_{PD,relative1}=\frac{\|\mathbf{PD}_{appoximate}-\mathbf{PD}_{exact}\|}{\max{w(\mov\oplus-\fix)}}\], where
$w$ is the width of an object in a given direction $\mathbf{n}$,
defined as the distance between supporting planes normal to
$\mathbf{n}$. Since the PD is the minimum distance from the origin
to the boundary of the Minkowski sums, $\max{w(\mov\oplus-\fix)}$
is an upper bound on the PD. However, since this metric depends on
orientation, it can widely vary. An alternative metric which is
unaffected by orientation is:
\[\mathbf{e}_{PD,relative2}=\frac{\|\mathbf{PD}_{approximate}-\mathbf{PD}_{exact}\|}{2\overline{v}(\mov)+2\overline{v}(\fix)}\],
where $\overline{v}$ denotes the average magnitude of the vertices
of an object, \ie
$\overline{v}=\frac{1}{N}\sum\limits_i\|\mathbf{v}_{i}\|$. If the
objects are spheres, $\mathbf{e}_{PD,relative1}
=\mathbf{e}_{PD,relative2}$, since the $\overline{v}$s would be
the radii of the spheres. If the PDs from Lien's method are
considered to be exact, then the errors in the results from our
algorithm are listed in Table \ref{tab:relative_error}.

\begin{table}[htb]
 \tbl{{\bf Average PD errors}}{
\centering 
\begin{tabular}{c||c|c}
  \hline
  Model & Mean error (\%) & Median error (\%) \\
  \hline
  \hline
  Bunny1 & 0.500 & 0.165 \\
  \hline
  Distorted-Torus & 1.165 & 0.066 \\
  \hline
\end{tabular}}\label{tab:relative_error}
 \end{table}

\subsubsection{Predefined Path Scenario}

\begin{figure}[htb] \centering
{\epsfig{file=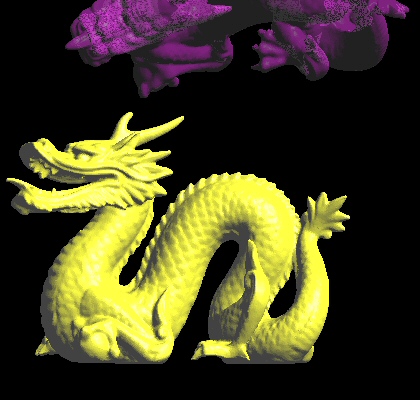,height=2.5cm}}
{\epsfig{file=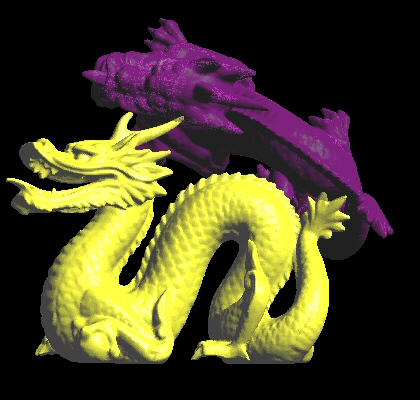,height=2.5cm}}
{\epsfig{file=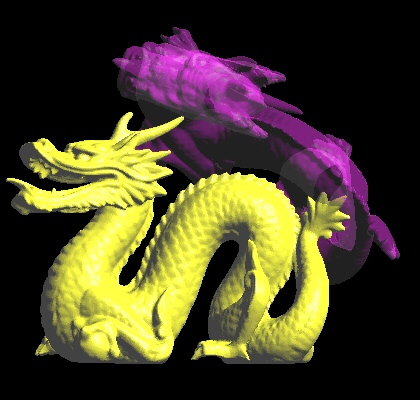,height=2.5cm}}
\caption{{\bf Path for the Dragon model generated by a physics
simulator.}}\label{fig:dragon_fall}
\end{figure}

\begin{figure}[htb] \centering
{\epsfig{file=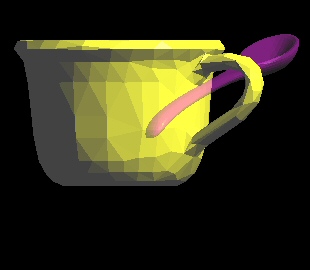,height=2.3cm}}
{\epsfig{file=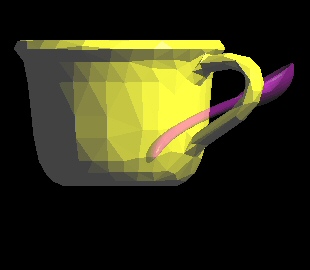,height=2.3cm}}
{\epsfig{file=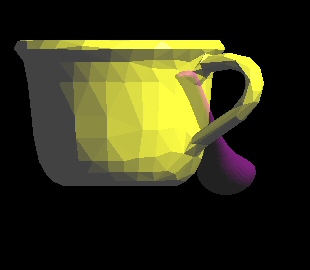,height=2.3cm}}
\caption{{\bf Path of the Spoon model relative to Cup generated by
a physics simulator.}}\label{fig:spoon_cup_hand}
\end{figure}

\begin{figure}[htb]
\centering \subfigure[Number of Contact
Features]{\epsfig{file=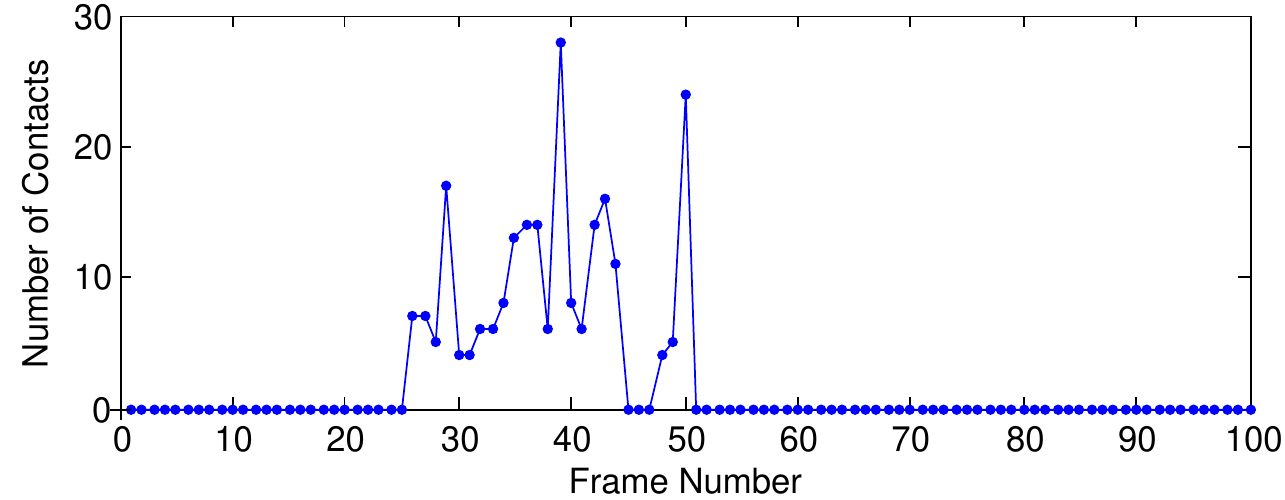,width=8cm}}
\subfigure[Number of
Iterations]{\epsfig{file=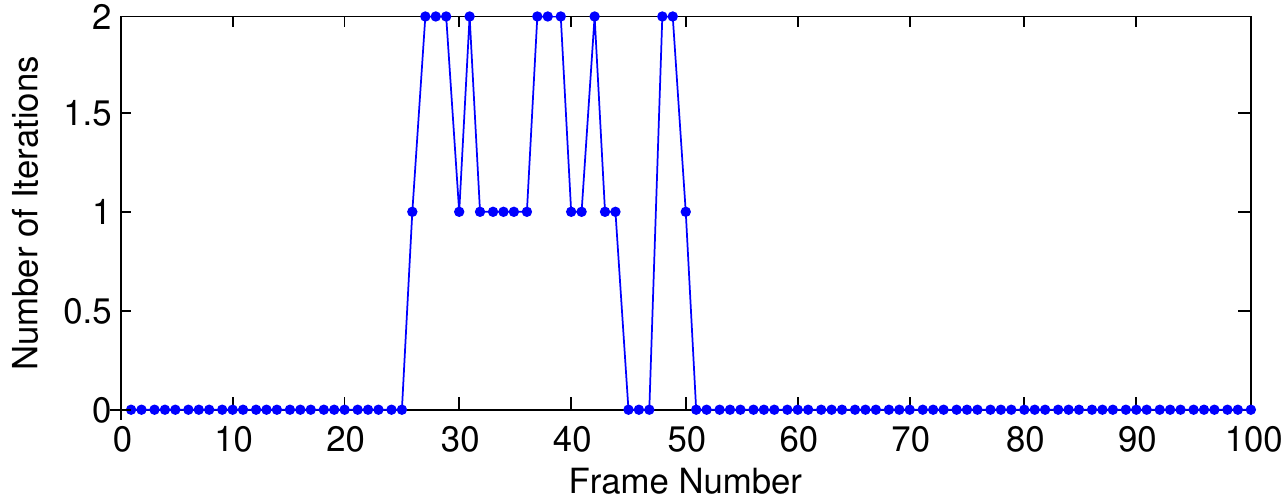,width=8cm}}
\subfigure[Computation Time
(msec)]{\epsfig{file=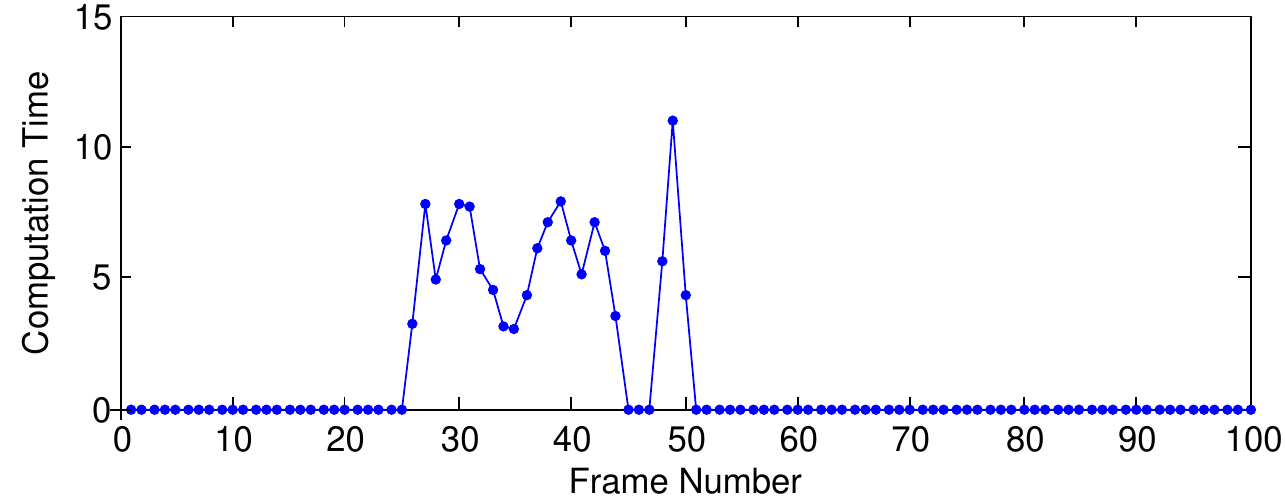,width=8cm}}
\caption{\bf Graphs of PD computation using two colliding Dragon
models moving on predefined
paths.}\label{fig:graph_dragon_dynamics}
\end{figure}

Using Havok$^{\mathrm{TM}}$\footnote{\url{http://www.havok.com}},
a rigid body dynamics simulator, we generated paths for the Dragon
(Fig. \ref{fig:dragon_fall}), and Spoon and Cup  models (Fig.
\ref{fig:spoon_cup_hand}), and tested our PD algorithm along these
paths. Fig. \ref{fig:graph_dragon_dynamics} shows the number of
contact feature pairs, the number of iterations, and the
computational time for the Dragon model. One Dragon is fixed in
space and a copy falls under gravity, as shown  in Fig.
\ref{fig:dragon_fall}. The centroid difference explained in Sec.
\ref{sec:dir} is used to find an initial collision-free
configuration.

\begin{figure}[htb] \centering
\subfigure[]{\epsfig{file=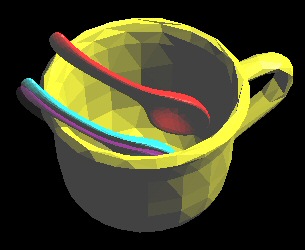,height=2.1cm}}
\subfigure[]{\epsfig{file=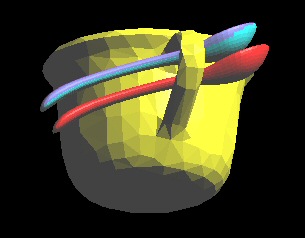,height=2.1cm}}
\subfigure[]{\epsfig{file=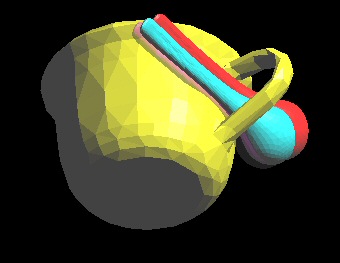,height=2.1cm}}
\caption{{\bf PD results using maximally clear configurations for
Spoon and Cup models moving on predefined
paths.}}\label{fig:clear_conf}
\end{figure}

\begin{figure*}[htb] \centering
\subfigure[]{\epsfig{file=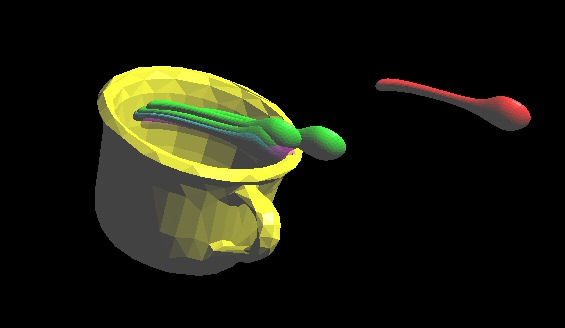,height=2.3cm}}
\subfigure[]{\epsfig{file=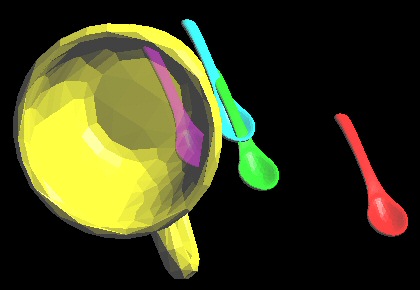,height=2.3cm}}
\subfigure[]{\epsfig{file=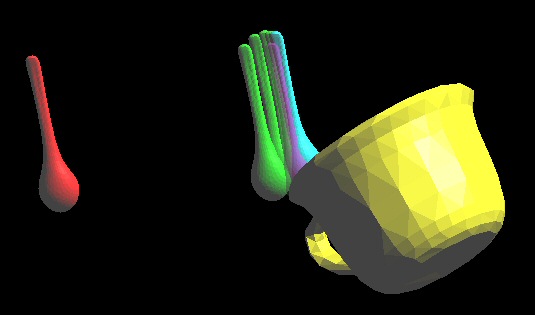,height=2.3cm}}
\subfigure[]{\epsfig{file=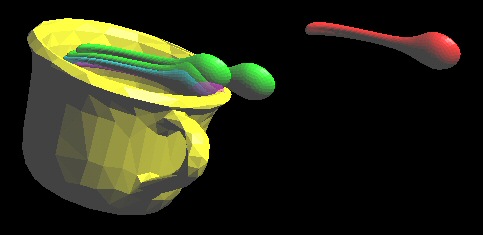,height=2.3cm}}
\subfigure[]{\epsfig{file=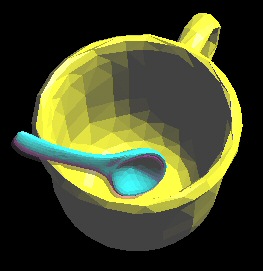,height=2.05cm}}
\subfigure[]{\epsfig{file=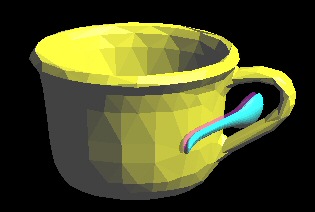,height=2.05cm}}
\subfigure[]{\epsfig{file=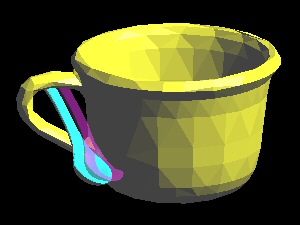,height=2.05cm}}
\subfigure[]{\epsfig{file=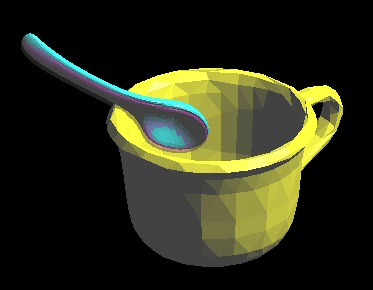,height=2.05cm}}
\subfigure[]{\epsfig{file=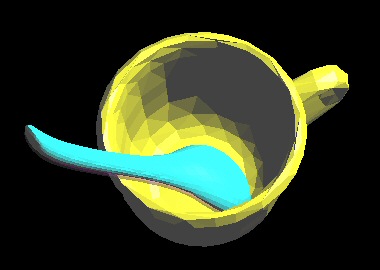,height=2.05cm}}
\subfigure[]{\epsfig{file=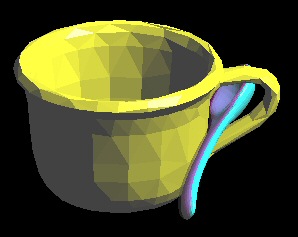,height=2.05cm}}
\caption{{\bf PD results for the Spoon and Cup models moving on
predefined paths.} Obstacles are shown in yellow, input
in-collision configuration in semitransparent magenta, initial
collision-free configurations in red, contact configurations from
in- and out-projections in green, and the configuration translated
by the PD is shown in cyan. In (a) $\sim$ (d), the centroid
difference was used to find a collision-free configurations. In
(e)$\sim$ (j), motion coherence was used.}\label{fig:spoon_cup}
\end{figure*}

Fig.~\ref{fig:spoon_cup} shows scenes from the motion of the Spoon
and Cup models along a predefined path. Collision-free
configurations were found using the centroid difference (a$\sim$d)
and using motion coherence (e$\sim$j).
Fig.~\ref{fig:clear_conf} shows the results when maximally clear
configurations were used for the same scene.
Fig.~\ref{fig:pre_free}(a) shows the maximally clear configuration
for the Cup. The centroid difference can generate a poor
collision-free configuration when the underling geometry and
topology are complicated, as demonstrated in
Fig.~\ref{fig:spoon_cup}(b). But in the same situations, the
maximally clear configurations are very satisfactory, as
demonstrated in Fig.~\ref{fig:spoon_cup}(e)$\sim$(j). The
performance of our algorithm for predefined paths is summarized in
Table \ref{tab:dynamics}.

\begin{figure*}[htb] \centering
{\epsfig{file=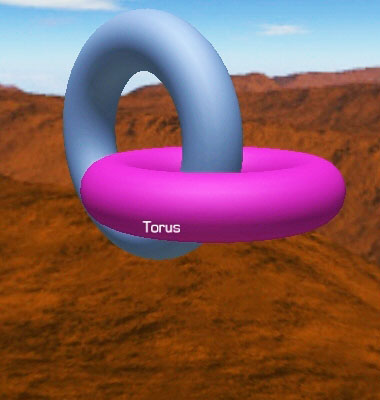,height=2.3cm}}
{\epsfig{file=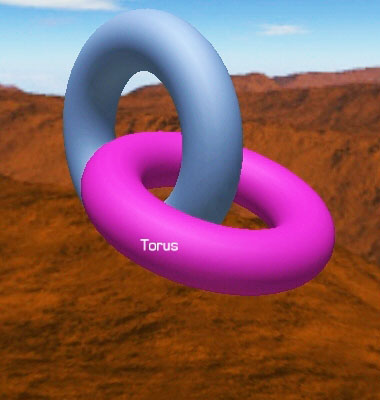,height=2.3cm}}
{\epsfig{file=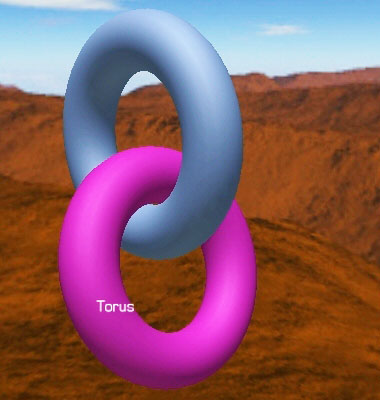,height=2.3cm}}
{\epsfig{file=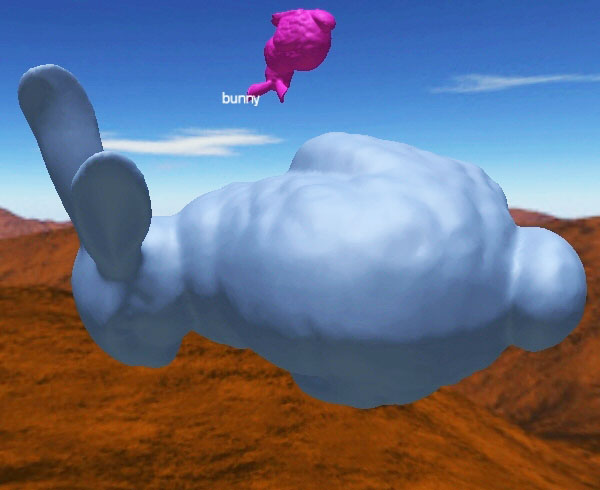,height=2.3cm}}
{\epsfig{file=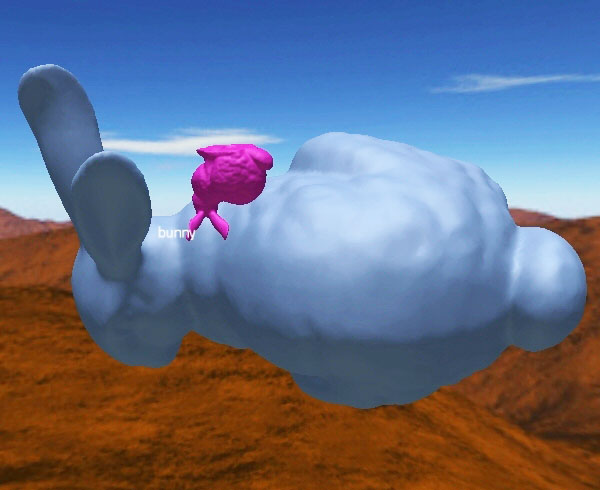,height=2.3cm}}
{\epsfig{file=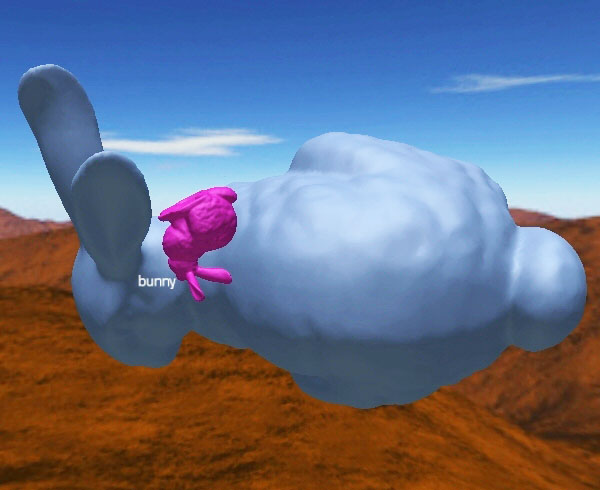,height=2.3cm}}
{\epsfig{file=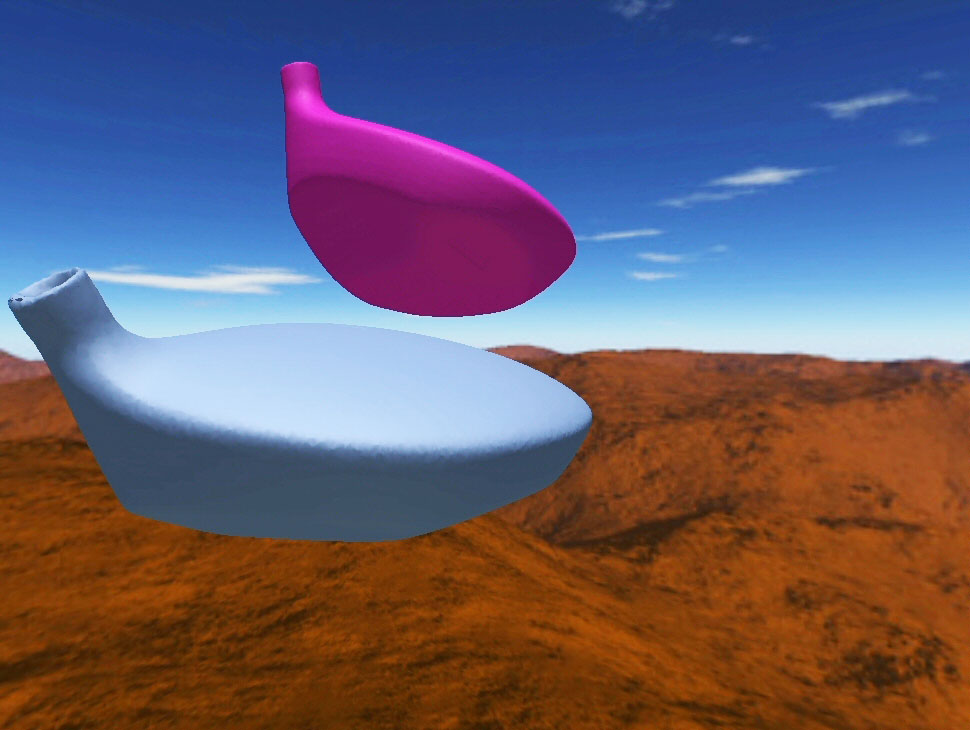,height=2.3cm}}
{\epsfig{file=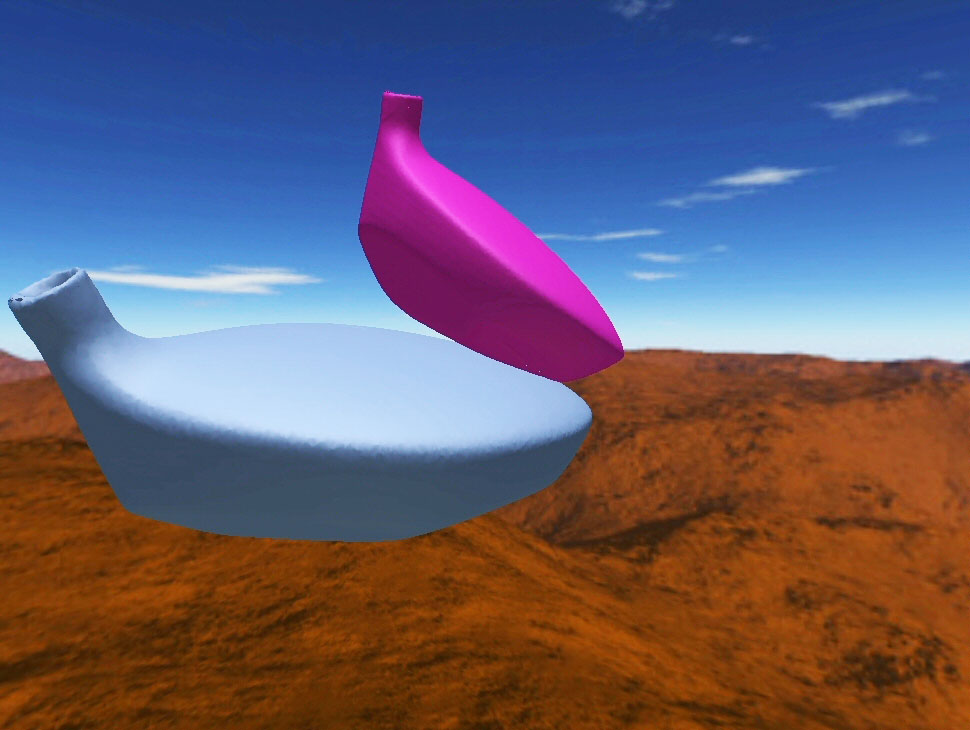,height=2.3cm}}
{\epsfig{file=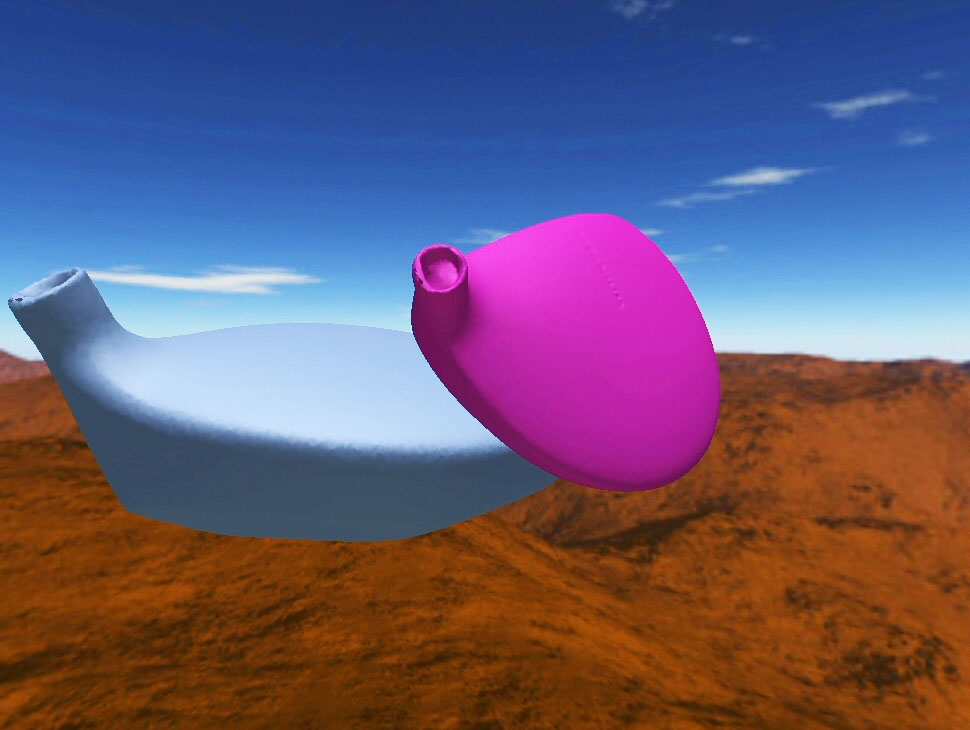,height=2.3cm}}
{\epsfig{file=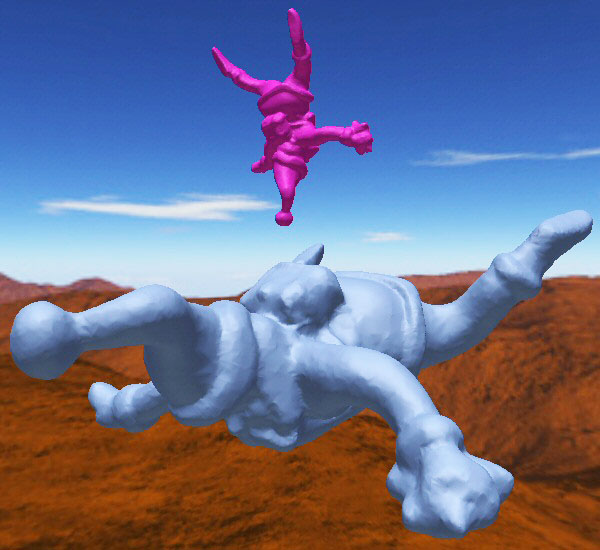,height=2.3cm}}
{\epsfig{file=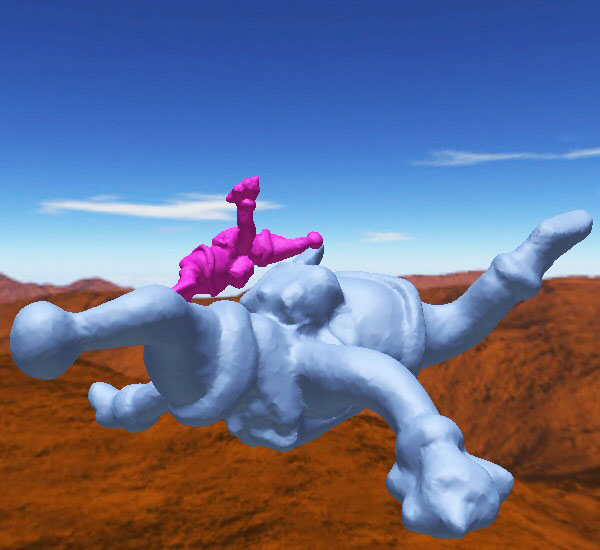,height=2.3cm}}
{\epsfig{file=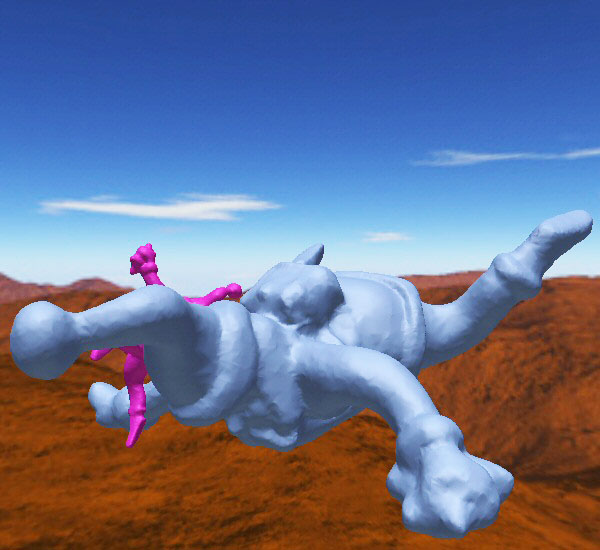,height=2.3cm}}
\caption{\bf Dynamic simulations results of pairs of Torus,
Bunny2, Golf-club and Santa models using impulse-based dynamics
\protect\cite{Guendelmann03sig}. }\label{fig:pulsk}
\end{figure*}

\begin{table}[htb]
 \tbl{{\bf PD performance on predefined paths}}{
\centering 
\begin{tabular}{c||c|c|c}
  \hline
  Models & Time (msec) & No. of Contacts & No. of Iterations \\
  \hline

  \hline
  Spoon and Cup & 0.96 & 4.49 & 1.02 \\

  \hline
  Buddha & 3.50 & 5.07 & 1.29 \\

  \hline
  Dragon & 5.84 & 10.32 & 1.45 \\

  \hline
\end{tabular}}\label{tab:dynamics}
 \end{table}

\subsubsection{Dynamics Scenario}\label{sec:dyn}

We integrated our PD algorithm into rigid-body dynamics
simulations involving the Torus, Bunny2, Golf-club and Santa
models (see Fig.~\ref{fig:pulsk}). We used impulse-based dynamics
\cite{Guendelmann03sig} to simulate rigid-body dynamics based on
the local PD method presented in Sec. \ref{sec:localpd}. The local
PDs are used to stablize the simulation and to resolve contacts
and collisions.
In this scenario, most collisions do not create deep penetrations
since collisions are resolved immediately. As a result, the
average PD computation is shorter, for example,  than that
required for a random configuration. We use maximally clear
configurations, motion coherence and centroid differences to find
initial collision-free configurations.  A performance summary
 is given in Table \ref{tab:pulsk}. Note that additional time is required for local PD computations.

\begin{table}[htb]
 \tbl{\bf Performance in the dynamics scenario.}{
\centering 
\begin{tabular}{c||c|c|c}
  \hline
  Models & Global PD (msec) & Local PD (msec) & Total (msec) 
   \\
  \hline
  \hline
  Torus & 3.13 & 1.04 & 4.17 
  \\
  \hline
  Bunny2 & 5.43 & 1.78 & 7.21 
  \\
  \hline
  Golf-club & 4.87 & 1.67 & 6.54 
  \\
  \hline
  Santa & 9.14 & 2.90 & 12.04 
  \\
  \hline
\end{tabular}}\label{tab:pulsk}
 \end{table}

\subsection{Discussion}

There are some limitations to our algorithm. An object with high geometric and topological complexities may require a large number of iterations to have a convergent solution. Furthermore, our method only
approximates an upper bound, which can depend on the initial
collision-free configuration, as Fig.~\ref{fig:spoon_cup}(b)
shows. Thus the quality of the initial collision-free
configuration is critical. We have presented several methods of
generating an initial collision-free configuration, and now
summarize our experiences:

\begin{enumerate}

\item Maximally clear configurations are suitable for strongly
non-convex objects of high genus, such as the Grate, Torus and Cup
in Figs.~\ref{fig:bisec_grate} and \ref{fig:clear_conf}, since a
collision-free configuration can be located inside a concavity.

\item Sampling-based search is useful for intricate or
interlocking objects such as those shown in Fig.
\ref{fig:bisec_grate}, but it is time-consuming.

\item In a dynamics simulation, such as that shown in
Fig.~\ref{fig:pulsk}, motion coherence can be exploited to provide
a good candidate for an initial configuration by using the
configuration in the frame before a collision occurs. This can be
achieved by caching.

\item The centroid difference is good for near-convex objects,
unless they are deeply penetrating. In that case, we need to try
several directions of back-off.

\item If the preprocessing cost of finding a maximally clear
configurations is unaffordable, and no application-dependent
information is available, random sampling can be employed.
\end{enumerate}

In order to automate the selection of an initial configuration, we
suggest following the above sequence and choose the resulting
configuration closest to the input. However, none of these
strategies guarantees a bound on the PD, since our algorithm uses
local optimization; it may terminate far from the global optimum.
However, the results of Sec. \ref{sec:rand} show that these
strategies work very well in practice, even for very challenging
cases.

\subsection{Comparisons against Recent Approaches}
In this section, we make qualitative comparisons of our algorithm
against recent algorithms related to penetration depth computation
including \cite{Nawratil2009,Allard2010}.

Nawratil et al.'s method \cite{Nawratil2009} iteratively
calculates generalized penetration depth using kinematical
geometry and constrained optimization. However, their method has
the following differences compared to ours:
\begin{itemize}
\item Since the goal of their work is to compute a minimal rigid
body displacement  to separate overlapped objects by using a
kinematical mapping in $\mathbb{R}^{12}$,  their computational
complexity is much higher than ours both in theory and in
practice. It is unclear whether the application of this method to
translational penetration depth computation can perform at
interactive rates like ours. \item They also suggest two methods
for finding an initial collision-free configuration, but this
procedure takes a considerable amount of time to perform, e.g. a
second for a model consisting of a few thousands triangles, which
is too expensive for interactive applications.
\end{itemize}


Allard et al.'s work is also related to penetration resolution
\cite{Allard2010}. However, their method is different from ours in
the following sense:
\begin{itemize}

\item Their work on penetration resolution is based on dynamics
simulation, whereas ours is based on purely geometric formulation,
and they do not have any guarantees on collision resolution,
whereas ours does. \item Their work has no sense of optimality for
collision resolution, whereas ours does. \item Their work may
suffer from the image resolution imposed by the use of GPUs, but
ours does not. \item Their work is suitable for deformable objects
as well as rigid ones, whereas ours is suitable for rigid objects.
\end{itemize}

%% file: concl.tex
\section{Conclusions and Future Work}

We have presented an interactive algorithm for computing the
penetration depth of complicated polygon-soup models. Our method
approximates the local contact space and iteratively performs in-
and out-projections based on a Gauss-Seidel LCP solver and
translational continuous collision detection. We have proposed
several schemes for selecting an initial collision-free
configuration which utilizes motion coherence, centroid
difference, and maximally clear configurations. We showed the
effectiveness of our PD algorithms in various scenarios. We also
presented a method of computing a local PD, and integrated it with
a dynamics simulation.

{\bf Future Work:} We are interested in extending our algorithm to
$n$-body PD problems, in which we need to separate multiple
colliding bodies simultaneously. We would also like to extend our
interactive algorithm to address the generalized PD problem
\cite{gpd,fap,Nawratil2009} in which both translational and
rotational motions are possible. The key to making this problem
amenable to our current framework is to find a way of linearizing
the curved contact space. In addition, we are considering how our
method might benefit other applications, such as haptic rendering
and robot motion planning.